\documentclass[aps,prb,twocolumn,amsmath,amssymb,nofootinbib,superscriptaddress,floatfix]{revtex4-1}
\usepackage{amsmath}
\usepackage{amssymb}
\usepackage{amsthm}
\usepackage{graphicx}

\makeatletter
\usepackage{tikz}
\usetikzlibrary{arrows,matrix,calc,scopes,decorations.markings}

\newcommand{\cblue}[1]{\textcolor{black}{#1}}

\begin{document}

\title{Non-Abelian String and Particle Braiding in Topological Order:\\
 \normalsize{Modular SL(3,$\mathbb{Z}$) Representation and 3+1D Twisted Gauge Theory}} 
\author{Juven C. Wang} \email{juven@mit.edu}  
\affiliation{Department of Physics, Massachusetts Institute of Technology, Cambridge, MA 02139, USA}
\affiliation{Perimeter Institute for Theoretical Physics, Waterloo, ON, N2L 2Y5, Canada}

\author{Xiao-Gang Wen} \email{xwen@perimeterinstitute.ca}
\affiliation{Perimeter Institute for Theoretical Physics, Waterloo, ON, N2L 2Y5, Canada}
\affiliation{Department of Physics, Massachusetts Institute of Technology, Cambridge, MA 02139, USA}


\begin{abstract}
String and particle braiding statistics 
are examined in a class of 
topological orders 
described by discrete gauge theories with a gauge group $G$ and a 4-cocycle twist $\omega_4$ of $G$'s cohomology group $\mathcal{H}^4(G,\mathbb{R}/\mathbb{Z})$
in 3 dimensional space and 1 dimensional time (3+1D).  
We establish the topological spin and the spin-statistics relation for the closed strings, and their multi-string braiding statistics.
The 3+1D twisted gauge theory can be characterized by a representation of a modular transformation group SL$(3,\mathbb{Z})$. 
We express the SL$(3,\mathbb{Z})$ generators $\mathsf{S}^{xyz}$ and $\mathsf{T}^{xy}$ in terms of the gauge group $G$ and the 4-cocycle $\omega_4$.  
As we
compactify one of the spatial 
directions $z$ into a compact circle with a gauge flux $b$ inserted,  
we can use the generators $\mathsf{S}^{xy}$ and $\mathsf{T}^{xy}$ of an SL$(2,\mathbb{Z})$ subgroup to study the dimensional
reduction of the 3D topological order $\mathcal{C}^{3\text{D}}$ to a direct sum of degenerate states of 2D
topological orders $\mathcal{C}_b^{2\text{D}}$ in different flux $b$ sectors: $\mathcal{C}^{3\text{D}} = \oplus_b \mathcal{C}_b^{2\text{D}}$. 
The 2D topological orders $\mathcal{C}_b^{2\text{D}}$ are described by 2D gauge theories of the group $G$ twisted by the
3-cocycles $\omega_{3(b)}$, dimensionally reduced from the 4-cocycle $\omega_4$. 
We show that the SL$(2,\mathbb{Z})$ generators, $\mathsf{S}^{xy}$ and $\mathsf{T}^{xy}$,
fully encode
a particular 
type of three-string braiding statistics with a pattern that is the connected sum of two Hopf links. 
With certain 4-cocycle twists, we discover that, by threading a third string through two-string unlink 
into three-string Hopf-link 
configuration, Abelian two-string braiding statistics is promoted to non-Abelian three-string braiding statistics.
\end{abstract}

\maketitle

\tableofcontents

\section{Introduction}
In the 1986 Dirac Memorial Lectures, Feynman explained the braiding statistics of fermions by demonstrating the plate trick and the belt trick.\cite{Feynman:1987gs}
Feynman showed that the wavefunction of a quantum system obtains a mysterious $(-1)$ sign by exchanging two fermions, which is associated with 
the fact of requiring an extra $2\pi$ twist or rotation to go back to the original state. 
However, it is known that there is richer physics in deconfined topological phases of 2+1D and 3+1D spacetime.\cite{Polyakov:1987ez}
(Here $d+1$D is $d$-dimensional space and $1$-dimensional time, while $d$D is $d$-dimensional space.)
In 2+1D, there are ``anyons'' with exotic braiding statistics for point particles.\cite{Wilczek:1990ik}
In 3+1D, Feynman only had to consider bosonic or fermionic statistics for point particles, without worrying about anyonic statistics.
Nonetheless, there are string-like excitations, whose braiding process in 3+1D can enrich the statistics of deconfined topological phases.
In this work, we aim to systematically address the string and particle braiding statistics in deconfined gapped phases of 3+1D \emph{topological orders}.
\cblue{ 
Namely, we aim to know what statistical phase does the wavefunction of the whole system gain under the string and particle braiding process.}

Since the discovery of 2+1D topological orders\cite{Wtop,WNtop,Wrig}(see Ref.\onlinecite{Wen:2012hm} for an overview), 
we have now gained quite systematic ways to classify and characterize them, 
by using the induced representations of the mapping class group of $\mathbb{T}^2$ torus 
(the modular group SL$(2,\Z)$ and the gauge/Berry phase structure of ground states\cite{Wrig,KW9327,{Wilczek:1984dh}}) 
and the topology-dependent ground state degeneracy,\cite{Wrig,Wang:2012am,Kapustin:2013nva}  
using the unitary fusion
categories,\cite{LWstrnet,CGW1038,GWW1017,KK1251,Fuchs:2013gha,GWW1332,LW1384,{Kong:2014qka}} and using
simple current algebra,\cite{MR9162,BW9215,WW9455,LWW1024} pattern of zeros,
\cite{WW0809,BW0932,SL0604,BKW0608,SY0802,BH0802} and field theories.\cite{BW9045,R9002,FK9169,WZ9290,BM0535}  
Our better understanding of topologically ordered states also holds the promises of applying their rich quantum phenomena,
including fractional statistics\cite{Wilczek:1990ik} and non-Abelian anyons, for topological quantum computation.\cite{2008RvMP...80.1083N}


However, our understanding of 3+1D topological orders is in its infancy and far from systematic. 
This motivates our work attempting to address:\\

\noindent
{\bf Q1}: ``\emph{How to} (\emph{at least partially}) \emph{classify and characterize 3D topological orders}?''\\

\noindent
By {\it classification}, we mean to count the number of distinct phases of topological orders and to give them a proper label.
By {\it characterization}, we mean to describe their properties in terms of physical observables.
Here our approach to study $d$D topological orders is to simply generalize the above 2D approach,
to use the ground state degeneracy (GSD) on $d$-torus $\mathbb{T}^d=(S^1)^d$, 
and the associated representations of the mapping class group of $\mathbb{T}^d$ (recently proposed in Ref.\onlinecite{{Kong:2014qka},{2014arXiv1401.0518M}}),  
\bea
\text{MCG}(\mathbb{T}^d)= \text{SL}(d,\Z). 
\eea 
\cblue{(Refer to Appendix \ref{2D_TO} and Reference cited therein for a brief review of the computation of 2D topological orders.)}
For
3D, the mapping class group SL$(3,\Z)$ is generated by the modular transformation $ \hat{\mathsf{S}}^{xyz}$ and $\hat{\mathsf{T}}^{xy}$:\cite{Coxeter}  
\begin{align} \label{eq:ST3D}
 \hat{\mathsf{S}}^{xyz} = \begin{pmatrix}
0 & 0 & 1 \\
1 & 0 & 0 \\
0& 1 & 0
\end{pmatrix},
\qquad
 \hat{\mathsf{T}}^{xy}= \begin{pmatrix}
1 & 1 & 0 \\
0 & 1 & 0 \\
0 & 0 & 1
\end{pmatrix}.
\end{align}

What are examples of 3D topological orders?
One class of them is described by a discrete gauge
theory with a finite gauge group $G$.  Another class is described by the {\it twisted gauge theory},\cite{Dijkgraaf:1989pz} a gauge theory $G$
with a 4-cocycle twist $\om_4 \in \cH^4(G,\R/\Z)$ of $G$'s fourth cohomology group.  But the twisted gauge
theory characterization of 3D topological orders is not one-to-one: different
pairs $(G,\om_4)$ can describe the same 3D topological order.
In this work, we will use $ \hat{\mathsf{S}}^{xyz}$ and $\hat{\mathsf{T}}^{xy}$ of SL$(3,\Z)$ to characterize 
the topological {\it twisted discrete gauge theory} with finite gauge group $G$, which 
has topology-dependent ground state degeneracy. 
The twisted gauge theories describe a large class of 3D gapped quantum liquids in condensed matter. 
Although we will study the SL$(3,\Z)$ modular data of the ground state sectors of gapped phases, these data can capture
the gapped excitations such as particles and strings. (This strategy is widely-used especially in 2D.) 
There are {\bf two main issues} that we will focus on addressing. 
The first is the {\bf dimensional reduction} from 3D to 2D of SL$(3,\Z)$ modular transformation and cocycles to study 3D topological order.
The second is the {\bf non-Abelian three-string braiding statistics} from a twisted discrete gauge theory of an Abelian gauge group.\\

\begin{figure}[!t] 
\begin{center} 
\includegraphics[scale=0.36]{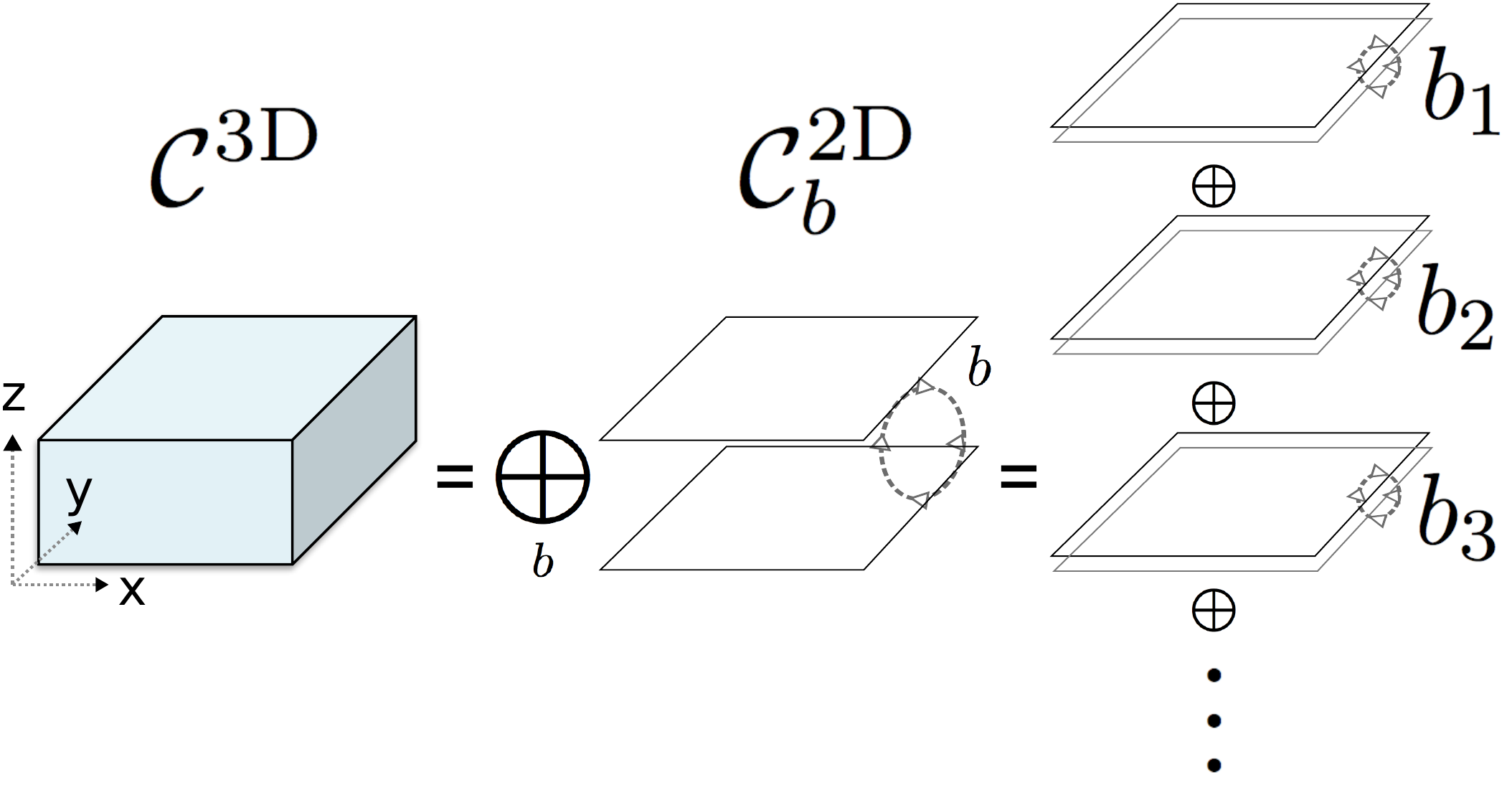} \end{center}
\caption{
A 3D topological order $\cC^{3\tD}$ can be regarded as  the direct sum of 2D topological orders $\cC^{2\tD}_b$ 
in different sectors $b$, as $\cC^{3\tD} = \oplus_b \cC^{2\tD}_b$, when we compactify a spatial direction $z$ into a circle. 
{\bf This idea is general and applicable to $\cC^{3\tD}$ without a gauge theory description.}
However, when $\cC^{3\tD}$ allows a gauge group $G$ description, the $b$ stands for a group element
(or the conjugacy class for the non-Abelian group) of $G$. 
Thus $b$ acts as a gauge flux 
along the arrow - - -$\vartriangleright$ in the compact direction $z$. 
Thus, $\cC^{3\tD}$ becomes the direct sum of different $\cC^{2\tD}_b$ under distinct gauge fluxes $b$.
} 
\label{fig:3Dto2D} 
\end{figure}

\noindent
($\star$1) {\bf Dimensional Reduction from 3D to 2D: for SL$(3,\Z)$ modular $\mathsf{S}$, $\mathsf{T}$ matrices and cocycles} -
For the first issue, our general philosophy is the following:\\

\noindent
``Since 3D topological orders are foreign and unfamiliar to us, we will \emph{dimensionally reduce 3D topological orders to several sectors of 2D topological orders} in the \emph{Hilbert space
of ground states} (\emph{not in the real space}, see Fig.\ref{fig:3Dto2D}). 
Then we will be able to \emph{borrow the more familiar 2D topological orders to understand 3Ds}.''\\

\noindent
We will compute the matrices $\mathsf{S}^{xyz}$ and $\mathsf{T}^{xy}$ that
generate the SL$(3,\Z)$ representation in the quasi-(particle or string)-excitations basis of 3+1D topological order.  
We find an explicit expression of $\mathsf{S}^{xyz}$ and $\mathsf{T}^{xy}$, in terms of the gauge
group $G$ and the 4-cocycle $\om_4$, for both Abelian and non-Abelian gauge
groups.  
(A calculation 
using a different novel approach, 
the universal wavefunction overlap for the normal untwisted gauge theory, is studied in \Ref{MW14}.)
We note that SL$(3,\Z)$ contains a subgroup SL$(2,\Z)$, which is
generated by $\hat{\mathsf{S}}^{xy}$ and $\hat{\mathsf{T}}^{xy}$, where  
\begin{align} \label{eq:Sxy}
\hat{\mathsf{S}}^{xy} = \begin{pmatrix}
0 & -1 & 0 \\
1& 0 & 0 \\
 0& 0 & 1
\end{pmatrix}.
\end{align}
In the most generic cases of topological orders (potentially {\it without a gauge group description}), the 
matrices $\mathsf{S}^{xy}$ and $\mathsf{T}^{xy}$ can still be block
diagonalized as the sum of several sectors in the quasi-excitations basis, each sector carrying an index of $b$,
\begin{align} \label{eq:3Dto2DST}
 \mathsf{S}^{xy}=\oplus_b \mathsf{S}^{xy}_b, \ \ \ \  \
 \mathsf{T}^{xy}=\oplus_b \mathsf{T}^{xy}_b,
\end{align}
The pair $(\mathsf{S}^{xy}_b,\mathsf{T}^{xy}_b)$, generating an SL$(2,\Z)$ representation,
describes a 2D topological order $\cC^{2\tD}_b$.
This leads to a dimension reduction of the 3D topological order $\cC^{3\tD}$:
\begin{align} \label{eq:C3DtoC2D}
 \cC^{3\tD} = \oplus_b \cC^{2\tD}_b .
\end{align}
In the more specific case, when the topological order allows a gauge group $G$ description which we focus on here, 
we find that the $b$ stands for a gauge flux for group $G$ (Namely, $b$ is a group element for an Abelian $G$, 
while $b$ is a conjugacy class for a non-Abelian $G$). 

The physical picture of the above dimensional reduction is the following (see Fig.\ref{fig:3Dto2D}): 
If we compactify one of the 3D spatial directions (say the $z$ direction) into a small circle,
the 3D topological order $\cC^{3\tD}$ can be viewed as a direct sum of 2D topological orders
$\cC^{2\tD}_b$ with (accidental) degenerate ground states at the lowest energy.

\onecolumngrid
\begin{center} 
\begin{figure}[!ht] 
\includegraphics[scale=0.46]{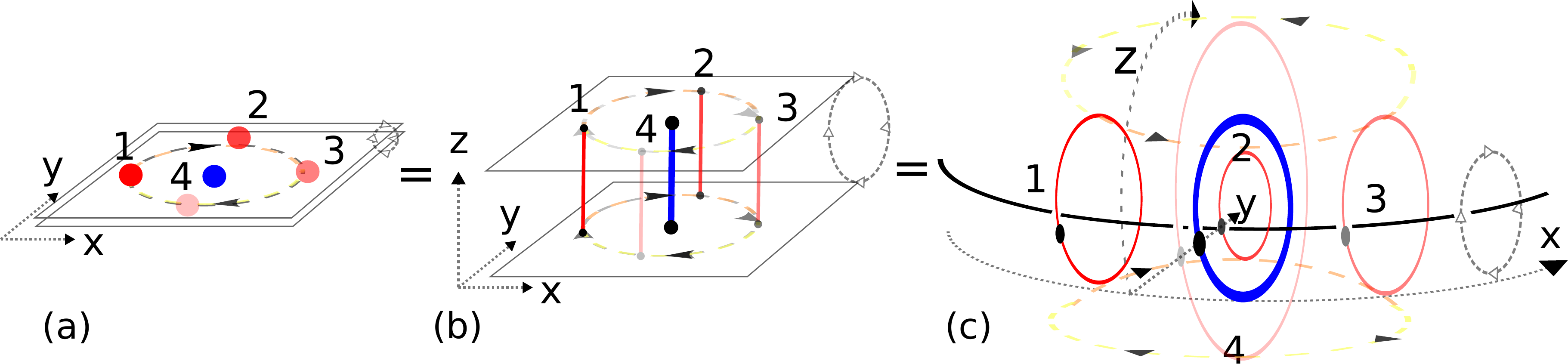} 
\caption{
Mutual braiding statistics following the path $1 \to 2\to 3\to 4 $ along the time evolution (see Sec.\ref{Sec:IIIc3string}): 
(a) From a 2D viewpoint of dimensional reduced $\cC^{2\tD}_b$, the 
$2\pi$ braiding 
of two particles is shown.
(b) The compact $z$ direction extends two particles to two closed (red, blue) strings.
(c) An equivalent 3D view, the $b$ flux (along the arrow - - -$\vartriangleright$)  
is regarded as the monodromy caused by a third (black) string.
We identify the coordinates 
$x,y$ and a compact $z$ to see that a full-braiding process
is the {\it one (red) string going inside to the loop of another (blue) string, and then going back from the outside.}
{For Abelian topological orders, the mutual braiding process between two excitations (A and B)
in Fig.\ref{fig:3strings_2D_3D_xy}(a) yields a statistical Abelian phase 
${e^{\ti \theta_{\text{(A)(B)} }} \propto {\sfS}^{xy}_{\text{(A)(B)}}}$ proportional to the 2D's $\mathsf{S}^{xy}$ matrix. 
The dimensional-extended equivalent picture Fig.\ref{fig:3strings_2D_3D_xy}(c) implies that the loop-braiding 
yields a phase ${e^{\ti \theta_{\text{(A)(B)}, b}} \propto {\sfS}^{xy}_{b\;\text{(A)(B)}}}$ of Eq.(\ref{eq:Sxyb}) (up to a choice of canonical basis), where $b$ is the flux of the black string.
We clarify that in both (b) and (c) our strings may carry both flux and charge. 
If a string carries only a pure charge, then it is effectively a point particle in 3D.
If a string carries a pure flux, then it is effectively a loop of a pure string in 3D.
If a string carries both charge and flux (as a dyon in 2D), then it is a loop with string fluxes attached with some charged particles in 3D.
Therefore our Fig.\ref{fig:3strings_2D_3D_xy}(c)'s string-string braiding actually represents several braiding processes: the particle-particle, particle-loop and loop-loop braidings, 
all processes are threaded with a background (black) string. 
}
} 
\label{fig:3strings_2D_3D_xy} 
\end{figure}
 \end{center}
\twocolumngrid

In this work, we will focus on a generic finite Abelian gauge group $G=\prod_i Z_{N_i} $ (isomorphic to products of cyclic groups) with
generic cocycle twists from the group cohomology.\cite{Dijkgraaf:1989pz} 
We examine the 3+1D twisted gauge theory twisted by 4-cocycle $\omega_4 \in \cH^4(G,\R/\Z)$, and 
reveal that it is a direct sum of 2+1D twisted gauge theories twisted by a dimensionally-reduced 3-cocycle  $\omega_{3(b)} \in \cH^3(G,\R/\Z)$ of
$G$'s third cohomology group, namely
\bea \label{eq:C3DtoC2Dom}
&& \cC^{3\tD}_{G,\omega_4} = \oplus_b \cC_{G_b,\omega_{3(b)}}^{2\tD}.   
\eea
Surprisingly, even for an Abelian group $G$, we find such a \emph{twisted Abelian gauge theory} can be dual to a twisted or untwisted \emph{non-Abelian gauge theory}. 
We study this fact for 3D as an extension of the 2D examples of Ref.\onlinecite{deWildPropitius:1995cf}. 
By this equivalence, 
we are equipped with (both untwisted and twisted) non-Abelian gauge theory to study its non-Abelian braiding statistics.

\noindent
($\star$2) {\bf Non-Abelian three-string braiding statistics} -
We are familiar with the 2D braiding statistics: 
there is only \emph{particle-particle braiding}, which yields
bosonic, fermionic or anyonic statistics by braiding a particle around another particle.\cite{Wilczek:1990ik}
We find that the 3D topological order introduces both particle-like and string-like excitations.  
We aim to address the question:\\

\noindent
{\bf Q2}: ``\emph{How to characterize the braiding statistics of strings and particles in 3+1D topological orders}?''\\

\noindent
The possible braiding statistics in 3D learned in the past literature are:\\
(i) \emph{particle-particle braiding} can only be bosonic or fermionic due to no nontrivial braid group in 3D for point particles.\\
(ii) \emph{particle-string braiding}, which is Aharonov-Bohm effect of $\Z_N$ gauge theory, where
a particle as $\Z_N$ charge braiding around a string (or a vortex line) as $\Z_N$ flux, obtaining a $e^{\ti \frac{2 \pi}{N}}$ phase of statistics.\cite{Wilczek:1990ik,{deWildPropitius:1995hk}} \\
(iii) \emph{string-string braiding}, where a closed string (a red loop), shown in Fig.\ref{fig:3strings_2D_3D_xy}(c) excluding the background black string, 
wrapping around a blue loop. 
The related idea known as loop-loop braiding forming the \emph{loop braid group} has been proposed mathematically.\cite{Baez:2006un}
(See also some earlier studies in Ref.\onlinecite{{Aneziris:1990gm},{Alford:1992yx}}.)

However, we will 
address that there are some extra \emph{new braiding statistics between three closed strings}:\\
(iv) \emph{{three-string braiding}}, shown in Fig.\ref{fig:3strings_2D_3D_xy}(c), 
where a closed string (a red loop) wrapping around another closed string (a blue loop) but the two loops both
are threaded by a third loop (the black string).
This braiding configuration is discovered recently by Ref.\onlinecite{WL1437}, also 
a related work in Ref.\onlinecite{JMR1462} for a twisted Abelian gauge theory. 

The new ingredient of our work on braiding statistics can be summarized as follows:
We consider the string and particle braiding of general twisted gauge theories with the most generic finite Abelian gauge group $G=\prod_u Z_{N_u}$, labeled by the data $({G,\omega_4})$. 
We provide a 3D to 2D dimensional reduction approach to realize the three-string braiding statistics of Fig.\ref{fig:3strings_2D_3D_xy}. 
We firstly show that the SL$(2,\Z)$ representations
$(\mathsf{S}^{xy}_b,\mathsf{T}^{xy}_b)$ fully encode this particular type of Abelian three-closed-string statistics shown in
Fig.\ref{fig:3strings_2D_3D_xy}. 
We further find that, for a twisted gauge theory with
an Abelian $(Z_N)^4$ group, certain 4-cocycles (named as Type IV 4-cocycles) will make the twisted theory to be a
non-Abelian theory. More precisely, {\bf while the two-string braiding statistics of unlink 
is Abelian,
the three-string braiding statistics of Hopf links, 
obtained from threading the two strings with a third string, will become non-Abelian.}
We also demonstrate that $(\mathsf{S}^{xy}_b$ 
encodes this 
three-string braiding statistics.

Our article is organized as follows. In Sec.\ref{sec:II}, \cblue{we address a third question:
{\bf Q3}: ``\emph{How to formulate or construct certain 3+1D topological orders on the lattice}?''}
We outline a lattice formulation of twisted gauge theories in terms of 
3D twisted quantum double models, which generalize the Kitaev's 2D toric code and quantum double models.
Our model is the lattice Hamiltonian formulation of Dijkgraaf-Witten theory,\cite{Dijkgraaf:1989pz} and we provide the \emph{spatial lattice} as well
as the \emph{spacetime lattice path integral} pictures.
In Sec.\ref{sec:RepST}, 
\cblue{we answer {\bf Q4}: ``\emph{What are the generic expressions of SL(3,$\Z$) modular data}?''}
We compute the modular SL$(3,\Z)$ representations of $\sfS$, $\sfT$ matrices, using both
the spacetime path integral approach and the Representation Theory approach.
In Sec.\ref{Sec:physicsST} and \ref{Sec:SL3ZMulti-String Braiding}, 
\cblue{we address:
{\bf Q5}: ``\emph{What is the physical interpretation of} SL(3,$\Z$) \emph{modular data in 3D}?''} 
We use the modular SL$(3,\Z)$ data to characterize the braiding-statistics of particles and strings.
In Sec.\ref{summary}, we discuss the link and knot patterns of string-braiding systematically, and end with a conclusion.
In addition to the main text, we organize the following information in 
the Supplemental Material: 
(i) group cohomology and cocycles; 
(ii) projective representation;
(iii) some examples of classification of topological orders;
(iv) direct calculations of $\sfS, \sfT$ using cocycle path integrals.\\


\noindent(NOTE: We adopt the name of {\it strings} 
for the vision to incorporate the excitations from both the closed strings (loops) and open strings. Such excitations can have fusion or braiding process.
In this work, however, we only focus on the closed string case.
Our notation for finite cyclic group is either $Z_N$ or $\mathbb{Z}_N$, though they are equivalent 
mathematically.
We denote $Z_N$ for the gauge group $G$, the discrete gauge $Z_N$ flux, or the $Z_N$ variables,   
while $\mathbb{Z}_N$ only for the classes of group cohomology or topological order classification. 
We denote ${\gcd(N_i,N_j)}\equiv N_{ij}$, $\gcd(N_i,N_j,N_k) \equiv N_{ijk}$, $\gcd(N_i,N_j,N_k,N_l) \equiv N_{ijkl}$,  
with gcd stands for the greatest common divisor. We also have $|G|$ as the order of the group, and $\R/\Z =$ U(1). 
We may use subindex $n$ for $\omega_n$ to indicate $n$-cocycle.
In principle, 
we will use {\bf types} to count the number of cocycles in cohomology groups. 
But we will use {\bf classes} to count the number of distinct phases in topological orders.
Normally {\bf the types overcount the classes}.
We use the hat symbol $ \hat{\mathsf{S}}^{}$ and $\hat{\mathsf{T}}^{}$ for the modular matrices acting on the real space in $x,y,z$ directions, 
so $\hat{\mathsf{S}}^{xyz}\cdot (x,y,z)=(z,x,y)$ and $\hat{\mathsf{T}}^{xy}\cdot (x,y,z)=(x+y,y,z)$;
while we denote the symbols $\mathsf{S},\mathsf{T}$ for modular matrices in the quasi-excitations basis.)

\section{Twisted Gauge Theory and Cocycles of Group Cohomology} \label{sec:II}

In this section, we aim to address the question:

\noindent
{\bf Q3}: ``\emph{How to formulate or construct certain 3+1D topological orders on the lattice}?''

We will consider 3+1D twisted discrete gauge theories. 
Our motivation to study the discrete gauge theory is that it is 
topological and exhibits Aharonov-Bohm phenomena (see \Ref{{Wilczek:1990ik},{deWildPropitius:1995hk}}). 
One approach to formulate a discrete gauge theory is the lattice gauge theory.\cite{Kogut:1979wt}
A famous example in both high energy and condensed matter communities 
is the $Z_2$ discrete gauge theory in 2+1D (or named as $Z_2$ toric code, $Z_2$ spin liquids, $Z_2$ topological order\cite{Wen:2004ym}). 
Kitaev's toric code and quantum double model\cite{Kitaev:1997wr} provides a simple Hamiltonian,
\bea  \label{eq:toric}
{H}=-\sum_v A_v -\sum_{p} B_p,
\eea
where a space lattice formalism is used, and $A_v$ is the vertex operator acting on the vertex $v$, $B_p$ is the plaquette (or face) term to ensure the
zero flux condition on each plaquette. Both $A_v, B_p$ consist of only Pauli spin operators for the $Z_2$ model. 
Such ground states of the Hamiltonian is found to be $Z_2$ gauge theory with $|G|^2=4$-fold topological degeneracy on the $\mathbb{T}^2$ torus.
Its generalization to a {\it twisted $Z_2$ gauge theory} is the $Z_2$ double-semions model, captured by the framework of Levin-Wen {\it string-net model}.\cite{LWstrnet,Wen:2004ym}

\subsection{Dijkgraaf-Witten topological gauge theory}

For a more generic twisted gauge theory, there is indeed another 
way using the spacetime lattice formalism to construct them 
by the Dijkgraaf-Witten topological gauge theory.\cite{Dijkgraaf:1989pz} There one can formulate the path integral $\mathbf{Z}$ (or partition function) 
of a $(d+1)$D gauge theory ($d$ dimensional space, 1 dimensional time) of a gauge group $G$ as, 
\bea \label{eq:path integral}
\mathbf{Z}& =&\sum_\gamma e^{\ti S[\gamma]}=\sum_\gamma e^{\ti 2\pi \langle \omega_{d+1}, \gamma(\cM_{\text{tri}})\rangle(\text{mod}{2\pi})} \nonumber\\
& =&\frac{|G|}{|G|^{N_v}} \frac{1}{|G|} \sum_{\{ g_{ab}\}} \prod_i (\omega_{d+1}{}^{\epsilon_i}(\{ g_{ab}\})) \mid_{v_{c,d} \in T_i}
\eea
where we sum over all mapping $\gamma: \cM \to BG$, from the spacetime manifold $\cM$ to $BG$, the classifying space of $G$.
In the second equality, we triangulate $\cM$ to $\cM_{\text{tri}}$ with the edge $[v_a v_b]$ connecting the vertex $v_a$ to the vertex $v_b$. 
The action $ \langle \omega_{d+1}, \gamma(\cM_{\text{tri}})\rangle$ evaluates the cocycles $\omega_{d+1}$ on the spacetime $(d+1)$-complex $\cM_{\text{tri}}$. 
By the relation between the topological cohomology class of $BG$ and the cohomology group of $G$: 
$H^{d+2}(BG,\Z)=\cH^{d+1}(G,\R/\Z)$,\cite{Dijkgraaf:1989pz,Hung:2012dx}
we can simply regard $\omega_{d+1}$ as the ${d+1}$-cocycles of the cohomology group $\cH^{d+1}(G,\R/\Z)$ (see more details in Appendix \ref{App:cocycles}).
The group elements $g_{ab}$ are assigned at the edge $[v_a v_b]$. 
The $|G|/|G|^{N_v}$ factor is to mod out the redundant gauge equivalence configuration, with the number of vertices $N_v$.
Another extra $|G|^{-1}$ factor mods out the group elements evolving in the time dimension.
The cocycle $\omega_{d+1}$ is evaluated on all the $d+1$-simplex $T_i$ (namely a $d+2$-cell) triangulation of the spacetime complex.
In the case of our 3+1D, we have 4-cocycle $\omega_{4}$ evaluated at the 4-simplex (or 5-cell) as 
\bea
&&{ \foursimplex{0}{1}{2}{3}{4}}  \label{eq:4simplex} 
=\omega_4{}^{\epsilon}(g_{01},g_{12}, g_{23}, g_{34}). \label{}
\eea
Here the cocycle $\omega_{4}$ satisfies cocycle condition: $\delta \omega_{4}=1$, which ensures the
path integral $\mathbf{Z}$ on the 4-sphere ${S}^4$ (the surface of the 5-ball) will be trivial as 1. This is a feature of topological gauge theory.
The $\epsilon$ is the $\pm$ sign of the orientation of 4-simplex,
which is determined by the sign of the volume determinant of 4-simplex evaluated by $\epsilon=\sgn(\det(\vec{01}, \vec{02}, \vec{03}, \vec{04}))$. 

\begin{figure}[!h] 
\includegraphics[scale=0.36]{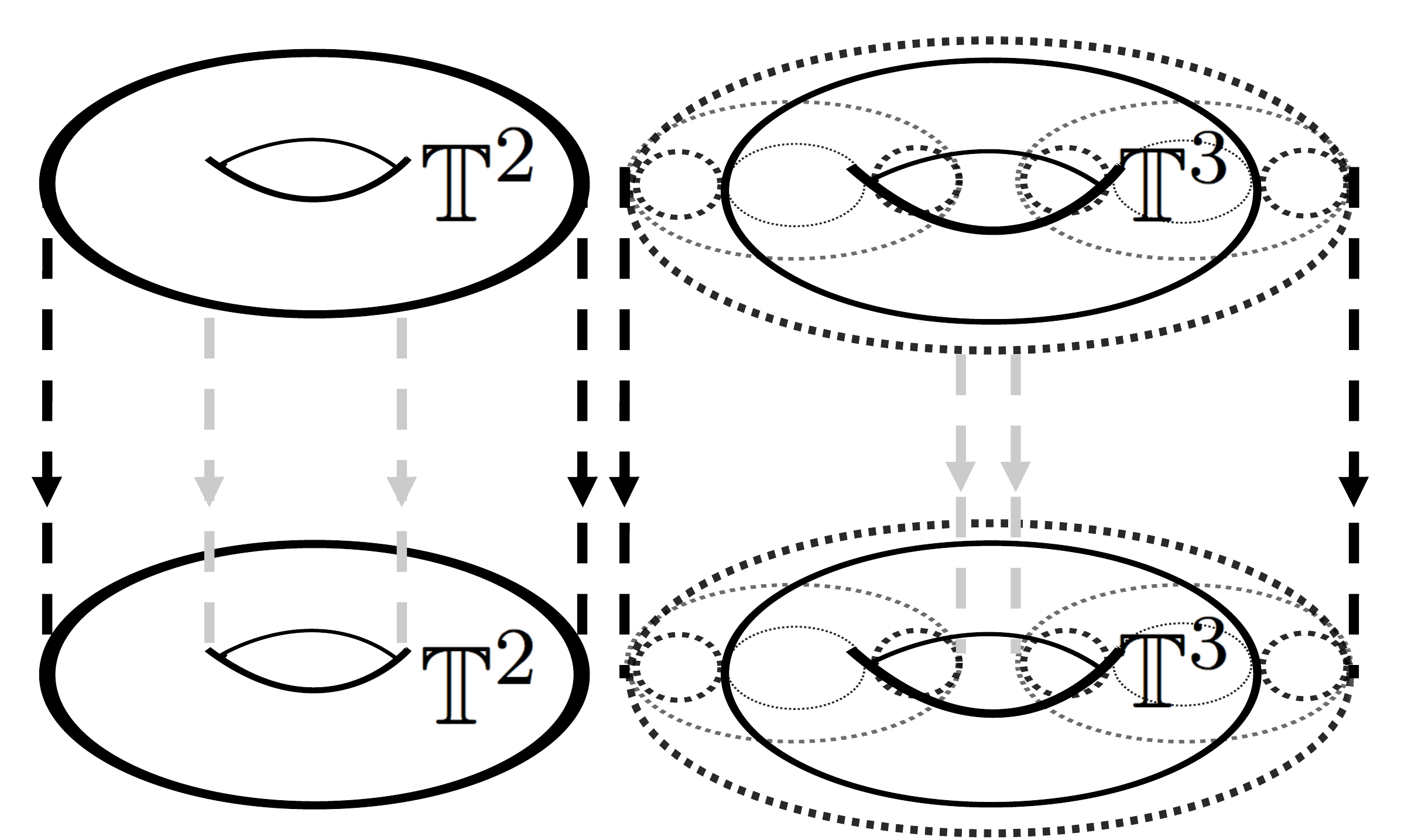}
\caption{
The illustration for ${\sfO}_{\text{(A)(B)}}= \langle  \Psi_{\text{A}}  |  \hat{\sfO}  | \Psi_{\text{B}}\rangle$.
The evolution from an initial state configuration $|\Psi_{in} \rangle$ on the spatial manifold (from the top) 
along the time direction (the dashed line - - -)  to the final state $|\Psi_{out} \rangle$ (on the bottom). 
For the spatial $\mathbb{T}^{d}$ torus, the mapping class group MCG$(\mathbb{T}^{d})$ is the modular SL$(d,\Z)$ transformation.
We show schematically the time evolution on the spatial $\mathbb{T}^{2}$, and $\mathbb{T}^{3}$ (shown as $\mathbb{T}^{2}$ attached a $S^1$ circle on each point).
} 
\label{torus2D3D} 
\end{figure}

{
We utilize Eq.(\ref{eq:path integral}) to calculate the path integral amplitude from an initial state configuration $|\Psi_{in} \rangle$ on the spatial manifold evolving
along the time direction  to the final state $|\Psi_{out} \rangle$, see Fig.\ref{torus2D3D}.
In general, the calcuation can be done for the mapping class group MCG on any spatial manifold $\cM_{space}$ as MCG$(\cM_{space})$. 
Here we focus on $\cM_{space}=\mathbb{T}^3$ and MCG$(\mathbb{T}^3)=$ SL$(3,\Z)$, as the modular transformation.
We first note that $|\Psi_{in} \rangle= \hat{\mathsf{O}} | \Psi_{\text{B}} \rangle$, such a generic SL$(3,\Z)$ transformation $\hat{\mathsf{O}}$ 
under SL$(3,\Z)$ representation can be absolutely generated by $\hat{\sfS}^{xyz}$ and $\hat{\sfT}^{xy}$ of Eq.(\ref{eq:ST3D}),\cite{Coxeter}
thus $\hat{\mathsf{O}}=\hat{\mathsf{O}}(\hat{\sfS}^{xyz},\hat{\sfT}^{xy})$ as a function of $\hat{\sfS}^{xyz},\hat{\sfT}^{xy}$.
The calculation of the modular SL$(3,\Z)$ transformation from $| \Psi_{in}\rangle$ to $| \Psi_{out}\rangle=| \Psi_{\text{A}} \rangle$ 
by filling the 4-cocycles $\omega_4$ into the spacetime-complex-triangulation renders the amplitude of the matrix element ${\sfO}_{\text{(A)(B)}}$:  
\bea \label{eq:Oamp}
&&{\sfO}({\sfS}^{xyz},{\sfT}^{xy})_{\text{(A)(B)}}= \langle  \Psi_{\text{A}}  | \hat{\sfO}(\hat{\sfS}^{xyz},\hat{\sfT}^{xy})  | \Psi_{\text{B}}\rangle,
\eea
both space and time are discretely triangulated, so this is a spacetime lattice formalism.}

\subsection{Canonical basis and the generalized twisted quantum double model $D^{\omega}(G)$ to 3D triple basis} \label{Sec:TQD} 
So far we answer the question \emph{{\bf Q3}} using the \emph{spacetime-lattice path integral}.
Our next goal is to \emph{construct its Hamiltonian on the space lattice},
and to find a good basis representing its quasi-excitations, 
such that we can efficiently read the information of ${\sfO}({\sfS}^{xyz},{\sfT}^{xy})$ in this \emph{canonical basis}.
We will outline the \emph{twisted quantum double model} generalized to 3D as the exactly soluble model in the next subsection, 
where the {canonical basis} can diagonalize its Hamiltonian.\\

\noindent
{\bf Canonical basis} - 
For a gauge theory with the gauge group $G$, one may naively think that a good basis for the amplitude Eq.(\ref{eq:Oamp}) is the group elements 
$| g_x, g_y, g_z \rangle$, with $g_i \in G$
as the flux labeling three directions of $\mathbb{T}^3$. 
However, this flux-only label $| g_x, g_y \rangle$ is known to be improper on $\mathbb{T}^2$ torus already - the canonical basis labeling particles in 2D
is $| \alpha, a \rangle$, requiring both the charge $\alpha$ (as the representation) and the flux $a$ (the group element or the conjugacy class of $G$).
We propose the proper way to label excitations for a 3+1D twisted discrete gauge theory for any finite 
group $G$
in the canonical basis requires one charge $\alpha$ and two fluxes $a,b$:
\bea \label{eq:3Dbasis}
|\alpha, a, b \rangle=\frac{1}{\sqrt{|G|}} \sum_{\substack{ {g_y \in C^a,g_z \in C^b}\\{g_x \in Z_{g_y} \cap Z_{g_z}  }} } \Tr[\widetilde{\rho}_\alpha^{g_y,g_z} (g_x) ] |  g_x, g_y, g_z \rangle.\;\;\;\;\;\;
\eea
which is the finite group discrete Fourier transformation on $|  g_x, g_y, g_z \rangle$. 
This is a 
generalization of the 2D result in \Ref{deWildPropitius:1995cf} and a very recent 3D Abelian case in \Ref{JMR1462}.
Here $\alpha$ is the charge of the representation (Rep) label, which is the $\sfC^{(2)}_{a,b}$ Rep of the centralizers $Z_a$, $Z_b$ of the conjugacy classes $C^a, C^b$.
(For Abelian $G$, the conjugacy class is the group element, and the centralizer is the full $G$.)
$\sfC^{(2)}_{a,b}$ Rep means an inequivalent unitary irreducible projective representation of $G$.
The $\widetilde{\rho}_{\alpha}^{a,b}(c)$ labels this inequivalent unitary irreducible projective $\sfC^{(2)}_{a,b}$ Rep of $G$.
The $\sfC^{(2)}_{a,b}$ is an induced 2-cocycle, dimensionally-reduced from the 4-cocycle $\omega_4$. We illustrate $\sfC^{(2)}_{a,b}$ in terms of geometric pictures in Fig.\ref{fig:Cab}.
The $\tilde{\rho}_\alpha^{g_y,g_z} (g_x)$ are determined by the $\sfC^{(2)}_{a,b}$ projective representation formula:
\bea
\label{eq:CabRep}
\widetilde{\rho}_{\alpha}^{a,b}(c)\widetilde{\rho}_{\alpha}^{a,b}(d)=\sfC^{(2)}_{a,b}(c,d)\widetilde{\rho}_{\alpha}^{a,b}(c  d). 
\eea
The trace term $\Tr[\widetilde{\rho}_\alpha^{g_y,g_z} (g_x) ]$ is named as {\it the character} in the math literature.
One can view the charge $\alpha_x$ along $x$ direction, the flux $a,b$ along the $y,z$. 
Other details and the calculations of $\sfC^{(2)}_{a,b}$ Rep with many examples can be found in Supplemental Material. 

\begin{figure}[h!]
\centering
  \begin{align}
&\sfC^{}_{a} (b,c): \begin{matrix}\CaRep\end{matrix} \label{fig:Cabc}\\
&\sfC^{(2)}_{a,b} (c,d): \begin{matrix}\CabRepL  \end{matrix} \underset{\DashedArrow}{t \; (d)} \begin{matrix}  \CabRepR \end{matrix} \label{fig:Cabcd},
  \end{align}
\caption{
The reduced 2-cocycle $\sfC^{}_{a} (b,c)$ from 3-cocycle $\omega_3$ in Eq.(\ref{fig:Cabc}), 
which triangulates a half of $\mathbb{T}^2$ and with a time interval $I$.
The reduced 2-cocycle $\sfC^{}_{a} (b,c)$ from 4-cocycle $\omega_4$ in Eq.(\ref{fig:Cabcd}), 
which triangulates a half of $\mathbb{T}^3$ and with a time interval $I$.
The dashed arrow $\DashedArrow$ stands for the time $t$ evolution.  }
\label{fig:Cab}
\end{figure} 

We firstly 
recall that, in 2D, a reduced 2-cocycle $\sfC^{}_{a} (b,c)$ comes from a slant product $ i_a \omega(b,c)$ of 3-cocycles,\cite{deWildPropitius:1995cf} 
which is geometrically equivalent to filling three 3-cocycles in a triangular prism of Eq.(\ref{fig:Cabc}).  This is known to
present the {\bf projective representation} 
$\widetilde{\rho}_{\alpha}^{a}(b)\widetilde{\rho}_{\alpha}^{a}(c)=\sfC^{}_{a}(b,c)\widetilde{\rho}_{\alpha}^{a}(bc)$,
because the induced 2-cocycle belongs to the second cohomology group $\cH^2(G,\R/\Z)$.\cite{deWildPropitius:1995cf,Chen:2011pg,Essin:2013rca,Mesaros:2012yd}
(See its explicit triangulation and a novel use of projective representation in Sec VI.B. of Ref.\onlinecite{Wang:2014tia}.)

Similarly, in 3D, a reduced 2-cocycle $\sfC^{}_{a} (b,c)$ from doing {\it twice} of slant products of 4-cocycles forming the geometry of Eq.(\ref{fig:Cabcd}), and renders
\bea \label{eq:2slant}
\sfC^{(2)}_{a,b}= i_b ( \sfC_a (c,d))=i_b ( i_a \omega(c,d)),
\eea 
present the $\sfC^{(2)}_{a,b}$-{projective representation} in Eq.(\ref{eq:CabRep}), 
which $\widetilde{\rho}_{\alpha}^{a,b}(c)$: $(Z_a, Z_b)$ ${\rightarrow}$ $\text{GL}\left(Z_a,Z_b\right)$ can be written
as a matrix in the general linear group ($\text{GL}$). This 3D generalization for the canonical basis in Eq.(\ref{eq:3Dbasis}) is not only natural, but also
consistent to 2D when we turn off the flux along $z$ direction (e.g. set $b=0$). which reduces 3D's $|\alpha, a, b \rangle$ to $|\alpha, a \rangle$ of the 2D case. \\

\noindent
{\bf Generalizing 2D twisted quantum double model $D^{\omega}(G)$ to 3D: twisted quantum triple model?} -- 
A natural way to combine the Dijkgraaf-Witten theory with Kitaev's quantum double model Hamiltonian approach will enable us to study the
Hamiltonian formalism for the twisted gauge theory, which is achieved 
in Ref.\onlinecite{Hu:2012wx},\onlinecite{Mesaros:2012yd} for 2+1D, 
named as the {\it twisted quantum double model}.
In 2D, the widely-used notation $D^{\omega}(G)$ implying the twisted quantum double model with 
its gauge group $G$ and its cocycle twist $\omega$.
It is straightforward to generalize their results to 3+1D. 

\begin{figure}[h!]
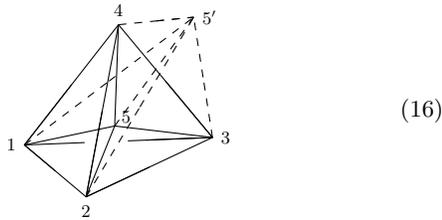

\centering
  \begin{align}
& \begin{matrix}\TQDthreeD{1}{2}{3}{4}{5'}{5} \end{matrix}
  \end{align}
\caption{The vertex operator $A_v$ for the generalized twisted quantum double model in 3D.
To evaluate $A_v$ operator acting on the vertex $5$, one effectively lifts $5$ to $5'$, and fill 4-cocycles $\omega$ into this geometry to compute the
amplitude $\mathbf{Z}$ in Eq.(\ref{eq:path integral}). For this specific 
3D spatial lattice surrounding vertex $5$ by $1,2,3,4$ neighbored vertices, 
there are four 4-cocycles $\omega$ filling
in the amplitude of $A_5^{[55']}$.}
\label{TQD3D_A}
\end{figure}

To construct the Hamiltonian on the 3D spatial lattice,
we follow \Ref{Hu:2012wx} with the form of twisted quantum double model Hamiltonian of Eq.(\ref{eq:toric}) and put the system on the $\mathbb{T}^3$ torus. 
However, some modification 
for 3D are adopted: the vertex operator $A_v=|G|^{-1} \sum_{[vv']=g \in G } A_v^g$ acting on the vertices of the lattice 
by lifting the vertex point $v$ to $v'$ living in an extra (fourth) dimension as Fig.\ref{TQD3D_A}, 
and one computes the 4-cocycle filling amplitude as $\mathbf{Z}$ in Eq.(\ref{eq:path integral}).
A plaquette operator $B_p^{(1)}$ still enforces the zero flux condition on each 2D face (a triangle $p$) spanned by three edges of a triangle.
This will ensure the zero flux on each face (along the Wilson loop of a 1-form gauge field).
Moreover, zero flux conditions are required if higher form gauge flux are presented. For example,
for 2-form field, one shall add an additional $B_p^{(2)}$ to ensure the zero flux on a 3-simplex (a tetrahedron $p$).
Thus, $\sum_p B_p^{}$ in Eq.(\ref{eq:toric}) becomes $\sum_p B_p^{(1)}+ \sum_p B_p^{(2)} +\dots$

Analogous to \Ref{Hu:2012wx}, the local operators $A_v$, $B_p$ of the Hamiltonian have nice commuting properties:  
$ [A_v^{g}, A_u^{h}]=0$ { if } $v \neq u$, $ [A_v^{g}, B_p]= [B_p, B_p']=0$,
also $A_v^{g=[vv']} A_{v'}^h=A_v^{gh}$. 
Notice that $A_g$ defines a ground sate projection operator $\sfP_v=|G|^{-1} \sum_g A^g_v$ if we consider a $\mathbb{T}^3$ torus triangulated in a cube with only a point $v$ 
(all eight points are identified).
It can be shown that both $A_g$ and $\sfP$ as projection operators projecting other states to the ground state $|\alpha, a, b \rangle$,
and $\sfP |\alpha, a, b \rangle=|\alpha, a, b \rangle$ and 
$A_v |\alpha, a, b \rangle \propto |\alpha, a, b \rangle$. 
Since $[A_v^{g}, B_p]=0$, one can simultaneously diagonalize the Hamiltonian Eq.(\ref{eq:toric}) by this {\it canonical basis} $|\alpha, a, b \rangle$ 
as the {\it ground state basis}.

A similar 3D model has been studied recently in \Ref{JMR1462}. 
There the zero flux condition is imposed in both the vertex operator as well as the plaquette operator.
Their Hilbert space thus is more constrained than \Ref{Hu:2012wx} and ours. 
However, in the ground state sector, we expect the physics is the same.
It is less clear to us whether such a name, {\bf twisted quantum double model} and its notation $D^{\omega}(G)$, are still proper usages in 3D or higher dimensions. 
With quantum double basis $| \alpha, a\rangle$ in 2D generalized to a triple basis $| \alpha, a, b\rangle$ in 3D, it allures us to call it
a {\bf twisted quantum triple model} in 3D. It awaits mathematicians and mathematical physicists to explore more details in the future.
\subsection{Cocycle of $\cH^4(G,\R/\Z)$ and its dimensional reduction} 
To 
study the twisted gauge theory of a finite Abelian group, 
we now provide its explicit data of cohomology group and 4-cocycles.\cite{Supplemental} 
Here $\mathcal{H}^{d+1}(G,\R/\Z)=\mathcal{H}^{d+1}(G,\tU(1))$ by $\R/\Z=\tU(1)$, as the ${(d+1)}$th-cohomology group of G over G module U(1).
Each class in $\mathcal{H}^{d+1}(G,\R/\Z)$ corresponds to a distinct $(d+1)$-cocycle.
The different 4-cocycles will label the distinct topological terms of 3+1D twisted gauge theories. (However, 
different topological terms may share the same data for topological orders, such as the same modular data $\sfS^{xyz}$ and $\sfT^{xy}$.
Thus different topological terms may describe the same topological order.)  
The 4-cocycles $\omega_4$ are 4-cochains, but additionally satisfy the cocycle condition $\delta \omega=1$. 
The $4$-cochain is a mapping $\omega_{4}^{}(a,b,c,d)$:  $(G)^4 \to \tU(1)$, which inputs
$a,b,c,d \in G$, and outputs a $\tU(1)$ phase.
Furthermore, distinct 4-cocycles are not identified by any 4-coboundary $\delta \Omega_3$.
(Namely, distinct cocycles $\omega_4$ and $\omega'_4$ do {\it not} satisfy $\omega_4/ \omega'_4 = \delta \Omega_3$, for any 3-cochain $\Omega_3$.)
The $4$-cochain satisfies the group multiplication rule:
$
(\omega_{4}\cdot\omega_{4}')(a,b,c,d)= \omega_{4}(a,b,c,d)\cdot \omega_{4}'(a,b,c,d), 
$
thus forms a group $\text{C}^4$,
the 4-cocycle  further forms its subgroup $\text{Z}^4$,
and the 4-coboundary further forms a $\text{Z}^4$'s subgroup $\text{B}^4$ (since $ \delta^2 =1$). In short,
$
\text{B}^4 \subset \text{Z}^4 \subset \text{C}^4.
$
The fourth cohomology group is a kernel $\text{Z}^4$ (the group of 4-cocycle) mod out the image $\text{B}^4$ (the group of 4-coboundary) relation:
$
\cH^4(G,\R/\Z)= \text{Z}^4 /\text{B}^4.
$
We derive the fourth cohomology group of a generic finite Abelian $G=\prod^k_{i=1} Z_{N_i}$ as
\bea
\label{eq:H4}
\mathcal{H}^4(G,\R/\Z) 
=  \prod_{1 \leq i < j < l<m \leq k} 
         (\Z_{N_{ij}})^2
         \times (\Z_{N_{ijl}})^2 
         \times \Z_{N_{ijlm}}.\;\;\; \;\;\;\;
\eea
We construct generic 4-cocycles (not identified by 4-coboundaries) for each type, 
summarized in Table \ref{table1}.
\begin{widetext}
\begin{center}
\begin{table} [!h]
\begin{tabular}{|c||c|c|c|}
\hline
\;$\mathcal{H}^4(G,\R/\Z)$\; &  4-cocycle name &  4-cocycle form  & Induced 3-cocycle $\sfC_b(a,c,d)$  \\[0mm]  \hline \hline 
$\Z_{N_{ij}}$ & Type II 1st $p_{{ \tII(ij)}}^{(1st)} $  & 
${\omega_{4,{ \tII}}^{(1st,ij)} (a,b,c,d)=  \exp \big( \frac{2 \pi \ti p_{{ \tII(ij)}}^{(1st)} }{ (N_{ij} \cdot N_j  )   }    (a_i b_j )( c_j +d_j - [c_j+d_j  ]) \big)}$ & Type I, II of $\mathcal{H}^3(G,\R/\Z)$ \\[2mm] \hline
$\Z_{N_{ij}}$ & Type II 2nd $p_{{ \tII(ij)}}^{(2nd)} $ & 
${\omega_{4,{ \tII }}^{(2nd,ij)} (a,b,c,d) =  \exp \big( \frac{2 \pi \ti p_{{ \tII(ij)}}^{(2nd)} }{ (N_{ij} \cdot N_i  )  }   (a_j b_i )( c_i +d_i - [c_i+d_i  ])  \big) } $ & Type I, II of $\mathcal{H}^3(G,\R/\Z)$  \\[2mm] \hline
$\Z_{N_{ijl}}$ & Type III 1st $p_{{ \tIII(ijl)}}^{(1st)} $ & 
${\omega_{4,{\tIII}}^{(1st,ijl)}(a,b,c,d) = \exp \big( \frac{2 \pi \ti p_{{ \tIII(ijl)}}^{(1st)} }{ (N_{ij} \cdot N_l )  }  (a_i b_j )( c_l +d_l - [c_l+d_l  ]) \big) }$ & two Type IIs of $\mathcal{H}^3(G,\R/\Z)$ \\[2mm] \hline
$\Z_{N_{ijl}}$ & Type III 2nd $p_{{ \tIII(ijl)}}^{(2nd)} $ & 
${\omega_{4,{\tIII}}^{(2nd,ijl)}(a,b,c,d) = \exp \big( \frac{2 \pi \ti p_{{ \tIII(ijl)}}^{(2nd)} }{ (N_{li} \cdot N_j )  }  (a_l b_i )( c_j +d_j - [c_j+d_j  ]) \big) }$ & two Type IIs of $\mathcal{H}^3(G,\R/\Z)$  \\[2mm] \hline
$\Z_{N_{ijlm}}$ & Type IV  $p_{{ \tIV (ijlm)}}^{} $ &
 $\omega_{4,{ \tIV}}^{(ijlm)} (a,b,c,d)=\exp \big( \frac{2 \pi \ti p_{{ \tIV}{(ijlm)}}^{}}{ N_{ijlm} }  a_i b_j c_l d_m \big)$  &   Type III of $\mathcal{H}^3(G,\R/\Z)$ \\[1.mm] \hline
\end{tabular}
\caption{ 
The cohomology group $\mathcal{H}^4(G,\R/\Z)$ and 4-cocycles $\omega_4$ for a generic finite Abelian group $G=\prod^k_{i=1} Z_{N_i}$.
The first column shows the types 
in $\mathcal{H}^4(G,\R/\Z)$ of Eq.(\ref{eq:H4}). The second column shows the topological term indices for 3+1D {\it twisted gauge theory}. 
(When all indices $p_{\dots}=0$, it becomes the {\it normal untwisted gauge theory}.)
The third column shows explicit 4-cocycle functions $\omega_{4}^{}(a,b,c,d)$:  $(G)^4 \to \tU(1)$. Here $a=(a_1,a_2,\dots,a_k)$, with $a\in G$ and $a_i \in Z_{N_i}$. 
(Same notations for $b,c,d$.) We define the mod $N_j$ relation by $[c_j+d_j  ] \equiv c_j+d_j  {\pmod {N_j}}$.
The last column shows the induced 3-cocycles from the slant product $\sfC_b(a,c,d) \equiv i_b \omega_4(a,c,d)$ in terms of 
Type I, II, III 3-cocycles of $\mathcal{H}^3(G,\R/\Z)$ listed in Table \ref{tableA1}.
}
\label{table1}
\end{table}
\end{center}
\end{widetext}

We name 
the Type II 1st and Type II 2nd 4-cocycles for those with topological term indices:
${p_{{ \tII(ij)}}^{(1st)} } \in \Z_{N_{ij}}$ and ${p_{{ \tII(ij)}}^{(2nd)} } \in \Z_{N_{ij}}$ of Eq.(\ref{eq:H4}).
There are Type III 1st and Type III 2nd 4-cocycles for topological term indices:
${p_{{ \tIII(ijl)}}^{(1st)} } \in \Z_{N_{ijl}}$ and ${p_{{ \tIII(ijl)}}^{(2nd)} } \in \Z_{N_{ijl}}$. 
There is 
also Type IV 4-cocycle topological term 
index: $p_{{ \tIV}{(ijlm)}}\in \Z_{N_{ijlm}}$.

Since we earlier prelude the relation Eq.(\ref{eq:C3DtoC2D}), $\cC^{3\tD} = \oplus_b \cC^{2\tD}_b$, 
between 3D topological orders (described by 4-cocycles) as the direct sum of sectors of 2D topological orders (described by 3-cocycles),
it is suggestive to see how the dimensionally-reduced 3-cocycle from 4-cocycles can hint the $\cC^{2\tD}_b$ theory of 2D.
The slant product $\sfC_b(a,c,d) \equiv i_b \omega_4(a,c,d)$ are organized in the last column of Table \ref{table1}.
Luckily, the Type II, III $\omega_4$ have a simpler form of $\sfC_b(a,c,d) = \omega_4(a,b,c,d)/\omega_4(b,a,c,d)$, while the reduced form of
Type IV $\omega_4$ is more involved.\cite{Supplemental} 

\begin{figure}[h!]
\centering
  \begin{align} \label{eq:Cb3}
&\sfC_b(a,c,d): \begin{matrix}\CbThreeRepL  \end{matrix} \underset{\DashedArrow}{t \; (d)}  \begin{matrix}\CbThreeRepR  \end{matrix} 
  \end{align}
\caption{The geometric interpretation of the induced 3-cocycle $\sfC_b(a,c,d) \equiv i_b \omega_4(a,c,d)$ from 4-cocycle $\omega_4$.
The combination of Eq.(\ref{eq:Cb3}) (with four 4-cocycles filling) times the contribution of Eq.(\ref{fig:Cabc}) (with three 3-cocycles filling)
will produce Eq.(\ref{fig:Cabcd}) with twelve 4-cocycles filling.}
\label{fig:Cb3}
\end{figure}

\begin{figure}[!h] 
\includegraphics[scale=0.38]{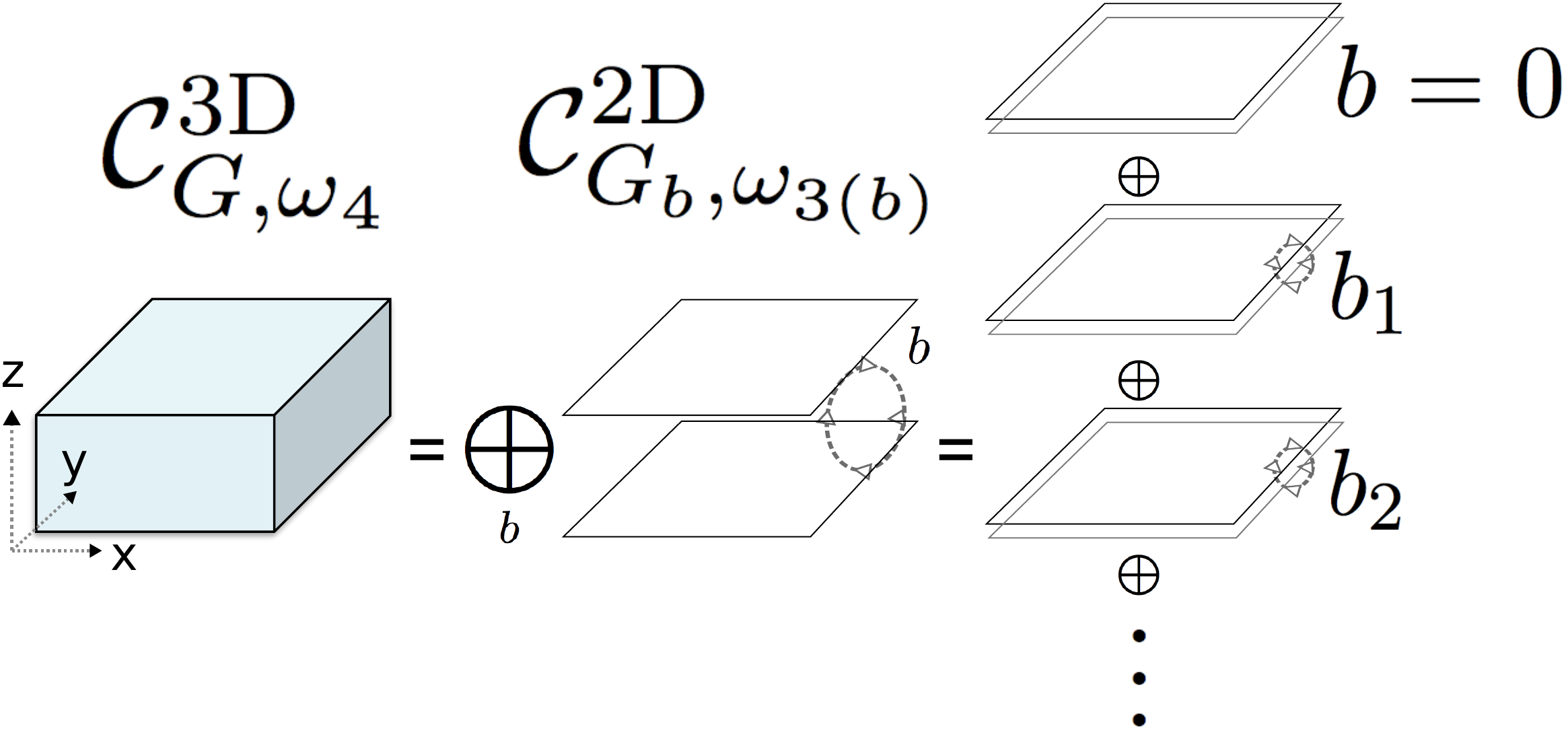}
\caption{
Combine the reasoning in Fig.\ref{fig:Cb3} and Fig.\ref{fig:3Dto2D}, we obtain the physical meaning of 
dimensional reduction from a 3+1D twisted gauge theory as a 3D topological order
 to several sectors of 2D topological orders: $\cC^{3\tD}_{G,\omega_4} = \oplus_b \cC_{G,\omega_{3(b)}}^{2\tD}$.
Here $b$ stands for the gauge flux (Wilson line operator) of gauge group $G$. 
Here $\omega_3$ are dimensionally reduced 3-cocycles from 4-cocycles $\omega_4$.
Notice there is a zero flux $b=0$ sector with $\cC_{G,\text{(untwist)}}^{2\tD} =\cC_{G{}}^{2\tD}$. }
\label{3Dto2D_0flux} 
\end{figure}

This indeed promisingly suggests the relation in Eq.(\ref{eq:C3DtoC2Dom}), $\cC^{3\tD}_{G,\omega_4} = \oplus_b \cC_{G,\omega_{3(b)}}^{2\tD}$ with $G_b=G$ the original group. If we view $b$ as the gauge flux along the $z$ direction, and compactify $z$ into a circle, then a single winding around $z$ acts as a monodromy defect carrying
the gauge flux $b$ (group elements or conjugacy classes).\cite{Wen:2013ue,{Santos:2013uda},Wang:2014tia} This implies a geometric picture in Fig.\ref{3Dto2D_0flux}.

One can tentatively write down a relation,
\bea \label{eq:C3DtoC2Domb}
\cC^{3\tD}_{G,\omega_4} =  \cC_{G,1\text{(untwist)}}^{2\tD} \oplus _{b \neq 0}  \cC_{G,\omega_{3(b)}}^{2\tD},
\eea 
There is a zero flux $b=0$ sector $\cC_{G,1\text{(untwist)}}^{2\tD} $ (with $\omega_3=1$) where the 2D gauge theory with $G$ is untwisted. 
There are other direct sums of $\cC_{G,\omega_{3(b)}}^{2\tD}$ with nonzero $b$ flux insertion has twisted $\omega_{3(b)}$. 

However, 
different cocycles can represent the same topological order with the equivalent modular data,
in the next we should exam this Eq.(\ref{eq:C3DtoC2Domb}) more carefully not in terms of cocycles, but in terms of the modular data $\sfS^{xyz}$ and $\sfT^{xy}$.


\section{Representation for $\sfS^{xyz}$ and $\sfT^{xy}$}  \label{sec:RepST}

The modular transformation $\hat{\sfS}^{xy}$, $\hat{\sfT}^{xy}$, $\hat{\sfS}^{xyz}$ of Eq.(\ref{eq:ST3D}),(\ref{eq:Sxy}) acts on the 3D real space (see Fig.\ref{fig:SL3Ztransform}) by
\bea
&&\hat{\mathsf{S}}^{xy}\cdot (x,y,z)=(-y,x,z),\\  
&&\hat{\mathsf{T}}^{xy}\cdot (x,y,z)=(x+y,y,z),\\
&&\hat{\mathsf{S}}^{xyz}\cdot (x,y,z)=(z,x,y).
\eea
\noindent
{\bf Q4}: ``\emph{What are the generic expressions of SL(3,$\Z$) modular data}?'' 

In Sec \ref{STcocycle}, we will first apply the \emph{cocycle approach} using the \emph{spacetime path integral} with 
SL$(3,\Z)$ transformation acting along the time evolution to formulate the SL$(3,\Z)$ modular data,
and then in Sec \ref{Rep} use the more powerful \emph{Representation} (Rep) \emph{Theory} to find out general expressions of those data in terms of $(G,\omega_4)$.

\begin{figure}[h!]
\centering
  \begin{align}
  &\hat{\sfS}^{xy}: \begin{matrix}\SLTWOtransformSLeft\end{matrix}
  \underset{\DashedArrow}{t} 
  \begin{matrix}\SLTWOtransformSRight\end{matrix}\\
   &\hat{\sfT}^{xy}: \begin{matrix}\SLTWOtransformTL\end{matrix}
  \;\;\underset{\DashedArrow}{t} 
  \begin{matrix}\SLTWOtransformTR\end{matrix}\\
    &\hat{\sfS}^{xyz}: \begin{matrix}\SLTHREEtransformSLeft\end{matrix}
    \underset{\DashedArrow}{t} 
      \begin{matrix}\SLTHREEtransformSR\end{matrix} 
  \end{align}
\caption{
The modular transformation SL$(2,\Z)$ is generated by $\hat{\sfS}^{xy}$ and $\hat{\sfT}^{xy}$,
while the SL$(3,\Z)$ is generated by $\hat{\sfS}^{xyz}$ and $\hat{\sfT}^{xy}$.
The dashed arrow $\DashedArrow$ stands for the time evolution (as in Fig.\ref{torus2D3D}) from $|\Psi_{in} \rangle$ to $|\Psi_{out} \rangle$ under $\hat{\sfS}^{xy}$, $\hat{\sfT}^{xy}$, $\hat{\sfS}^{xyz}$ respectively.
The $\hat{\sfS}^{xy}$ and $\hat{\sfT}^{xy}$ transformations on
a $\mathbb{T}^3$ torus's $x$-$y$ plane with $z$ direction untouched is equivalent to its transformations on a $\mathbb{T}^2$ torus).}
\label{fig:SL3Ztransform}
\end{figure}

\subsection{Path Integral and Cocycle approach} \label{STcocycle}

{
The cocycles approach uses the spacetime lattice formalism, where we triangulate the spacetime complex of a 4-manifold $\cM=\mathbb{T}^3 \times I$,
(a $\mathbb{T}^3$ torus times a time interval $I$) of Fig.\ref{fig:SL3Ztransform} into 4-simplices.
We then apply the path integral $\mathbf{Z}$ in Eq.(\ref{eq:path integral}) and the amplitude form in Eq.(\ref{eq:Oamp}) to obtain
\bea 
&&{\sfT}^{xy}_{\text{(A)(B)}}= \langle   \Psi_{\text{A}}  |  \hat{\sfT}^{xy} | \Psi_{\text{B}}\rangle,\\
&&{\sfS}^{xy}_{\text{(A)(B)}}= \langle  \Psi_{\text{A}}  |  \hat{\sfS}^{xy} | \Psi_{\text{B}}\rangle,\\
&&{\sfS}^{xyz}_{\text{(A)(B)}}= \langle \Psi_{\text{A}}  |   \hat{\sfS}^{xyz} | \Psi_{\text{B}}\rangle,\\
&&\text{GSD}=\Tr[\sfP] =\sum_A \langle  \Psi_{\text{A}}  |  \sfP | \Psi_{\text{A}}\rangle. \label{eq:GSDpathZ}
\eea
Here $| \Psi_{\text{A}}\rangle$ and $| \Psi_{\text{B}}\rangle$ are ground state basis on $\mathbb{T}^d$ torus, for example, they are
$|\alpha, a\rangle$ (with $\alpha$ charge and $a$ flux) in 2+1D and $|\alpha, a, b\rangle$ (with $\alpha$ charge and $a, b$ fluxes) in 3+1D.
We also include the data of ground state degeneracy (GSD), where the $\sfP$ is the projection operator to ground states discussed in Sec.\ref{Sec:TQD}.
In the case of d-D GSD on $\mathbb{T}^d$ (e.g. 3D GSD on $\mathbb{T}^3$), we simply compute the $\mathbf{Z}$ amplitude filling in $\mathbb{T}^d \times S^1= \mathbb{T}^{d+1}$.
There is no short cut here except doing explicit calculations.\cite{Supplemental}  
}
\subsection{Representation Theory approach} \label{Rep}

The cocycle approach in Sec.\ref{STcocycle} provides nice physical intuitions on the modular transformation process. 
However, the calculation is tedious. There is a powerful approach simply using Representation Theory, we will present the 
general formula of $\hat{\sfS}^{xys}$, $\hat{\sfT}^{xy}$, $\hat{\sfS}^{xy}$ data in terms of $(G, \omega_4)$ directly.
We outline the three steps:\\ 
(i) Obtain the Eq.(\ref{eq:2slant})'s $\sfC^{(2)}_{a,b}$ by doing the slant product twice from 4-cocycle $\omega_4$,
or triangulating Eq.\ref{fig:Cabc} in Fig\ref{fig:Cab}. \\ 
(ii) Derive $\widetilde{\rho}_{\alpha}^{a,b}(c)$ of $\sfC^{(2)}_{a,b}$-{projective representation} in Eq.(\ref{eq:CabRep}), 
which $\widetilde{\rho}_{\alpha}^{a,b}(c)$ 
is a general linear matrix.\\
(iii) 
Write the modular data in the canonical basis $|\alpha, a,b\rangle$, $|\beta, c,d\rangle$ of Eq.(\ref{eq:3Dbasis}). 
  
  After some long computations,\cite{Supplemental} we find 
the most general formula $\sfS^{xyz}$ for a group $G$ (both Abelian or non-Abelian) with cocycle twist $\omega_4$:
\begin{widetext} 
\bea \label{Eq.Sxyz} 
&& {\sfS^{xyz}_{(\alpha, a,b)(\beta, c,d)}=\frac{1}{|G|}  \langle \alpha_x, a_y, b_z  | \sum_w \sfS^{xyz}_w | \beta_{x'}, c_{y'}, d_{z'}\rangle }
 { =\frac{1}{|G|}\sum_{\substack{{g_y\in C^a \cap Z_{g_z} \cap  Z_{g_x},}\\{g_z\in C^b \cap C^c,}\\{g_x\in Z_{g_y} \cap  Z_{g_z} \cap C^d }   }} 
  \text{Tr}\widetilde{\rho}^{g_y, g_z}_{\alpha_x}(g_x)^{*}  \; \text{Tr}\widetilde{\rho}^{g_z,g_x}_{\beta_{y}}(g_y)^{} \delta_{g_x,h_{z'}  } \delta_{g_y,h_{x'}} \delta_{g_z,h_{y'}}}. \;\;\;\;\;\;
\eea
\end{widetext}
Here $C^a,C^b,C^c,C^d$ are conjugacy classes of the group elements $a,b,c,d \in G$. In the case of non-Abelian $G$, we should regard $a,b$ as 
its conjugacy class $C^a,C^b$ in $|\alpha, a,b\rangle$. $Z_g$ means the centralizer of the conjugacy class of $g$.
For Abelian $G$, it simplifies to 
\bea 
&& {\sfS^{xyz}_{(\alpha, a,b)(\beta, c,d)}= 
 { \frac{1}{|G|}  
  \text{Tr}\widetilde{\rho}^{a,b}_{\alpha}(d)^{*}  \; \text{Tr}\widetilde{\rho}^{b,d}_{\beta}(a)^{} \delta_{b,c} }
 \equiv \frac{1}{|G|}  \sfS^{\alpha,\beta}_{d,a,b} \delta_{b,c}} \;\;\;\;\;\;\nonumber \\
&&= { 
 { \frac{1}{|G|}  
  \text{Tr}\widetilde{\rho}^{a_y,b_z}_{\alpha_x}(d_{z'})^{*}   \text{Tr}\widetilde{\rho}^{b_z,d_{z'}}_{\beta_{x'}}(a_y)^{} \delta_{b_z,c_{y'}} }
 \equiv \frac{1}{|G|}  \sfS^{\alpha_x,\beta_{y}}_{d_x,a_y,b_z} \delta_{b_z,c_{y'}}}.
 \nonumber
\eea
We denote $\beta_{x'}=\beta_{y}$, $d_{z'}=d_x$ due to the coordinate identification under $\hat{\mathsf{S}}^{xyz}$.
The assignment of the directions of gauge fluxes (group elements) are clearly expressed in the second line. 
We may use the first line expression for simplicity.

We also provide the most general formula of $\sfT^{xy}$ in $|\alpha, a,b\rangle$ basis:
\bea \label{eq:Txy} 
&& {{ \sfT^{xy}=\sfT^{a_y,b_z}_{\alpha_x} =\frac{\text{Tr} \widetilde{\rho}^{a_y,b_z}_{\alpha_x}(a_y)}{\text{dim}({\alpha})} }} \equiv \exp(\ti \Theta^{a_y,b_z}_{\alpha_x}).
\eea
Here ${\text{dim}({\alpha})}$ means the dimension of the representation, equivalently the rank of the
matrix of $\widetilde{\rho}^{a,b}_{\alpha_x}(c)$.
Since SL$(2,\Z)$ is a subgroup of SL$(3,\Z)$, we can express the SL$(2,\Z)$'s $\sfS^{xy}$ by SL$(3,\Z)$'s $\sfS^{xyz}$ and $\sfT^{xy}$
(an expression for both the real spatial basis and the canonical basis): 
\bea \label{eq:Sxydecompose}
&& \sfS^{xy}=((\sfT^{xy})^{-1} \sfS^{xyz})^3 (\sfS^{xyz}\sfT^{xy})^2  \sfS^{xyz} (\sfT^{xy})^{-1}.
\eea

For Abelian $G$, and when $\sfC^{(2)}_{a,b}(c,d)$ is a 2-coboundary (cohomologically trivial), the dimensionality of Rep is $\text{dim}(\text{Rep}) \equiv \text{dim}(\alpha)=1$, the $\sfS^{xy}$ is simplified:
\bea \label{Eq:SxyAbSimp}
&& 
{ \sfS^{xy}_{(\alpha,a,b)({\beta,c,d})}=  { \frac{1}{|G|}  } 
{ \frac{ \text{tr}\widetilde{\rho}^{a,b}_{\alpha}(a c^{-1})^{*}   }{\text{tr} \widetilde{\rho}^{a,b}_{\alpha}(a)}}
  \frac{  {  \; {\text{tr} \widetilde{\rho}^{c, d}_{\beta}(a c^{-1})}  }  }{\text{tr}\widetilde{\rho}^{c,d}_{\beta}(c)^{} }
  \delta_{b,d} }. \;\;\;\;\;\;
\eea

We can verify our above results by firstly computing the cocycle path integral approach in Sec.\ref{STcocycle}, and substitute from the flux basis to the canonical basis
by Eq.(\ref{eq:3Dbasis}). We have made several consistent checks, by comparing our $\hat{\sfS}^{xy}$, $\hat{\sfT}^{xy}$, $\hat{\sfS}^{xyz}$ to: (1) the known 2D case for the untwisted theory of a non-Abelian group,\cite{deWildPropitius:1995cf} (2) the recent 3D case for the untwisted theory of a non-Abelian group,\cite{MW14}
(3) the recent 3D case for the twisted theory of a Abelian group.\cite{JMR1462} And our expression works for all cases: the (un)twisted theory of (non-)Abelian group. 
More detailed 
calculations are reserved in Supplemental Material (Appendix \ref{AppendixST}).

\subsection{Physics of $\sfS$ and $\sfT$ in 3D}  \label{Sec:physicsST}
\begin{figure}[!h] 
\includegraphics[scale=0.25]{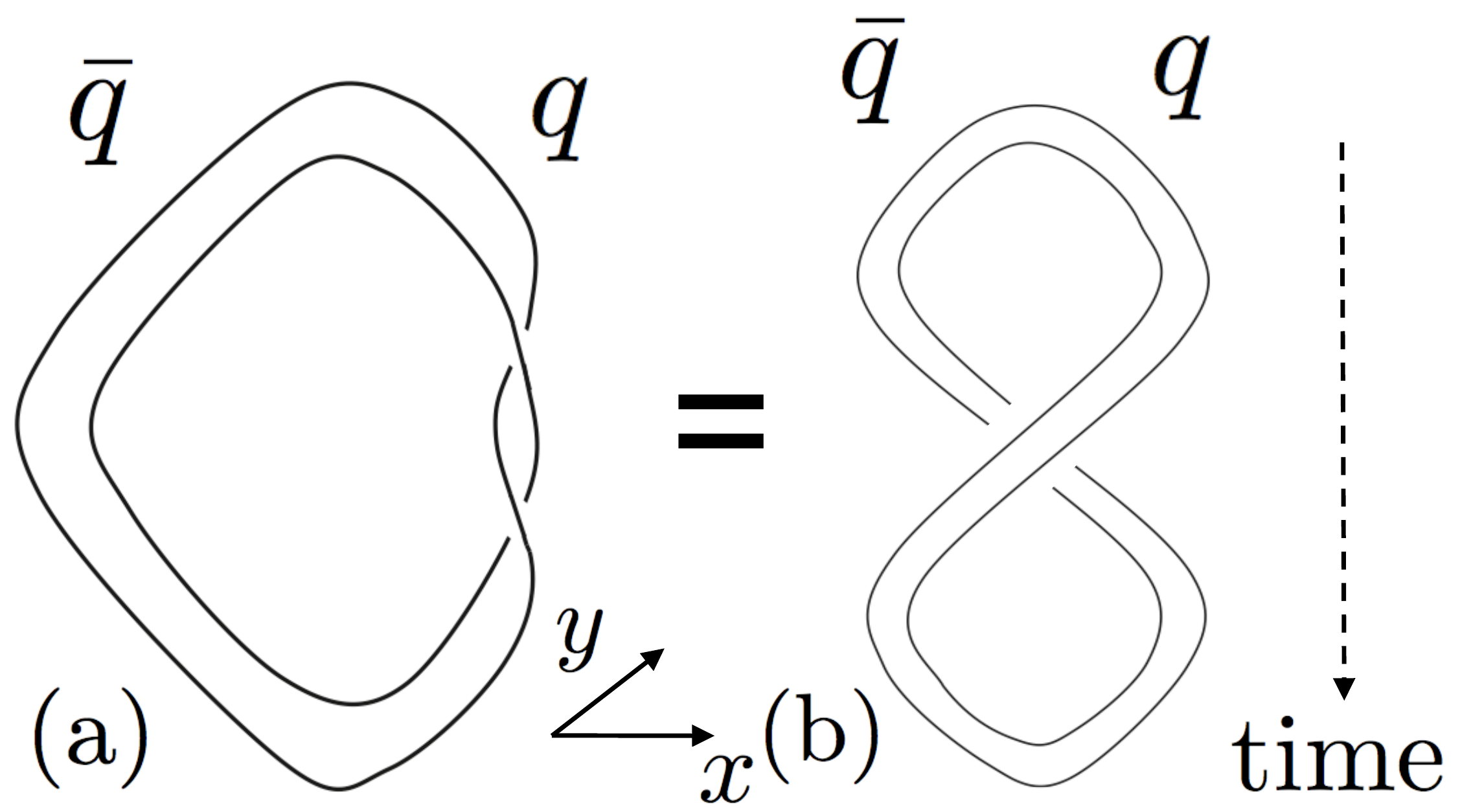}
\caption{
Both process (a) and (b) starts from creating a pair of particle $q$ and anti-particle $\bar{q}$, but the wordlines evolve along time to the bottom differently.
The process (a) produces a phase $e^{\ti 2 \pi s}$ due to $2\pi$ rotation of q, with spin $s$. 
The process (b) produces a phase $e^{\ti \Theta}$ 
due to the 
exchange statistics.
The homotopic equivalence by deformation implies $e^{\ti 2 \pi s}=e^{\ti \Theta}$.
}
\label{fig:spin_statistics} 
\end{figure}

The ${\sfS}^{xy}$ and ${\sfT}^{xy}$ in 2D are known to have precise physical meanings. 
At least for Abelian topological orders, there is no ambiguity that 
${\sfS}^{xy}$ in the quasiparticle basis provides the mutual statistics of two particles (winding one around the other by $2\pi$), while
${\sfT}^{xy}$ in the quasiparticle basis provides the self statistics of two identical particles (winding one around the other by $\pi$).
Moreover, the intimate spin-statistics relation   
shows that the statistical phase $e^{\ti \Theta}$ gained by interchanging 
two identical particles is equal to the spin $s$ by $e^{\ti 2 \pi s}$. 
Fig.\ref{fig:spin_statistics} illustrates the {\it spin-statistics relation}.\cite{Finkelstein:1968hy} Thus, people also call ${\sfT}^{xy}$ in 2D as the {\it topological spin}.
Here we ask:\\ 
\noindent
{\bf Q5}: ``\emph{What is the physical interpretation of} SL(3,$\Z$) \emph{modular data in 3D}?''

Our approach again is by dimensional reduction 
of Fig.\ref{fig:3Dto2D}, via Eq.(\ref{eq:3Dto2DST}) and Eq.(\ref{eq:C3DtoC2D}): 
$\mathsf{S}^{xy}=\oplus_b \mathsf{S}^{xy}_b$, $\mathsf{T}^{xy}=\oplus_b \mathsf{T}^{xy}_b$, $ \cC^{3\tD} = \oplus_b \cC^{2\tD}_b$,
reducing the 3D physics to the direct sum of 2D topological phases in different flux sectors,
so we can retrieve the familiar physics of 2D to interpret 3D.\\

For our case with a gauge group 
description, the $b$ (subindex of $\mathsf{S}^{xy}_b$, $\mathsf{T}^{xy}_b$, $\cC^{2\tD}_b$)
labels the gauge flux (group element or conjugacy class $C^b$) winding around the compact $z$ direction in Fig.\ref{fig:3Dto2D}.
This $b$ flux can be viewed as the by-product of a {\it monodromy defect} causing a {\it branch cut} 
(\emph{a symmetry twist}\cite{Wen:2013ue,Wang:2014tia,{Santos:2013uda},{Wang:2014pma}}), 
such that the wavefunction will gain a {\it phase} by winding around the compact $z$ direction.
Now we further regard the $b$ flux as a {\it string threading around} in the background, so that winding around this background string (e.g. the black string threading in
Fig.\ref{fig:3strings_2D_3D_xy}(c),\ref{fig:Txy}(c),\ref{fig:Sxyhalf}(c)) gains the $b$ flux effect if there is a nontrivial winding on the compact direction $z$.
The arrow - - -$\vartriangleright$ along the compact $z$ schematically indicates such a $b$ flux effect from the background string threading.
\subsubsection{${\sfT}^{xy}_b$ and topological spin of a closed string}
We apply the above idea to interpret ${\sfT}^{xy}_b$, shown in Fig.\ref{fig:Txy}. From Eq.(\ref{eq:Txy}), we have
${ {\sfT}^{xy}_b=\sfT^{a_y,b_z}_{\alpha_x}}$ ${\equiv \exp(\ti \Theta^{a_y,b_z}_{\alpha_x})}$ with a fixed $b_z$ label for a given $b_z$ flux sector.
For each $b$, ${\sfT}^{xy}_b$ acts as a familiar 2D ${\sfT}$ matrix $\sfT^{a_y}_{\alpha_x}$, which provides the topological spin of a quasiparticle $(\alpha,a)$
with charge $\alpha$ and flux $a$, in Fig.\ref{fig:Txy}(a).

From the 3D viewpoint, however, this $|\alpha, a\rangle$ particle is actually a closed string 
compactified along the compact $z$ direction.
Thus, in Fig.\ref{fig:Txy}(b), the self-$2\pi$ rotation of the topological spin of a quasiparticle $|\alpha, a\rangle$ is indeed the self-$2\pi$ rotation of a framed closed string.
(Physically we understand that there the string can be {\it framed} with arrows, because of the inner texture of the string excitations are allowed in a condensed matter system,
due to defects or the finite size lattice geoemtry.)
Moreover, from an equivalent 3D view in Fig.\ref{fig:Txy}(c), we can view the self-$2\pi$ rotation of a framed closed string as the {\it self-$2\pi$ flipping} of a framed closed string, which
flips the string inside-out and then outside-in back to its original status. 
This picture works for both 
$b=0$ zero flux sector as well as 
$b \neq 0$ under the background string threading. {\bf We thus propose that ${ {\sfT}^{xy}_b}$ as the topological spin of a framed closed string,
threaded by a  background string carrying a monodromy $b$ flux.}

\onecolumngrid
\begin{widetext}
\begin{center} 
\begin{figure}[!h] 
\includegraphics[scale=0.5]{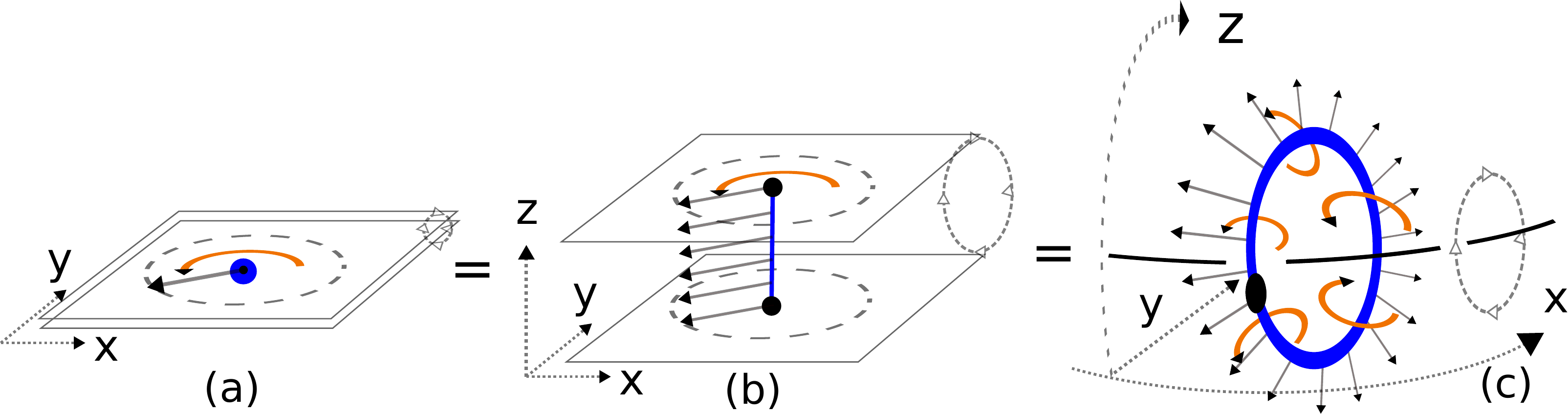}
\caption{
Topological spin of (a) a particle by $2\pi$-self rotation in 2D, (b) a framed closed-string by $2\pi$-self rotation in 3D with a compact $z$,
(c) a closed-string (blue) by $2\pi$-self flipping, threaded by a background (black) string creating monodromy $b$ flux (along the arrow - - -$\vartriangleright$), under a single Hopf link $2^2_1$ configuration.
All above equivalent pictures describe the physics of topological spin in terms of  ${\sfT}^{xy}_b$. 
{For Abelian topological orders, the spin of an excitation (say A) in Fig.\ref{fig:Txy}(a) yields an
Abelian phase 
${e^{\ti \Theta_{\text{(A)} }} = {\sfT}^{xy}_{\text{(A)(A)}}}$ proportional to the diagonal of 2D's $\mathsf{T}^{xy}$ matrix. 
The dimensional-extended equivalent picture Fig.\ref{fig:Txy}(c) implies that the loop-flipping
yields a phase ${e^{\ti \Theta_{\text{(A)},b }} = {\sfT}^{xy}_{b\;\text{(A)(A)}}}$ of Eq.(\ref{eq:Txy}) (up to a choice of canonical basis), where $b$ is the flux of the black string.}
} 
\label{fig:Txy} 
\end{figure}
\begin{figure}[!h] 
\includegraphics[scale=0.5]{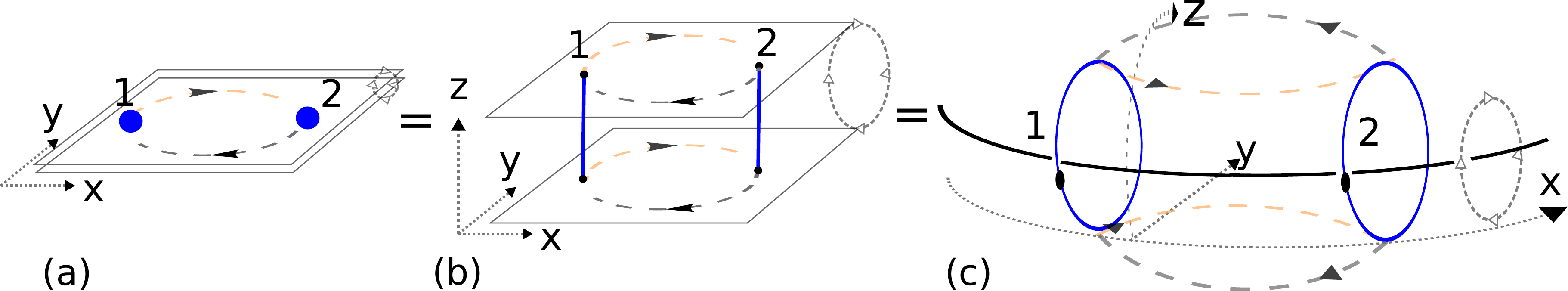}  
\caption{
Exchange statics of (a) two identical particles at positions 1 and 2 by a $\pi$ winding (half-winding), 
(b) two identical strings by a $\pi$ winding 
in 3D with a compact $z$,
(c) two identical closed-strings (blue) with a $\pi$-winding 
around, both threaded by a background (black) string creating monodromy $b$ flux, under 
the Hopf links $2^2_1 \# 2^2_1$ configuration. Here figures (a)(b)(c) describe the equivalent physics in 3D with a compact $z$ direction.
The physics of exchange statics of a closed string turns out to be related to the topological spin of Fig.\ref{fig:Txy},
discussed in Sec.\ref{spin-statistics}. 
} 
\label{fig:Sxyhalf} 
\end{figure}
 \end{center}
\end{widetext}
\twocolumngrid

\subsubsection{${\sfS}^{xy}_b$ and three-string braiding statistics} \label{Sec:IIIc3string}

Similarly, we apply the same philosophy to do 3D to 2D reduction for ${\sfS}^{xy}_b$, each effective 2D treading with a distinct gauge flux $b$. 
We can obtain $\mathsf{S}^{xy}_b$ from Eq.(\ref{eq:Sxydecompose}) with SL$(3,\Z)$ modular data.
Here we will focus on interpreting $\mathsf{S}^{xy}_b$ in the Abelian topological order.
Writing $\mathsf{S}^{xy}_b$ in the canonical basis $|\alpha, a,b\rangle$, $|\beta, c,d\rangle$ of Eq.(\ref{eq:3Dbasis}),
we find that, true for Abelian topological order
\bea \label{eq:Sxyb}
\mathsf{S}^{xy}_b 
= \sfS^{xy}_{(\alpha,a,b)({\beta,c,d})} \equiv \frac{1}{|G|} \sfS^{2\tD\;\alpha,\beta}_{a,c\;(b)} \delta_{b,d}.
\eea
As we predict the generality in Eq.(\ref{eq:3Dto2DST}), the $\mathsf{S}^{xy}_b$ here is diagonalized with the $b$ and $d$ identified (as the z-direction flux created by the background string threading).
For a given fixed $b$ flux sector, the only free indices are $|\alpha, a\rangle$ and $|\beta, c\rangle$, all collected in $\sfS^{2\tD\;\alpha,\beta}_{a,c\;(b)}$.
(Explicit data will be presented in Sec.\ref{TypeII,III twist})
Our interpretation is shown in Fig.\ref{fig:3strings_2D_3D_xy}.
From a 2D viewpoint, $\mathsf{S}^{xy}_b$ gives the full $2\pi$ braiding statistics data 
of two quasiparticle $|\alpha, a\rangle$ and $|\beta, c\rangle$  excitations in Fig.\ref{fig:3strings_2D_3D_xy}(a). 
However, from the 3D viewpoint, the two particles are actually two closed strings compactified along the compact $z$ direction.
Thus, the full-$2\pi$ braiding of two particles in Fig.\ref{fig:3strings_2D_3D_xy}(a) becomes that of two closed-strings in Fig.\ref{fig:3strings_2D_3D_xy}(b).
More explicitly, an equivalent 3D view in Fig.\ref{fig:3strings_2D_3D_xy}(c), we identify the coordinates $x,y,z$ carefully to see such a full-braiding process
is that {\it one (red) string going inside to the loop of another (blue) string, and then going back from the outside.}

The above picture again works for both $b=0$ zero flux sector as well as $b \neq 0$ under the background string threading.
When $b \neq 0$, the third (black) background string in Fig.\ref{fig:3strings_2D_3D_xy}(c) treading through the two (red, blue) strings. 
The third (black) string creates the monodromy defect/branch cut on the background, 
and carrying $b$ flux along $z$ acting on two (red, blue) strings which have nontrivial winding on the third string.
This three-string braiding has been firstly emphasized in a recent paper,\cite{WL1437} 
here we make further connection to the data ${ {\sfS}^{xy}_b}$ and understand its physics in a 3D to 2D under $b$ flux sectors.

{\bf We have shown and proposed that ${ {\sfS}^{xy}_b}$ can capture the physics of three-string braiding statistics with
two strings threaded by a third background string causing $b$ flux monodromy, where the three strings have
the linking configuration as the connected sum of two Hopf links $2^2_1\# 2^2_1$.}

\subsubsection{Spin-Statistics relation for closed strings} \label{spin-statistics}

Since a spin-statistics relation for 2D particles is shown by Fig.\ref{fig:spin_statistics}. 
We may wonder, by using our 3D to 2D reduction picture, whether a {\it spin-statistics relation for a closed string} holds?

To answer this question, we should compare the {\it topological spin} picture of $\mathsf{T}^{xy}_b=\sfT^{a_y,b_z}_{\alpha_x}$ ${\equiv \exp(\ti \Theta^{a_y,b_z}_{\alpha_x})} $
to the {\it exchange statistic} picture of two closed strings in Fig.\ref{fig:Sxyhalf}. 
Fig.\ref{fig:Sxyhalf} essentially takes a {\it half-braiding} of the $\mathsf{S}^{xy}_b$ process of Fig.\ref{fig:3strings_2D_3D_xy},
and considers doing half-braiding on the same excitations in $|\alpha, a,b\rangle=|\beta, c,d\rangle$.
In principle, one can generalize the framed worldline picture of particles in Fig.\ref{fig:spin_statistics} to the framed worldsheet picture of closed-strings.
(ps. The framed worldline is like a worldsheet, the framed worldsheet is like a worldvolume.)
Such an interpretation shows that the topological spin of Fig.\ref{fig:Txy} and the exchange statistics of Fig.\ref{fig:Sxyhalf} carry the same data, namely
\bea \label{Spin-Statistics3D}
 {\sfT}^{xy}_b=\sfT^{a_y,b_z}_{\alpha_x} =(\sfS^{2\tD\;\alpha_x,\alpha_x}_{a_y,a_y\;(b_z)})^{\frac{1}{2}} \text{ or } (\sfS^{2\tD\;\alpha_x,\alpha_x}_{a_y,a_y\;(b_z)})^{\frac{1}{2}*} 
\eea
from the data of Eq.(\ref{eq:Txy}),(\ref{eq:Sxyb}). The equivalence holds, up to a (complex conjugate ${}^*$) sign caused by the orientation of the rotation and the exchange.

In Sec.(\ref{TypeII,III twist}), we will show, 
for the twisted gauge theory of Abelian topological orders, such an interpretation Eq.\ref{Spin-Statistics3D} is correct and agrees with our data. 
We shall name this as the {\bf spin-statistics relation for a closed string}.

In this section, we have obtained the explicit formulas of  $\mathsf{S}^{xyz}$,  $\mathsf{T}^{xy}$,  $\mathsf{S}^{xy}$ in Sec.\ref{STcocycle},\ref{Rep}, 
and as well as  captured the physical meanings of $\mathsf{S}^{xy}_b$,  $\mathsf{T}^{xy}_b$ in Sec.\ref{spin-statistics}.
Before concluding, we note that the full understanding of ${\sfS}^{xyz}$ seems to be intriguingly related to the 3D nature. 
It is {\it not} obvious to us that the use of 3D to 2D reduction can capture all physics of ${\sfS}^{xyz}$. 
We will come back to comment this issue in the Sec.\ref{summary}.

\section{SL$(3,\mathbb{Z})$ Modular Data and Multi-String Braiding} \label{Sec:SL3ZMulti-String Braiding}

\subsection{Ground state degeneracy and Particle, String types} \label{sec:GSD} 

We now proceed to study the topology-dependent ground state degeneracy (GSD), modular data $\mathsf{S}$, $\mathsf{T}$ of 3+1D twisted gauge theory 
with finite group $G=\prod_i Z_{N_i}$. 
We shall comment that the GSD on $\mathbb{T}^2$ of 2D topological order counts the number of quasi-particle excitations, 
which from the Representation (Rep) Theory is simply counting the number of charges $\alpha$ and  fluxes $a$ forming the quasi-particle basis $|\alpha, a \rangle$ spanned the ground state Hilbert space.
{\bf In 2D, GSD counts the number of types of quasi-particles} (or anyons) as well as {\bf the number of basis $|\alpha, a \rangle$.}
For higher dimension,  GSD on $\mathbb{T}^d$ of $d$-D topological order still counts {\bf the number of canonical basis $|\alpha, a, b, \dots \rangle$},
however, may over count the number of types of {\bf particles (with charge), strings (with flux)}, etc excitations. 
From a untwisted $Z_N$ field theory perspective, the fluxed string may be described by a 2-form $B$ field, and the charged particle
may be described by a 1-form $A$ field, with a BF action $\int BdA$.
As we can see the fluxes $a,b$ are over-counted.\\

\begin{figure}[!h] 
\includegraphics[scale=0.35]{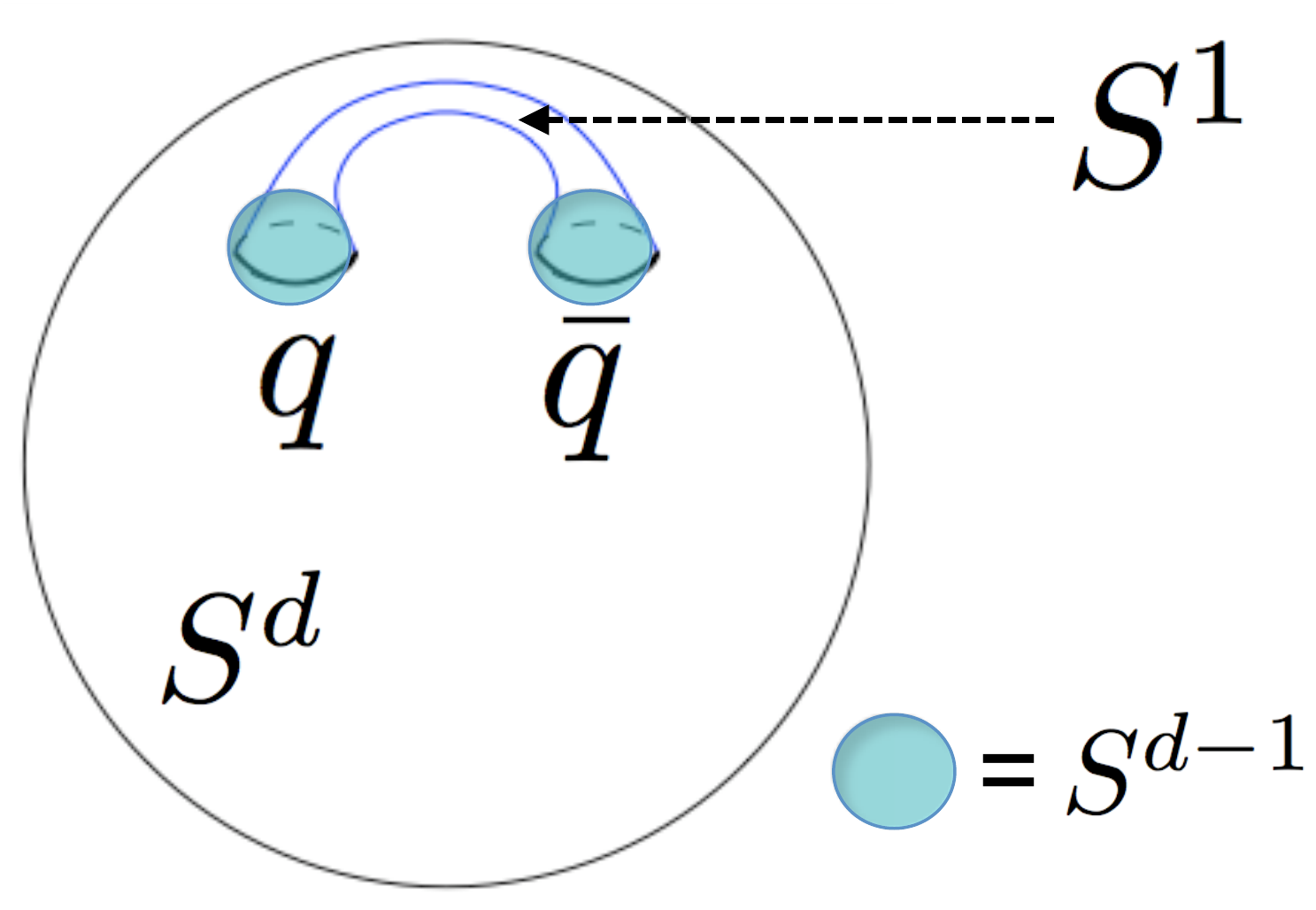}
\caption{
The number of particle types = GSD on $S^{d-1} \times S^1$
}
\label{fig:qpSphere} 
\end{figure}

We suggest that to count the number of types of particles of $d$-D is equivalent to Fig.\ref{fig:qpSphere} process, 
where we dig a ball $B^d$ with a sphere $S^{d-1}$ around the particle $q$, which resides on $S^d$. And we connect it through a $S^1$ tunnel to its anti-particle $\bar{q}$.
This process causes creation-annihilation from vacuum, and counts how many types of $q$ sectors is equivalent to:
\bea \label{eq:num_part}
\text{the number of particle types}=\text{GSD on } S^{d-1} \times I.
\eea
with $I \simeq S^1$ for this example. For spacetime integral, one evaluates Eq.(\ref{eq:GSDpathZ}) on $\cM=S^{d-1} \times S^1 \times S^1$.

For counting closed string excitations, one may naively use $\mathbb{T}^2$ to enclose a string as analogous to use $\mathbb{S}^2$ to enclose a particle in 3D.
Then, one may deduce 
$
\text{the number of string types}=\text{GSD on } \mathbb{T}^2 \times S^1\overset{?}{=}\mathbb{T}^3
$, 
and that of spacetime integral on $\mathbb{T}^4$, as we already mentioned earlier which is {\it incorrect} and overcounting.
We suggest, 
\begin{equation} 
\label{eq:num_string}
\text{the number of string types}=\mathsf{S}^{xy}, \mathsf{T}^{xy}\text{'s number of blocks},\;\;\;\;
\end{equation}
which block is labeled by $b$ as the form of Eq.\ref{eq:3Dto2DST}.
We will show the counting by Eq.(\ref{eq:num_part}), (\ref{eq:num_string}) in explicit examples in the next.

\subsection{Abelian examples: 3D twisted $Z_{N_1} \times Z_{N_2} \times Z_{N_3}$ gauge theories with Type II, III 4-cocycles}  \label{TypeII,III twist} 
We firstly study the most generic 3+1D finite Abelian twisted gauge theories with Type II, III 4-cocycle twists in Table \ref{table1}. 
It is general enough for us to consider $G=Z_{N_1} \times Z_{N_2} \times Z_{N_3}$ with non-vanished gcd $N_{ij}, N_{ijl}$.
The Type II, III (both their 1st and 2nd kinds) twisted gauge theory have GSD$=|G|^3$ on the spatial $\mathbb{T}^3$ torus. 
As such the canonical basis $|\alpha, a, b \rangle$ of the ground state sector labels the charge ($\alpha$ along $x$) and two fluxes (a, b along $y$, $z$), each of the three has $|G|$ kinds.
Thus, naturally from the Rep Theory viewpoint, we have GSD$=|G|^3$.
However, as mentioned in Sec.\ref{sec:GSD}, the $|G|^3$ overcounts the number of strings and particles. By using Eq.(\ref{eq:num_part}),(\ref{eq:num_string}),
we find there are $|G|$ types of particles and $|G|$ types of strings. The canonical basis $|\alpha, a, b \rangle$ (GSD on $\mathbb{T}^3$) counts twice the flux sectors.

In Table \ref{tableSxyzZN123}, we show their $\mathsf{S}^{xyz}$ by computing Eq.(\ref{Eq.Sxyz}), where we denote $a=(a_1,a_2,a_3,\dots)$, with $a_j \in Z_{N_j}$, and the same notation for other $b,c,d$ fluxes:
\begin{widetext}
\begin{center}
\begin{table} [!h]
\begin{tabular}{|c||c|c|c|}
\hline
\;$\mathcal{H}^4(G,\R/\Z)$\; &  4-cocycle &  $\sfS^{\alpha,\beta}_{d,a,b} $  & Induced $\sfS^{xy}_{b}$  \\[0mm]  \hline \hline 
$\Z_{N_{12}}$ & Type II 1st   & 
$\exp \big(  \underset{k}{\sum}  \frac{2 \pi \ti  }{ N_k   }  \; (\beta_k {a}_k-\alpha_k {d}_k )  \big) \cdot 
\exp \big( \frac{2 \pi \ti p_{{ \text{II}(12)}}^{(1st)} }{ (N_{12} \cdot N_2  )   }  \;  (a_1 d_2+a_2 d_1 ) b_2-2 a_2 b_1 d_2  \big)$ & Type I, II of $\mathcal{H}^3$ \\[2mm] \hline
$\Z_{N_{12}}$ & Type II 2nd & 
${\exp \big( \underset{k}{\sum}\frac{2 \pi \ti  }{ N_k   }  \; (\beta_k {a}_k-\alpha_k {d}_k )  \big) \cdot 
\exp \big( \frac{2 \pi \ti p_{{ \text{II}(12)}}^{(2nd)} }{ (N_{12} \cdot N_1  )   }  \;  (a_1 d_2+a_2 d_1 ) b_1-2 a_1 b_2 d_1 \big)} $ & Type I, II of $\mathcal{H}^3$  \\[2mm] \hline
$\Z_{N_{123}}$ & Type III 1st & 
${\exp \big( \underset{k}{\sum} \frac{2 \pi \ti  }{ N_k   }  \; (\beta_k {a}_k-\alpha_k {d}_k )  \big) \cdot 
\exp \big(  \frac{2 \pi \ti p_{{ \text{III}(123)}}^{(1st)} }{ (N_{12} \cdot N_3 )  }  \;  (a_1 b_2-a_2 b_1 )  d_3+(b_2 d_1-b_1 d_2 )  a_3 \big)}$ & two Type IIs of $\mathcal{H}^3$ \\[2mm] \hline
$\Z_{N_{123}}$ & Type III 2nd  & 
${\exp \big( \underset{k}{\sum} \frac{2 \pi \ti  }{ N_k   }  \; (\beta_k {a}_k-\alpha_k {d}_k )  \big) \cdot 
\exp \big(  \frac{2 \pi \ti p_{{ \text{III}(123)}}^{(2nd)} }{ (N_{31} \cdot N_2 )  }  \;  (a_3 b_1-a_1 b_3 )  d_2+(b_1 d_3-b_3 d_1 )  a_2 \big)}$ & two Type IIs of $\mathcal{H}^3$  \\[2mm] \hline
\end{tabular}
\caption{ $\mathsf{S}^{xyz}=\sfS^{xyz}_{(\alpha, a,b)(\beta, c,d)} \equiv\frac{1}{|G|}  \sfS^{\alpha,\beta}_{d,a,b} \delta_{bc}$ modular data of 3+1D twisted gauge theories with
$G=Z_{N_1} \times Z_{N_2} \times Z_{N_3}$.
In the last column, the $\mathcal{H}^3$ is the shorthand of $\mathcal{H}^3(G,\R/\Z)$; the induced $\sfS^{xy}_{b}$ is shown in Table \ref{tableSxyZN123}.
}
\label{tableSxyzZN123}
\end{table}
\end{center}
\end{widetext}
Here we 
extract the $\sfS^{\alpha,\beta}_{d,a,b}$ part of $\mathsf{S}^{xyz}$ ignoring the $|G|^{-1}$ factor:
\bea \label{eq:Sxyzdab}
\mathsf{S}^{xyz} 
=\sfS^{xyz}_{(\alpha, a,b)(\beta, c,d)} \equiv\frac{1}{|G|}  \sfS^{\alpha,\beta}_{d,a,b} \delta_{b, c}.
\eea
The $\sfS$--matrix reads ${g_x}_k=d_k$, ${g_y}_k=a_k$ in Eq.(\ref{Eq.Sxyz}).
In Table \ref{tableTxyZN123}, we show ${\sfT}^{xy}$. Here for Abelian $G$, with $\sfC^{(2)}_{a,b}(c,d)$ is a 2-coboundary (cohomologically trivial) thus $\text{dim(Rep)}=1$, 
we compute $\sfS^{xy}$ by Eq.(\ref{Eq:SxyAbSimp}) and that reduces to Eq.(\ref{eq:Sxyb})
$\mathsf{S}^{xy}_b= (\sfS^{xy})_{(\alpha,a,b)({\beta,c,d})}\equiv \frac{1}{|G|} \sfS^{2\tD\;\alpha,\beta}_{a,c\;(b)} \delta_{b,d}$.
In Table \ref{tableSxyZN123}, we show $\sfS^{xy}$ in terms of $\sfS^{2\tD\;\alpha,\beta}_{a,c\;(b)}$ for simplicty.
\begin{widetext}
\begin{center}
\begin{table} [!h]
\begin{tabular}{|c||c|c|c|}
\hline
\;$\mathcal{H}^4(G,\R/\Z)$\; &  4-cocycle &  $\sfT^{a,b}_{\alpha}$  & Induced  $\sfT^{xy}_{b}$  \\[0mm]  \hline \hline 
$\Z_{N_{12}}$ & Type II 1st   & 
$\exp \big( \underset{k}{\sum}    \frac{2 \pi \ti  }{ N_k   }  \; \alpha_k \cdot a_k  \big)
\cdot \exp \big( \frac{2 \pi \ti p_{{ \text{II}(12)}}^{(1st)} }{ (N_{12} \cdot N_2  )   }  \;  (a_2 b_1-a_1 b_2 ) (a_2)  \big)$ & Type I, II of $\mathcal{H}^3(G,\R/\Z)$ \\[2mm] \hline
$\Z_{N_{12}}$ & Type II 2nd & 
${\exp \big(  \underset{k}{\sum} \frac{2 \pi \ti  }{ N_k   }  \; \alpha_k \cdot a_k  \big) 
\cdot \exp \big( \frac{2 \pi \ti p_{{ \text{II}(12)}}^{(2nd)} }{ (N_{12} \cdot N_1  )   }  \;  (a_1 b_2-a_2 b_1 ) (a_1)  \big) } $ & Type I, II of $\mathcal{H}^3(G,\R/\Z)$  \\[2mm] \hline
$\Z_{N_{123}}$ & Type III 1st & 
${\exp \big(  \underset{k}{\sum} \frac{2 \pi \ti  }{ N_k   }  \; \alpha_k \cdot a_k  \big)
\cdot \exp \big( \frac{2 \pi \ti p_{{ \text{III}(123)}}^{(1st)} }{ (N_{12} \cdot N_3  )   }  \;  (a_2 b_1-a_1 b_2 ) (a_3)  \big)}$ & two Type IIs of $\mathcal{H}^3(G,\R/\Z)$ \\[2mm] \hline
$\Z_{N_{123}}$ & Type III 2nd  & 
${\exp \big(  \underset{k}{\sum} \frac{2 \pi \ti  }{ N_k   }  \; \alpha_k \cdot a_k \big)
\cdot \exp \big( \frac{2 \pi \ti p_{{ \text{III}(123)}}^{(2nd)} }{ (N_{31} \cdot N_2  )   } \;  (a_1 b_3-a_3 b_1 ) (a_2)  \big) }$ & two Type IIs of $\mathcal{H}^3(G,\R/\Z)$  \\[2mm] \hline
\end{tabular}
\caption{ ${\sfT}^{xy}$ modular data of the 3+1D twisted gauge theories with
$G=Z_{N_1} \times Z_{N_2} \times Z_{N_3}$. We can view this in terms of the index $b$ for blocks of ${\sfT}^{xy}_b=\sfT^{a_y,b_z}_{\alpha_x}$, with the flux $b$ along the compact $z$ direction.
}
\label{tableTxyZN123}
\end{table}
\begin{table} [!h]
\makebox[\textwidth][c]{
\begin{tabular}{|c||c|}
\hline
$\omega_4$ &  
$\sfS^{2\tD\;\alpha,\beta}_{a,c\;(b)}    = 
{ { \text{tr}\widetilde{\rho}^{a,b}_{\alpha}(a^2 c^{-1})^{*}   }}
  {  {  \; {\text{tr} \widetilde{\rho}^{c, b}_{\beta}(a c^{-2})}  }  }$    \\[0mm]  \hline \hline 
II 1st   &  %
${\exp \big( \underset{k}{\sum} \frac{2 \pi \ti  }{ N_k   }  (\alpha_k (c_k-2a_k)+ \beta_k (a_k-2c_k))  \big)   
 \cdot \exp \big( \frac{2 \pi \ti p_{{ \text{II}(12)}}^{(1st)} }{ (N_{12} \cdot N_2  )   }  \;  b_1 (2 a_2 c_2-2a_2^2 -2c_2^2 )
+  b_2 (2a_1 a_2 +2c_1 c_2 -a_1 c_2  -a_2 c_1)   \big)} $ 
\\[2mm] \hline
II 2nd & 
${\exp \big( \underset{k}{\sum}  \frac{2 \pi \ti  }{ N_k   }  ( \alpha_k (c_k-2a_k)+ \beta_k (a_k-2c_k))  \big)   
 \cdot \exp \big( \frac{2 \pi \ti p_{{ \text{II}(12)}}^{(2nd)} }{ (N_{12} \cdot N_1  )   }  \;  b_2 (2 a_1 c_1-2a_1^2 -2c_1^2 )
+  b_1 (2a_1 a_2 +2c_1 c_2 -a_1 c_2  -a_2 c_1)   \big)} $ 
\\[2mm] \hline
III 1st & 
${\exp \big( \underset{k}{\sum}   \frac{2 \pi \ti  }{ N_k   }  ( \alpha_k (c_k-2a_k)+ \beta_k (a_k-2c_k) ) \big)   
 \cdot \exp \big( \frac{2 \pi \ti p_{{ \text{III}(123)}}^{(1st)} }{ (N_{12} \cdot N_3  )   }  \;  b_1 ( a_2 c_3+ a_3 c_2- 2a_2 a_3 -2c_2 c_3 )
+  b_2 (2a_1 a_3 +2c_1 c_3 -a_1 c_3  -a_3 c_1)   \big) }$ 
\\[2mm] \hline
III 2nd  & 
${\exp \big( \underset{k}{\sum}   \frac{2 \pi \ti  }{ N_k   }  (\alpha_k (c_k-2a_k)+ \beta_k (a_k-2c_k) ) \big)   
 \cdot \exp \big( \frac{2 \pi \ti p_{{ \text{III}(123)}}^{(2nd)} }{ (N_{31} \cdot N_2  )   }  \;  b_3 ( a_1 c_2+ a_2 c_1- 2a_1 a_2 -2c_1 c_2 )
+  b_1 (2a_3 a_2 +2c_3 c_2 -a_3 c_2  -a_2 c_3)   \big)  
}$ 
\\[2mm] \hline
\end{tabular}
}
\caption{ ${\sfS}^{xy}$ modular data of the 3+1D twisted gauge theories with $G=Z_{N_1} \times Z_{N_2} \times Z_{N_3}$.
There are two more columns ($\mathcal{H}^4(G,\R/\Z)$, induced ${\sfS}^{xy}_b$) not shown here, since the data simply duplicates Table \ref{tableSxyzZN123}'s first and fourth column. 
The basis chosen here is not canonical for excitations, in the sense that particle braiding around trivial vacuum still gain a non-trivial statistic phase. 
Finding the proper canonical basis for each $b$ block of ${\sfS}^{xy}_b$ can be done by Ref.\onlinecite{1303.0829}'s method.
}
\label{tableSxyZN123}
\end{table}
\end{center}
\end{widetext}

{\bf Several remarks follow}:\\
(1)  For an untwisted gauge theory (topological term $p_{..}=0$), which is the direct product of $Z_N$ gauge theory or $Z_N$ toric code, its statistics has the form 
$\exp \big(  \underset{k}{\sum}  \frac{2 \pi \ti  }{ N_k   }  \; (\beta_k {a}_k-\alpha_k {d}_k )  \big)$ and 
$\exp \big( \underset{k}{\sum}    \frac{2 \pi \ti  }{ N_k   }  \; \alpha_k \cdot a_k  \big)$. This shall be described by the BF theory of $\int B dA$ action. 
With $\alpha, \beta$ as the charge of particles (1-form gauge field $A$), $a,b$ as the flux of string(2-form gauge field $B$). This essentially describes
{\bf the braiding between a pure-particle and a pure-string}.
\\

\noindent
(2)  Both ${\sfS}^{xy}$, ${\sfT}^{xy}$ have block diagonal forms as ${\sfS}^{xy}_b$, ${\sfT}^{xy}_b$ respect to the $b$ flux (along $z$) correctly reflects what 
Eq.(\ref{eq:3Dto2DST}) preludes already.\\

\noindent
(3) ${\sfT}^{xy}$ is in SL$(3,\Z)$ canonical basis automatically and full-diagonal, but ${\sfS}^{xy}$ may not be in the canonical basis 
for each blocks of ${\sfS}^{xy}_b$, due to its SL$(2,\Z)$ nature. 
We can find the proper basis in each $b$ block by Ref.\onlinecite{1303.0829} method. Nevertheless, the eigenvalues of ${\sfS}^{xy}$ in Table \ref{tableSxyZN123} 
are still proper and invariant regardless any basis. \\

\noindent
(4) {\bf Characterization of topological orders}:  We can further compare the 3D ${\sfS}^{xy}_b$ data to SL$(2,\Z)$'s data of 2D ${\sfS}^{xy}$ of 
$\mathcal{H}^3(G,\R/\Z)$ in Table \ref{tableA1}. 
(see Appendix \ref{App:cocycles} for data) 
All of the dimensional reduction of these data 
(${\sfS}^{xy}_b$ in Table \ref{tableSxyzZN123}, \ref{tableSxyZN123},
and  ${\sfT}^{xy}_b$ in Table \ref{tableTxyZN123}) agree with 3-cocycle (induced from 4-cocycle $\omega_4$) in Table \ref{table1}'s last column.
Gathering all data, we conclude that Eq.(\ref{eq:C3DtoC2Domb}) becomes explicitly. For example, Type II twists for $G=(Z_2)^2$ as,
\bea
&& {\cC^{3\tD}_{{(Z_{2})^2},1}} =  4 \cC_{{(Z_{2})^2},1}^{2\tD}  \;\;\;\;\; \label{eq.Z2toric_3D}\\
&& {\cC^{3\tD}_{(Z_{2})^2},\omega_{4,\tII}} =  \cC_{(Z_{2})^2}^{2\tD} \oplus   \cC_{{(Z_{2})^2},\omega_{3,\tI}}^{2\tD} 
\oplus 2 \cC_{{(Z_{2})^2},\omega_{3,\tII}}^{2\tD} \;\;\;\;\; \label{eq.Z2twist_3D}
\eea
{Such a Type II $\omega_{4,\tII}$ can produce a $b=0$ sector of ($Z_2$ toric code $\otimes$ $Z_2$ toric code) of 2D as $\cC_{(Z_{2})^2}^{2\tD}$,
some $b \neq 0$ sector of ($Z_2$ double-semions $\otimes$ $Z_2$ toric code) 
as $\cC_{{(Z_{2})^2},\omega_{3,\tI}}^{2\tD}$ and another $b \neq 0$ sector $\cC_{{(Z_{2})^2},\omega_{3,\tII}}^{2\tD}$, for example.}
This procedure can be applied to other types of cocycle twists.\\

\noindent
(5) {\bf Classification of topological orders}: \\
We shall interpret the decomposition in Eq.(\ref{eq:C3DtoC2Domb}) as the implication of classification. 
Let us do the {\it counting of number of phases} in the simplest example of Type II, $G=Z_2\times Z_2$ twisted theory.
There are four types in $(p_{{ \text{II}(12)}}^{(1st)}, p_{{ \text{II}(12)}}^{2nd} ) \in \mathcal{H}^4(G,\R/\Z)=(\Z_2)^2$.
However, we find there are {\bf only two distinct topological orders} out of four. 
One is the trivial $(Z_2)^2$ gauge theory as Eq.(\ref{eq.Z2toric_3D}), the other is the nontrivial type as Eq.(\ref{eq.Z2twist_3D}).
There are two ways to see this, 
(i) from the full ${\sfS}^{xyz}$, ${\sfT}^{xy}$ data. 
(ii) viewing the sector of ${\sfS}^{xy}_b$, ${\sfT}^{xy}_b$ under distinct fluxes $b$, which is from a $\mathcal{H}^3(G,\R/\Z)$ perspective.
We should beware that in principle {\it tagging particles, strings or gauge groups is not allowed}, so one can identify many seemly-different orders by relabeling their excitations. 
We will give more examples of counting 2D, 3D topological orders in Appendix \ref{App:cocycles}. 

\noindent
(6)  {\bf Spin-statistics relation of closed strings} in Eq.(\ref{Spin-Statistics3D}) is verified 
 to be correct here, while we take the complex conjugate in Eq.(\ref{Spin-Statistics3D}).
 This is why we draw the orientation of Fig.\ref{fig:Txy},\ref{fig:Sxyhalf} oppositely.
Interpreting ${\sfT}^{xy}$ as the {\bf topological spin} also holds.\\

\noindent
(7) {\bf Cyclic relation for $\sfS^{xyz}$ in 3D}:
For all above data (Type II, Type III), 
there is a special cyclic relation 
when the charge labels are equal $\alpha=\beta$ for $\sfS^{\alpha,\beta}_{a,b,d}$
(e.g. for pure fluxes $\alpha=\beta=0$, namely for pure strings):
\bea \label{eq:cyclic_S_eq}
\sfS^{\alpha,\alpha}_{a,b,d} \cdot \sfS^{\alpha,\alpha}_{b,d,a} \cdot \sfS^{\alpha,\alpha}_{d,a,b}=1.
\eea
However, such a cyclic relation does not hold (even at the zero charge) for $\sfS^{2\tD\;\alpha,\beta}_{a,c\;(b)}$, namely 
$\sfS^{2\tD\;\alpha,\beta}_{a,c\;(b)} \cdot \sfS^{2\tD\;\alpha,\beta}_{c,b\;(a)} \cdot \sfS^{2\tD\;\alpha,\beta}_{b,a\;(c)} \neq 1$ in general.
Some other cyclic relations are studied recently in Ref.\onlinecite{{WL1437},{JMR1462}}, 
for which we have not yet made detailed comparisons but the perspectives may be different.
In Ref.\onlinecite{JMR1462}, their cyclic relation is determined by triple linking numbers associated with the membrane operators.  
In Ref.\onlinecite{WL1437}, their cyclic relation is related to the loop braiding of Fig.\ref{fig:3strings_2D_3D_xy}, which has its relevancy instead 
to $\sfS^{2\tD\;\alpha,\beta}_{a,c\;(b)}$, not our cyclic relation of $\sfS^{\alpha,\beta}_{a,b,d}$ for 3D. 
We will comment more about the difference and the subtlety of $\sfS^{xy}$ and $\sfS^{xyz}$ in Sec.\ref{summary:cyclic}.

\subsection{Non-Abelian examples: 3D twisted $(Z_n)^4$ gauge theories with Type IV 4-cocycle}  \label{sec:TypeIV4cocycle}

We now study a more interesting example, a generic 3+1D finite Abelian twisted gauge theory with Type IV 4-cocycle twists with $p_{ijlm}\neq 0$ in Table \ref{table1}. 
For generality, the formula we have also incorporated  Type IV twists together with the aforementioned Type II, III twists. 
So
all 4-cocycle twists will be discussed in this subsection.
Differ from the previous example in Sec.\ref{TypeII,III twist} of Abelian topological order with Abelian statistics,
we will show Type IV 4-cocycle $\omega_{4,\tIV}$  
will cause the {\it gauge theory becomes non-Abelian, having non-Abelian statistics even if the original $G$ is Abelian}.
Our inspiration roots in a 2D example for Type III 3-cocycle twist in Table \ref{tableA1}
 will cause a similar effect, discovered 
in Ref.\onlinecite{deWildPropitius:1995cf}.
In general, one can consider $G=Z_{N_1} \times Z_{N_2} \times Z_{N_3} \times Z_{N_4}$ with non-vanished gcd $N_{1234}$;
while we will focus on $G=(Z_n)^4$ with $N_{1234}=n$, with $n$ is prime for simplicity.
From $\cH^4(G,\R/\Z)=\Z_{n}^{21}$, we have $n^{21}$ types of theories, while $n^{20}$ are Abelian gauge theories, and 
$n^{20} \cdot (n-1)$ types with Type IV $\omega_4$ are endorsed with non-Abelian statistics.
\\

\noindent
{\bf Ground state degeneracy (GSD)-}

We compute the GSD of gauge theories with a Type IV twist on the spatial $\mathbb{T}^3$ torus, truncated from $=|G|^3=|n^4|^3=n^{12}$ to:
\bea
\text{GSD}_{\mathbb{T}^3,\tIV}= \big(n^8+n^9-n^5\big) &+&\big(n^{10} -n^{7} -n^{6} +n^{3}\big) \;\;\;\;\;\;\;\\
 \equiv \text{GSD}^{Abel}_{\mathbb{T}^3,\tIV} &+& \text{GSD}^{nAbel}_{\mathbb{T}^3,\tIV} \label{eq:T3TypeIVdec}
\eea
(We derive the above only for a prime $n$. The GSD truncation is less severe and is in between $\text{GSD}_{\mathbb{T}^3,\tIV}$ and $|G|^3$ for non-prime.)
As such the canonical basis $|\alpha, a, b \rangle$ of the ground state sector on $\mathbb{T}^3$ {\it no longer} have $|G|^3$ labels 
with the $|G|$ number charge and two pairs of $|G| \times |G|$ number of fluxes 
as Sec.\ref{TypeII,III twist}. This truncation is due to the nature of non-Abelian physics of Type IV $\omega_{4,\tIV}$ twisted.
We shall explain our notation in Eq.(\ref{eq:T3TypeIVdec}), the ($n$)$Abel$ means the contribution from (non-)Abelian excitations. 
From the Rep Theory viewpoint, we can recover the truncation back to $|G|^3$ by carefully reconstructing the {\it quantum dimension of excitations}. 
We obtain
\bea \label{eq:GSDtotalT3}
&& |G|^3= { \big(\text{GSD}^{Abel}_{\mathbb{T}^3,\tIV}\big) +\big(\text{GSD}^{nAbel}_{\mathbb{T}^3,\tIV}\big) \cdot n^2}  \\ 
&& ={ \{n^4+n^5-n\} \cdot n^4 \cdot (1)^2 }\nonumber\\
&& {+\{(n^{4})^2 -n^{5} -n^{4} +n\} \cdot n^2 \cdot (n)^2} \nonumber \\ 
&& = { \{  \text{Flux}^{Abel}_\tIV  \} \cdot n^4 \cdot (\text{dim}_1)^2+  \{ \text{Flux}^{nAbel}_\tIV  \} \cdot n^2 \cdot (\text{dim}_n)^2} \nonumber
\eea
The $\text{dim}_{m}$ means the dimension of Rep\,as $\text{dim(Rep)}$ is $m$, which is also the {\it quantum dimension} of excitations. Here we have 
a dimension $1$ for Abelian and $n$ for non-Abelian. 
In summary, we understand the decomposition precisely 
in terms of each (non-)Abelian contribution by
\bea \label{eq:GSDarray}
\left\{
 \begin{array}{l}
\text{flux sectors}=|G|^2=|n^4|^2= \text{Flux}^{Abel}_\tIV  +\text{Flux}^{nAbel}_\tIV \\ 
\text{GSD}_{\mathbb{T}^3,\tIV}=  \text{GSD}^{Abel}_{\mathbb{T}^3,\tIV}+ \text{GSD}^{nAbel}_{\mathbb{T}^3,\tIV} \\
\text{dim(Rep)}^2=1^2, n^2\\
\text{Numbers of charge Rep}= n^4, n^2.
\end{array}
\right.
\eea
 
Actually, the canonical basis $|\alpha, a, b \rangle$ (GSD on $\mathbb{T}^3$) still works,
the sum of Abelian $\text{Flux}^{Abel}_\tIV$ and non-Abelian  $\text{Flux}^{nAbel}_\tIV $ counts the flux number of $a,b$ as the unaltered $|G|^2$.
The charge Rep $\alpha$ is unchanged with a number of $|G|=n^4$ for Abelian sector with a rank-1 matrix (1-dim linear or projective) representation, 
however, the charge Rep $\alpha$ is truncated to a smaller number $n^2$ for non-Abelian sector also with a larger rank-n matrix (n-dim projective) representation.

Another view on $\text{GSD}_{\mathbb{T}^3,\tIV}$ can be inspired by a generic formula like Eq.(\ref{eq:3Dto2DST})
\bea
\text{GSD}_{\cM' \times S^1}=\oplus_b \text{GSD}_{b,{\cM'}}= \sum_b \text{GSD}_{b,\cM'}, 
\eea
where we sum over GSD in all different $b$ flux sectors, with $b$ flux along $S^1$. 
Here we can take $\cM' \times S^1=\mathbb{T}^3$ and $\cM'=\mathbb{T}^2$.
For non-Type IV (untwisted, Type II, III) $\omega_4$ case, 
we have $|G|$ sectors of $b$ flux and each has $\text{GSD}_{b,{\mathbb{T}^2}}=|G|^2$.
For Type IV $\omega_4$ case $G=(Z_n)^4$ with a prime $n$, we have 
\bea
&&\text{GSD}_{\mathbb{T}^3,\tIV} \nonumber\\
&&=|G|^2+(|G|-1)\cdot |Z_n|^2\cdot (1 \cdot  |Z_n|^3 + (|Z_n|^2-1) \cdot n) \nonumber \\
&&=n^8+(n^4-1)\cdot n^2\cdot (1 \cdot n^3 + (n^3-1) \cdot n).
\eea
As we expect, the first part is from the zero flux $b=0$, which is the normal untwisted 2+1D $(Z_n)^4$ gauge theory (toric code) as $\cC^{2D}_{(Z_n)^4}$
with $|G|^2=n^8$ on 2-torus.
The remained $(|G|-1)$ copies are inserted with nonzero flux ($b\neq 0$) as $\cC^{2D}_{(Z_n)^4,\omega_3}$ with Type III 3-cocycle twists of Table \ref{tableA1}.
In some case but not all cases, $\cC^{2D}_{(Z_n)^4,\omega_3}$ is $\cC^{2D}_{{(Z_n)}_{\text{untwist}} \times (Z_n)^3_{\text{twist}},\omega_3}$.
In either case, the $\text{GSD}_{b,{\mathbb{T}^2}}$ for $b \neq 0$ has the same decomposition always equivalent to a untwisted $Z_n$ gauge theory with $GSD_{\mathbb{T}^2}=n^2$ direct product with
\bea
&&\text{GSD}_{\mathbb{T}^2,\omega_{3,\tIII}} =(1 \cdot n^3 + (n^3-1) \cdot n)\\ 
&&  \equiv \text{GSD}^{Abel}_{\mathbb{T}^2,\omega_{3,\tIII}} + \text{GSD}^{nAbel}_{\mathbb{T}^2,\omega_{3,\tIII}}, 
\eea
which we generalize the result derived for 2+1D Type III $\omega_3$ twisted theory with $G=(Z_2)^3$ in Ref.\onlinecite{deWildPropitius:1995cf} to $G=(Z_n)^3$ of a prime $n$.
One can repeat the counting for 2+1D as Eq.(\ref{eq:GSDtotalT3})(\ref{eq:GSDarray}), see Appendix \ref{App:cocycles}.

To summarize, from the GSD counting, we already foresee there exist {\bf non-Abelian strings in 3+1D Type IV twisted gauge theory, 
with a quantum dimension $n$.} Those non-Abelian strings (fluxes) carries $\text{dim(Rep)}=n$ non-Abelian charges. Since charges are sourced by particles,
those {\bf non-Abelian strings are not pure strings but attached with non-Abelian particles.} 
(For a projection perspective from 3D to 2D, a nonAbelain string of $\cC^{3\tD}$ is a non-Abelain dyon with both charge and flux of $\cC^{2\tD}_b$.)\\

\noindent
{\bf Modular $\sfT^{xy}$ of 3D-}\\

We shall compute $\sfT^{xy},\sfS^{xyz}$ using the formula derived in Sec.\ref{Rep} for Type IV $\omega_4$ theory (for generality, we also include the twists by Type II, III $\omega_4$). Due to the large GSD and the quantum dimension of non-Abelian nature,
we focus on a simplest example $G=(Z_2)^4$ theory to have the smallest amount of data. 
By $\cH^4(G,\R/\Z)=\Z_{2}^{21}$, we have $2^{21}$ types of theories, where 
$2^{20}$ types with Type IV are endorsed with non-Abelian statistics. 
(While $2^{20}$ types are Abelian gauge theories of non-Type IV have their $\sfT,\sfS$ data in Sec.\ref{TypeII,III twist}.)
For $G=(Z_2)^4$, there are still $\text{GSD}_{\mathbb{T}^3,\tIV}=1576$.
Thus both $\sfT$ and $\sfS$ are matrices with the rank 1576.
$\sfT^{xy}$ has 1576 components along diagonal.\\

For $G=(Z_2)^4$, we firstly define a quantity ${\eta_{g_1,g_2,g_3}}$ of convenience 
from the $\mathsf{C}^{(2)}_{a,b}(c,d)$ in Eq.(\ref{eq:2slant}), 
\bea
{\eta_{g_1,g_2,g_3}} \equiv\left\{
    \begin{array}{lr}
      0,& \text{ if } \mathsf{C}^{(2)}_{g_1,g_2}(g_3,g_3)=+1\\
      1,& \text{ if } \mathsf{C}^{(2)}_{g_1,g_2}(g_3,g_3)=-1
    \end{array}
    \right.
\eea
Below the $p_{lm}$, $p_{lmn}$ are the shorthand of Type II, III (both 1st, 2nd) topological term labels, the $ p_{lm}  f_{{}_{lm}}(a,b,c)$, $p_{lmn} f_{{}_{lmn}}(a,b,c)$ 
abbreviate the function forms in the exponents of Type II, III $\omega^4$ in Table \ref{table1}. 
Namely, we regard their 4-cocycle $\omega_4(a,b,c,d)$ as a trivial 2-cocycle $\sfc_{a,b}(c,d)$ written as 
$\sfc_{a,b}(c,d) = \frac{ \eta_{a,b}(c) \eta_{a,b}(d) }{ \eta_{a,b}(c+d) }$, where $\eta_{a,b}(c)$ is a 1-cochain:
$\eta_{a,b}(c)=\exp({\ti p_{lm}  f_{{}_{lm}}(a,b,c)})=\exp({ \frac{2 \pi \ti}{N_{lm} N_m} p_{lm}  a_l b_m c_m)})$ for Type II case.
$\eta_{a,b}(c)=\exp({\ti p_{lmn} f_{{}_{lmn}}(a,b,c)})=\exp({ \frac{2 \pi \ti}{N_{lm} N_n} p_{lmn}  a_l b_m c_n)})$ for Type III case.
We derive ${ \sfT^{xy}=\sfT^{a_y,b_z}_{\alpha_x}}$ of Eq.(\ref{eq:Txy}) 
in Table \ref{TxyZ2f}.

\onecolumngrid
\begin{widetext}
\begin{table}[!h]
\makebox[\textwidth][c]{
\begin{tabular}{|c||c|} 
\hline
Excitations $(\alpha, a,b)$\; & \;  $\mathsf{T}^{a,b}_{\alpha}$ \\ \hline
$(\alpha,F(j_{Abel})),$ \; & \; $\exp \big(\sum_{k=1}^4 \pi \ti \;\alpha_k a_k \big)$  \;  $\to$ e.g. $\pm 1$\\ \hline
$(((\pm)_a,(\pm)_b),F(j_{nAbel}))$ \; & \; $e^{\ti \frac{\pi}{2} ( \underset{l<m<n}{\underset{l,m,n\in\{1,2,3,4\}}{\sum}} p_{lm} f_{{}_{lm}}(a,b,a) +p_{lmn} f_{{}_{lmn}}(a,b,a) )} (\pm)_a (\pm)_b (\ti)^{\eta_{a,b,a}} $  \;  $\to$ e.g. $\pm 1$ { or }  $\pm \ti$\\ 
\hline
 \end{tabular}
}
\caption{ SL$(3,\Z)$ modular data ${ \sfT^{xy}=\sfT^{a_y,b_z}_{\alpha_x}}$ for $(Z_2)^4$ theory with Type IV $\omega^4$.
The formula of $\sfT^{xy}$ is separated to two sets: the first set with 736 components (from the sector $\text{GSD}^{Abel}_{\mathbb{T}^3,\tIV}$) and 
another 840 components
(from the sector $\text{GSD}^{nAbel}_{\mathbb{T}^3,\tIV} $). 
$F=(a_i,b_i)$ are fluxes with 8 components, $(a_1,a_2,a_3,a_4) \in (Z_2)^4$ and $(b_1,b_2,b_3,b_4) \in (Z_2)^4$.
The number of distinct fluxes in $F(j_{Abel})$ is 46($= \text{Flux}^{Abel}_\tIV$), the number of distinct fluxes $F(j_{nAbel})$ is 210($= \text{Flux}^{nAbel}_\tIV$).
This table contains {\it all } $2^{20}$ kinds of ${ \sfT^{xy}}$ for the non-Abelian theories in $\cH^4(G,\R/\Z)=\Z_{2}^{21}$ (half of $2^{21}$). 
$((\pm)_a,(\pm)_b)$ pair makes up the numbers of charge Rep $n^2=2^2$ in Eq.(\ref{eq:GSDarray}). Details of the rank-2 matrix Rep is shown in Appendix \ref{App:cocycles}.
}\label{TxyZ2f}
\end{table}

\noindent
{\bf Modular $\sfS^{xyz}$ of 3D-}

The $\sfS^{xyz}$ matrix has $1576 \times 1576$ components. We organize $\sfS^{xyz}$ into four blocks, denoting $(n)Abel$ for (non)Abelian  with 736 (840) components.
Defining $\mathsf{S}^{xyz}_{(\alpha,a,b)(\beta,c,d)} \equiv \frac{1}{|G|}\mathsf{S}^{\alpha,\beta}_{a,b,d} \delta_{b,c}$, we obtain 
\begin{equation} \label{eq:SnAb3DZ2f}
\mathsf{S}^{xyz}=\frac{1}{|G|}\bBigg@{3.5}( \mkern-10mu
\begin{tikzpicture}[baseline=-.65ex]
\matrix[
  matrix of math nodes,
  column sep=1ex,
] (m)
{
\mathsf{S}_{Abel,Abel} & \mathsf{S}_{Abel,nAbel}\\
\mathsf{S}_{nAbel,Abel} & 
\mathsf{S}_{nAbel,nAbel}  \\
};
\draw[dashed]
  ([xshift=1. ex]m-1-1.north east) -- ([xshift=.5 ex]m-2-1.south east);
\draw[dashed]
  ([xshift=-4.ex,yshift=.ex] m-1-1.south west) -- ([xshift=4. ex,yshift=.ex]m-1-2.south east);
\node[above] at ([xshift=-1.ex, yshift=.2ex]m-1-1.north) {$\scriptstyle (\beta_1,\beta_2,\beta_3,\beta_4,\;c,d)$};  
\node[above] at ([xshift=.5 ex, yshift=.2ex]m-1-2.north) {$\scriptstyle ((\pm)_c,(\pm)_d,\; c, d)$};
\node[above] at ([xshift=-1.5ex, yshift=2.2ex]m-1-1.north) {\small{\text{736 components}}};  
\node[above] at ([xshift=.5 ex, yshift=2.2ex]m-1-2.north) {\small{\text{ 840 components}}};
\node[right,overlay] at ([xshift=3.ex,yshift=.4ex]m-1-2.east) {$\scriptstyle (\alpha_1,\alpha_2,\alpha_3,\alpha_4,\; a,b)$};
\node[right,overlay] at ([xshift=3.ex]m-2-2.east) {$\scriptstyle ((\pm)_a,(\pm)_b,\;a,b)$};
\end{tikzpicture}\mkern-16mu
\bBigg@{3.5})
\end{equation}
\bea \label{eq:SnAb3DZ2fAr}
\left\{
    \begin{array}{l l}
\mathsf{S}_{Abel,Abel} =& 1 \cdot \exp( \underset{k}{\sum}\frac{2\pi \ti}{N_k}(-\alpha_k d_k+\beta_k a_k)) \cdot \delta_{b,c} =
(-1)^{(-\alpha_k d_k+\beta_k a_k)} \cdot \delta_{b,c},\\[1.5mm]
\mathsf{S}_{Abel,nAbel} =& 2\cdot (-1)^{(-\alpha_k d_k)}  
\cdot e^{\ti \frac{\pi}{2} ( \underset{l<m<n}{\underset{l,m,n\in\{1,2,3,4\}}{\sum}} p_{lm} f_{{}_{lm}}(b,d,a) +p_{lmn} f_{{}_{lmn}}(b,d,a) )} 
(\pm 1)_b (\pm 1)_d
\cdot (\ti)^{\eta_{b,d,a}} \delta_{a\in\{\mathbf{1},b,d, bd\}}
\cdot \delta_{b,c},\;\;\;\;\;\;\;\;\\[1.5mm]
\mathsf{S}_{nAbel,Abel} =& 2\cdot (-1)^{(\beta_k a_k)}  
\cdot e^{-\ti \frac{\pi}{2} ( \underset{l<m<n}{\underset{l,m,n\in\{1,2,3,4\}}{\sum}} p_{lm} f_{{}_{lm}}(a,b,d) +p_{lmn} f_{{}_{lmn}}(a,b,d) )} 
(\pm 1)_a (\pm 1)_b
\cdot (\ti)^{\eta_{a,b,d}} \delta_{d\in\{\mathbf{1},a,b, ab\}}
\cdot \delta_{b,c},\;\;\;\;\;\;\;\;\\[1.5mm]
 \mathsf{S}_{nAbel,nAbel} =& 4 
\cdot e^{-\ti \frac{\pi}{2} ( \underset{l<m<n}{\underset{l,m,n\in\{1,2,3,4\}}{\sum}} p_{lm} f_{{}_{lm}}(a,b,d) +p_{lmn} f_{{}_{lmn}}(a,b,d) )} 
\cdot e^{\ti \frac{\pi}{2} ( \underset{l<m<n}{\underset{l,m,n\in\{1,2,3,4\}}{\sum}} p_{lm} f_{{}_{lm}}(b,d,a) +p_{lmn} f_{{}_{lmn}}(b,d,a) )} \\
&\cdot (\pm 1)_a (\pm 1)_b (\pm 1)_c (\pm 1)_d  \cdot (-\ti)^{\eta_{a,b,d}} \cdot (\ti)^{\eta_{b,d,a}} \cdot \delta_{a\in\{b,d, bd\}} \cdot \delta_{d\in\{a,b, ab\}} \cdot \delta_{b,c}.
 \end{array}
  \right.
\eea
\end{widetext}
\twocolumngrid
The $\exp( \underset{k}{\sum}\frac{2\pi \ti}{N_k}(-\alpha_k d_k+\beta_k a_k))$ factor in the top-left block shows the pure-particle pure-string braiding 
of untwisted $Z_N$ gauge theory (no $\omega_4$ dependence).
We define $\delta_{a\in\{b,d, bd\}} =1$ if ${a\in\{b,d, bd\}}$ , otherwise  $\delta_{a\in\{b,d, bd\}} =0$.
Some other technical details follow: for $G=(Z_2)^4$,
the constraint $\delta_{a\in\{b,d, bd\}} \cdot \delta_{d\in\{a,b, ab\}}$ reduces to $\delta_{d\in\{a, ab\}}$. The survival nonzero $ \mathsf{S}_{nAbel,nAbel}$ are only in two kinds of forms, either $d=a$ or $d=ab$. 
\bea
\mathsf{S}_{nAbel,nAbel}=\left\{
    \begin{array}{l}
      \mathsf{S}^{\alpha,\beta}_{a,b,a} \delta_{b,c} \delta_{d,a}\\
      \mathsf{S}^{\alpha,\beta}_{a,b,ab} \delta_{b,c} \delta_{d,ab}
    \end{array}
    \right.
\eea

\noindent
{\bf Some remarks follow}:\\

\noindent
(1) {\bf Dimensional reduction from 3D to 2D sectors with $b$ flux}: From the above $\sfS^{xyz},\sfT^{xy}$, there is no difficulty to derive $\sfS^{xy}$ from Eq.(\ref{eq:Sxydecompose}).
From all these modular data $\sfS^{xy}_b, \sfT^{xy}_b$ data, we find consistency 
with the dimensional reduction of 3D topological order by 
comparing with induced 3-cocycle $\omega_3$ from $\omega_4$. 
Let us consider a single specific example, given the Type IV $p_{1234}=1$ and other zero Type II,III indices $p_{..}=p_{...}=0$,
\bea \label{eq:TypeIV3D2Dbranch}
&& \cC^{3\tD}_{(Z_2)^4, \omega_{4,\tIV}}  = \oplus_b \cC_b^{2\tD}   \\
&& =\cC^{2\tD}_{(Z_2)^4} \oplus  10 \; \cC^{2\tD}_{(Z_2) \times (Z_2)^3_{(ijl)},  \omega_{3,\tIII}^{(ijl)} }  \oplus 5 \cC^{2\tD}_{(Z_2)^4, \omega_{3,\tIII} \times \omega_{3,\tIII} \times \dots} \nonumber\\   
&& =\cC^{2\tD}_{(Z_2)^4} \oplus  10 \; \cC^{2\tD}_{(Z_2) \times (D_4)}  \oplus 5 \cC^{2\tD}_{ (Z_2)^4, \omega_{3,\tIII} \times \omega_{3,\tIII} \times \dots} \nonumber   
\eea
The $\cC^{2\tD}_{(Z_2)^4}$ again is 
the normal ${(Z_2)^4}$ gauge theory at $b=0$.
The 10 copies of $\cC^{2\tD}_{(Z_2) \times (D_4)}$ 
with a untwisted dihedral $D_4$ gauge theory ($|D_4|=8$) product with the normal ${(Z_2)}$ gauge theory.
The duality to $D_4$ theory in 2D can be expected,\cite{deWildPropitius:1995cf} see Table \ref{table:Z2cubeclass}.
(For a byproduct of our work, we go beyond Ref.\onlinecite{deWildPropitius:1995cf}
to give the complete classification of all twisted 2D $\omega_3$ of $G=(Z_2)^3$ and their 
corresponding topological orders and twisted quantum double $D^{\omega}(G)$ in Appendix.\ref{App:cocycles}.) 
The remained 5 copies $\cC^{2\tD}_{ (Z_2)^4, \omega_{3,\tIII} \times \omega_{3,\tIII} \times  \dots} $ must contain the twist on the full group $(Z_2)^4$, not just its subgroup. 
This peculiar feature suggests the following remark.\\

\noindent
(2) 
Sometimes there may exist a duality between a twisted Abelian gauge theory and a untwisted non-Abelian gauge theory,\cite{deWildPropitius:1995cf}
one may wonder whether one can find a dual non-Abelian gauge theory for  $\cC^{3\tD}_{(Z_2)^4, \omega_{4,\tIV}}$?
We find that, however, {\bf $\cC^{3\tD}_{(Z_2)^4, \omega_{4,\tIV}}$ cannot be dual to a normal gauge theory (neither Abelian nor non-Abelian),
but must be a  twisted (Abelian or non-Abelian) gauge theory}. The reason is more involved. 
Let us firstly recall the more familiar 2D case. One can consider $G=(Z_2)^3$ example with $\cC^{2\tD}_{(Z_2)^3, \omega_{3}}$, 
with $\cH^3(G,\R/\Z)=(\Z_{2})^{7}$. 
There are $2^6$ for non-Abelian types with Type III $\omega_3$ (the other $2^6$ Abelian without with Type III $\omega_3$).
We find the $64$ non-Abelian {\bf types} of 3-cocycles $\omega_3$ go to 5 {\bf classes} labeled $\omega_3[1]$, $\omega_3[{3d}]$,
$\omega_3[{3i}]$, $\omega_3[{5}]$ and $\omega_3[{7}]$, and their twisted quantum double model $D^\omega(G)$ are shown in Table \ref{table:Z2cubeclass}. 
The number in the bracket $[..]$ is related to the number of pairs of $\pm \ti$ in the $\sfT$ matrix and the $d/i$ stand for the linear dependence($d$)/independence($i$)
of fluxes generating cocycles. 
\onecolumngrid
\begin{widetext}
\begin{table}[!h]
\begin{tabular}{|c||c| c|} 
\hline
Class \; & \;  Twisted quantum double $D^\omega(G)$ & \; Number of Types\\ \hline
$\omega_3[1]$ \; & \; $D^{\omega_3{[1]}}(Z_2{}^3)$, $D(D_4)$ & \;  7\\ \hline
$\omega_3[{3d}]$ \; & \; $D^{\omega_3{[3d]}}(Z_2{}^3)$, $D^{\gamma^4}(Q_8)$ & \;  7\\ \hline
$\omega_3[{3i}]$ \; & \; $D^{\omega_3[{3i}]}(Z_2{}^3)$, $D^{}(Q_8)$, $D^{\alpha_1}(D_4)$, $D^{\alpha_2}(D_4)$ & \;  28\\ \hline
$\omega_3[{5}]$ \; & \; $D^{\omega_3{[5]}}(Z_2{}^3)$, $D^{\alpha_1\alpha_2}(D_4)$ & \; 21 \\ \hline
$\omega_3[{7}]$ \;  & \; $D^{\omega_3{[7]}}(Z_2{}^3)$ & \; 1 \\ \hline
 \end{tabular}
\caption{ $D^\omega(G)$, the twisted quantum double model of $G$ in 2+1D, and their 3-cocycles $\omega_{3}$(involving Type III) types in $\cC^{2\tD}_{(Z_2)^3, \omega_{3}}$. 
We classify the 64 types of 2D non-Abelian twisted gauge theories to 5 classes, which agree with Ref.\onlinecite{Goff2006}.
Each class has distinct non-Abelian statistics. Both dihedral group $D_4$ and quaternion group $Q_8$ are non-Abelian groups of order 8, as $|D_4|=|Q_8|=|(Z_2)^3|=8$.
$D^\omega(G)$ data can be found in Ref.\onlinecite{Goff2006}. Details are reserved to Appendix \ref{App:cocycles}.  
}
\label{table:Z2cubeclass}
\end{table}
\end{widetext}
\twocolumngrid
From Table \ref{table:Z2cubeclass}, we show that two classes of 3-cocycles for  $D^{\omega_3}(Z_2)^3$ of 2D can have dual descriptions by gauge theory of non-Abelian dihedral group $D_4$, quaternion 
group $Q_8$. 
However, the other three classes of 3-cocycles  for  $D^{\omega_3}(Z_2)^3$ do not have a dual (untwisted) non-Abelian gauge theory.

Now let us go back to consider 3D $\cC_{G,\omega_{4,\tIV}}^{3\tD}$, 
with $|Z_2|^4=16$. 
From Ref.\onlinecite{MW14}, we know 3+1D $D_4$ gauge theory has decomposition by its 5 centralizers.
Apply the rule of decomposition 
to other groups, it implies that for untwisted group $G$ in 3D $\cC^{3\tD}_{G}$, we can decompose it  
into sectors 
of $\cC^{2\tD}_{G_b,b}$, here $G_b$ becomes the {\bf centralizer} of the {\bf conjugacy class}(flux) $b$:
$
\cC^{3\tD}_{G} =\oplus_b \cC^{2\tD}_{G_b,b}.
$
Some useful information 
is: 
\bea
&&\cC^{3\tD}_{(Z_2)^4} = 16 \cC^{2\tD}_{(Z_2)^4}  \label{eq:Z2fbranch}\\
&&\cC^{3\tD}_{D_4} = 2\cC^{2\tD}_{D_4} \oplus 2\cC^{2\tD}_{(Z_2)^2} \oplus \cC^{2\tD}_{Z_4},\\ 
&&\cC^{3\tD}_{Z_2 \times D_4} = 4\cC^{2\tD}_{Z_2 \times  D_4} \oplus 4\cC^{2\tD}_{(Z_2)^3} \oplus 2 \cC^{2\tD}_{Z_2 \times Z_4}, \label{eq:D4Z2branch}\\
&&\cC^{3\tD}_{Q_8} = 2\cC^{2\tD}_{Q_8} \oplus 3 \cC^{2\tD}_{Z_4},\\
&&\cC^{3\tD}_{Z_2 \times Q_8} =4\cC^{2\tD}_{Z_2 \times Q_8} \oplus 6 \cC^{2\tD}_{Z_2 \times Z_4}.
\eea
and we find that no such decomposition is possible from $|G|=16$ group to match Eq.(\ref{eq:TypeIV3D2Dbranch})'s. Furthermore,
if there exists a non-Abelian $G_{nAbel}$ to have Eq.(\ref{eq:TypeIV3D2Dbranch}), those $(Z_2)^4$, $(Z_2) \times (D_4)$ or the twisted $(Z_2)^4$ 
must be the centralizers of $G_{nAbel}$. But one of the centralizers (the centralizer of the identity element as a conjugacy class $b=0$) of $G_{nAbel}$ must be $G_{nAbel}$ itself, which has already ruled out 
from Eq.(\ref{eq:Z2fbranch}),(\ref{eq:D4Z2branch}). Thus, we prove that {\bf $\cC^{3\tD}_{(Z_2)^4, \omega_{4,\tIV}}$ is not a normal 3+1D gauge theory 
(not ${Z_2 \times  D_4}$, neither Abelian nor non-Abelian) but must only be a twisted gauge theory.}
 

\begin{figure}[!h] 
\includegraphics[scale=0.4]{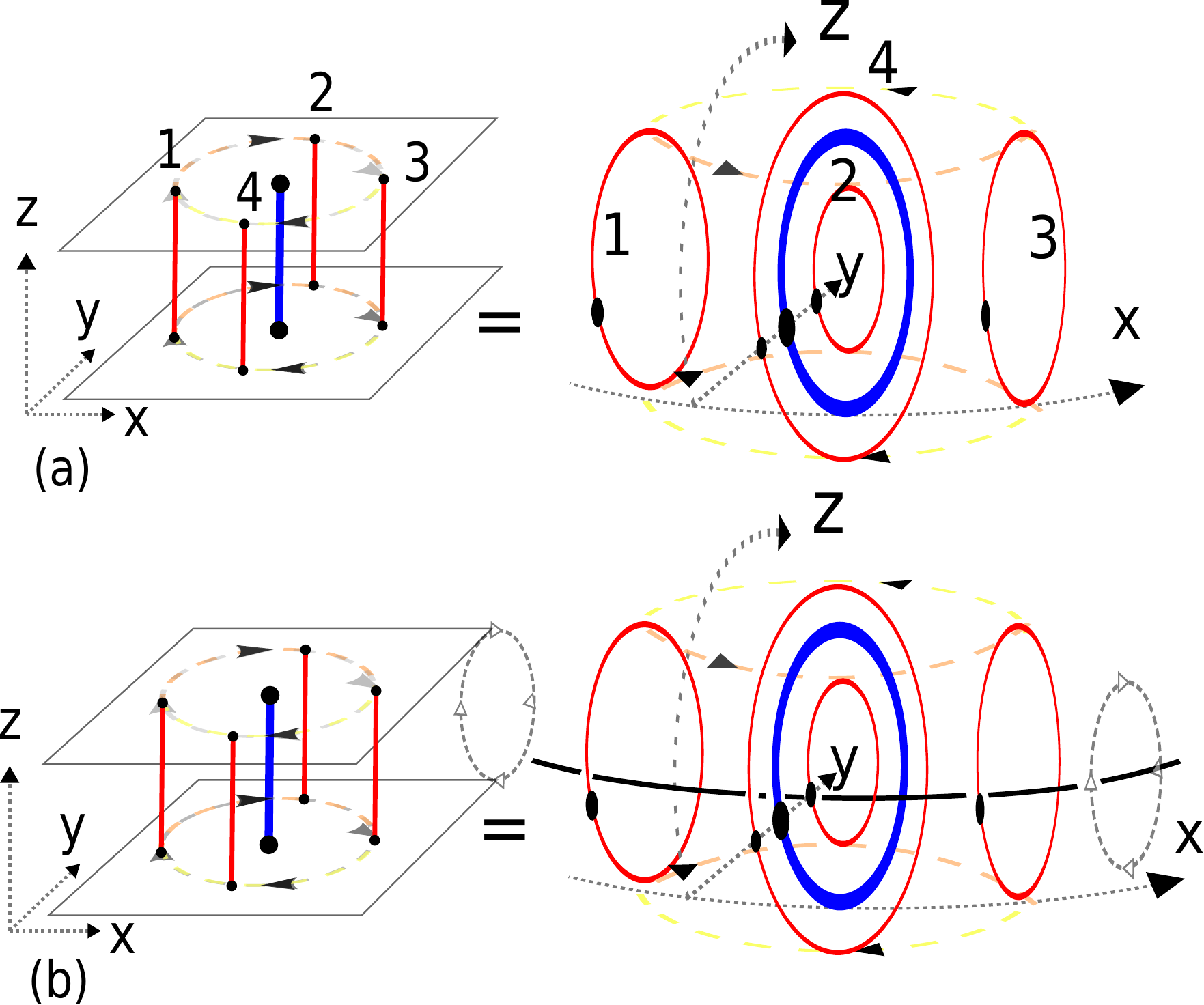}
\caption{
For 3+1D Type IV $\omega_{4,\tIV}$ twisted gauge theory $\cC_{G,\omega_{4,\tIV}}^{3\tD}$:
(a) {\bf Two-string statistics in unlink $0^2_1$ configuration is Abelian}. (The $b=0$ sector as $\cC^{2\tD}_G$.)
(b) {\bf Three-string statistics in two Hopf links $2^2_1 \# 2^2_1$ configuration is non-Abelian}.
(The $b \neq 0$ sector in  $\cC_b^{2\tD}=\cC_{G,\omega_{3,\tIII}}^{2\tD}$.) 
The $b \neq 0$ flux sector creates a monodromy effectively acting as the third (black) string threading the two (red,blue) strings.
}
\label{fig:3strings_2D_3D_Ab_to_nAb} 
\end{figure}

\noindent
(3) We discover that, see Fig.\ref{fig:3strings_2D_3D_Ab_to_nAb}, for any twisted gauge theory $\cC_{G,(\omega_{4,\tIV} \cdot \omega_{4,..})}^{3\tD}$ with Type IV 4-cocycle $\omega_{4,\tIV}$
(which non-Abelian nature is not affected by adding other Type II,III $\omega_{4,..}$), 
by {\bf threading a third string through two-string unlink $0^2_1$ 
into three-string Hopf links $2^2_1 \# 2^2_1$ configuration, Abelian two-string statistics is promoted to non-Abelian three-string statistics.}
We can see the physics from Eq.(\ref{eq:TypeIV3D2Dbranch}), the $\cC_b^{2\tD}$ is Abelian in $b=0$ sector; but non-Abelian in $b \neq 0$ sector. 
The physics of Fig.\ref{fig:3strings_2D_3D_Ab_to_nAb} is then obvious, 
by applying our discussion in Sec.\ref{Sec:physicsST} about the equivalence between string-threading and the $b\neq 0$ monodromy causes a branch cut.\\

\noindent
(4) {\bf Cyclic relation for non-Abelian $\sfS^{xyz}$ in 3D}: 
Interestingly, for the $(Z_2)^4$ twisted gauge theory with non-Abelian statistics,
we find that a similar cyclic relation Eq.(\ref{eq:cyclic_S_eq}) still holds as long as two conditions are satisfied: (i) the charge labels are equivalent $\alpha=\beta$
and (ii) $\delta_{a\in\{b,d, bd\}} \cdot \delta_{d\in\{a,b, ab\}} \cdot \delta_{b \in\{d, a, da\}} =1$.
However, Eq.(\ref{eq:cyclic_S_eq}) is modified with a factor depending on the dimensionality of Rep $\alpha$:
\bea \label{eq:cyclic_nAb_S_eq} 
\sfS^{\alpha,\alpha}_{a,b,d} \cdot \sfS^{\alpha,\alpha}_{b,d,a} \cdot \sfS^{\alpha,\alpha}_{d,a,b} \cdot |{\dim(\alpha)}|^{-3} =1.
\eea
This identity should hold for any Type IV non-Abelian strings.
This is a cyclic relation of 3D nature, 
 instead of a dimensional-reducing 2D nature for $\sfS^{2\tD\;\alpha,\beta}_{a,c\;(b)}$ of Fig.\ref{fig:3strings_2D_3D_xy}.  

\section{Conclusion} \label{summary} 

\subsection{Knot and Link configuration} \label{sec:knotandlinks}

\begin{figure}[!h] 
\includegraphics[scale=0.425]{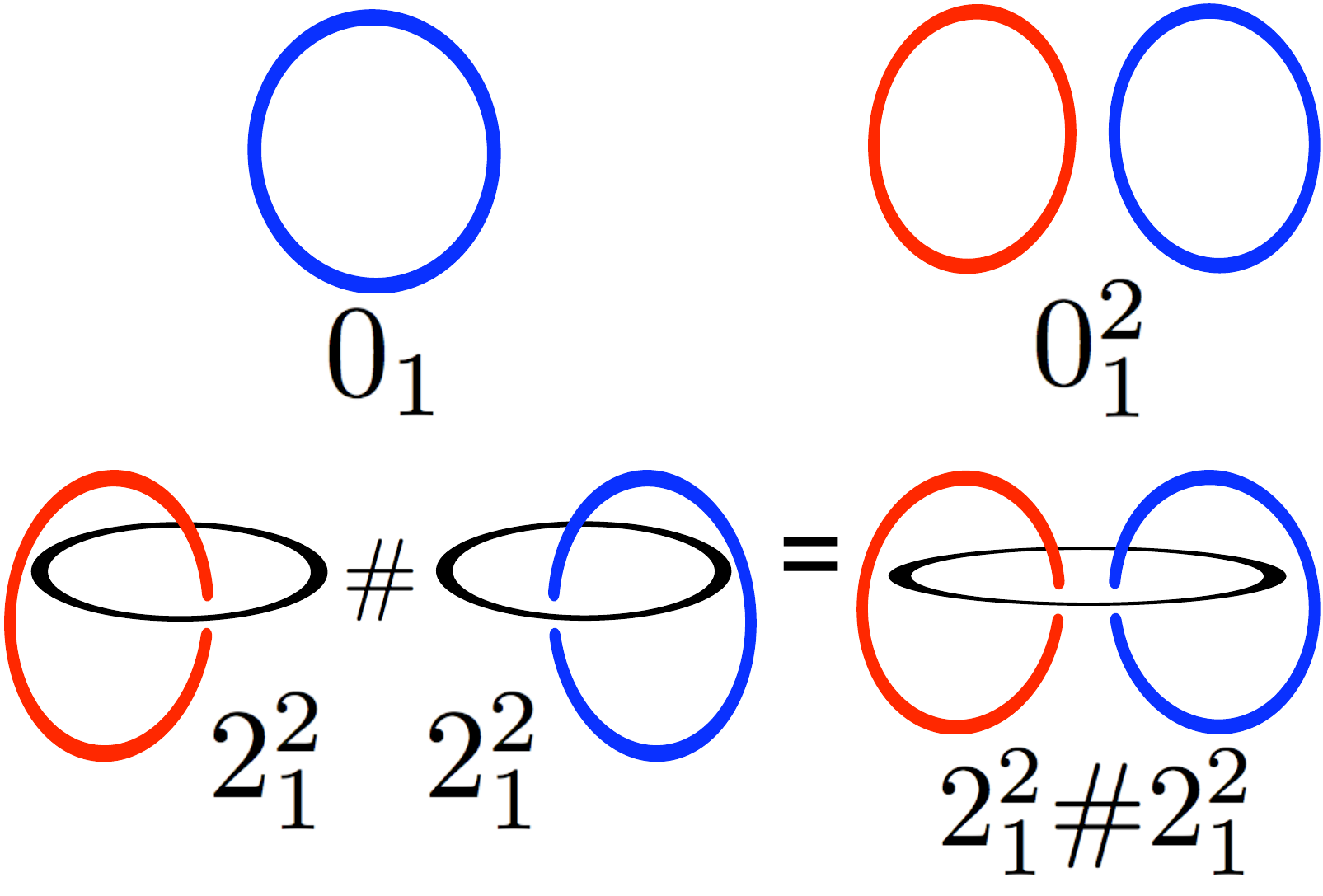}
\caption{
Under Alexander-Briggs notation, an unknot is $0_1$, 
two unknots can form an unlink $0^2_1$.
A Hopf link is $2^2_1$,
a connected sum of two Hopf links is $2^2_1 \# 2^2_1$.
}
\label{fig:unknot_unlink_Hopf_links} 
\end{figure}

Throughout our presentation, we have been indicating that the mathematics of knots and links may be helpful to organize our string-braiding patterns in 3D. 
Here we illustrate them more systematically. We will use Alexander-Briggs notation for the knots and links, see Fig.\ref{fig:unknot_unlink_Hopf_links}.

The knots and links for our string-braiding patterns are organized into Table \ref{table:link}.
We recall that, in Sec.\ref{Sec:physicsST}, the {\bf topological spin} for a closed string in the $b=0$ flux sector of $\cC_b^{2\text{D}}$
does a self-$2\pi$ flipping under the $0_1$ unknot configuration. 
Due to our {\bf spin-statistics relation} of a closed string, we can view the {topological spin} of $b = 0$ sector
as the  {\bf exchange statistics} of two identical strings in $0^2_1$ unlink configuration.
On the other hand, for the {\bf topological spin} in the $b \neq 0$ flux sector, 
we effectively thread a (black) string through the (blue) unknot, which forms a Hopf link $2^2_1$.
Meanwhile, we can view the topological spin of $b \neq 0$ sector
as the {\bf exchange statistics} of two identical strings treaded by a third (black) string in a connected sum of two Hopf links $2^2_1 \# 2^2_1$ configuration.
Furthermore, we can promote two-string Abelian statistics under the $0^2_1$ unlink  of $b=0$ sector
to three-string Abelian (in Sec.\ref{TypeII,III twist}) 
or non-Abelian statistics (in Sec.\ref{sec:TypeIV4cocycle}) under Hopf links $2^2_1 \# 2^2_1$ of $b \neq 0$ sector.

\begin{table}[!h]
\begin{tabular}{|c||c| c|} 
 \hline
$\cC_b^{2\text{D}}$ \; & \;  Physics of Strings & \; Knots and Links\\ \hline
 \; & \; topological spin ($\sfT$) & \; $0_1$ \\ \cline{2-3} 
 $b=0$\; & \; exchange statistics & \;   $0^2_1$ \\  \cline{2-3}  
 \; & \; 2-string braiding & \;   $0^2_1$ \\ \hline
 \; & \;  topological spin ($\sfT$) & \; $2^2_1$ \\ \cline{2-3} 
 $b \neq 0$\;  & \;  exchange statistics & \; $2^2_1 \# 2^2_1$ \\ \cline{2-3} 
  \;  & \;  3-string braiding & \; $2^2_1 \# 2^2_1, \dots$ \\ \hline
 \end{tabular}
\caption{ 
Various string-braiding patterns in terms of knots and links in Alexander-Briggs notation:
the topological spin of a loop, the exchange/braiding statistics of two loops without any background string inserted ($b =0$ sector) or with another background string inserted ($b \neq 0$ sector).
Here we effectively view the string braiding statistics of 3D topological order in terms of 2D sectors: $\mathcal{C}^{3\text{D}} = \oplus_b \cC_b^{2\text{D}}$. }
\label{table:link}
\end{table}

Nothing prevents us from considering more generic knot and link patterns for three-string or multi-string braiding.
Our reason is here -  
From the full modular SL$(3,\Z)$ group viewpoint, 
the ${\sfS}^{xyz}$ is a necessary generator to access the full data of the SL$(3,\Z)$ group.
However, we have learned that our 3D to 2D reduction by Eq.(\ref{eq:3Dto2DST}) using SL$(2,\Z)$ subgroup's data $\mathsf{S}^{xy}$ and $\mathsf{T}^{xy}$
already encode all the physics of braidings under the simplest knots and links in Fig.\ref{fig:unknot_unlink_Hopf_links} -
These include self-flipping topological spin and exchange/braiding statistics (Sec.\ref{Sec:physicsST},\ref{Sec:SL3ZMulti-String Braiding}).
It suggests that ${\sfS}^{xyz}$ contains more than these string-braiding configurations. 
In addition, there are more generic Mapping Class Groups $\text{MCG}(\cM_{space})$ beyond $\text{MCG}(\mathbb{T}^3)= \text{SL}(3,\Z)$,
which potentially encode more exotic multi-string braidings.

Indeed, as we already notice in Sec.\ref{Sec:SL3ZMulti-String Braiding}, the 3D $\sfS$ matrix essentially contains the information of three fluxes $(d,a,b)=(d_x,a_y,b_z)$
in Eq.(\ref{eq:Sxyzdab}), $\mathsf{S}^{xyz}=\sfS_{(\alpha, a,b)(\beta, c,d)} \equiv\frac{1}{|G|}  \sfS^{\alpha,\beta}_{d,a,b} \delta_{bc}$.
Since strings carry fluxes in 3D, 
this further suggests that we should look for the braiding involving with three strings,
where the 3-loop braiding has also been recently emphasized in Ref.\onlinecite{WL1437,JMR1462}.

\begin{figure}[!h] 
\includegraphics[scale=0.3]{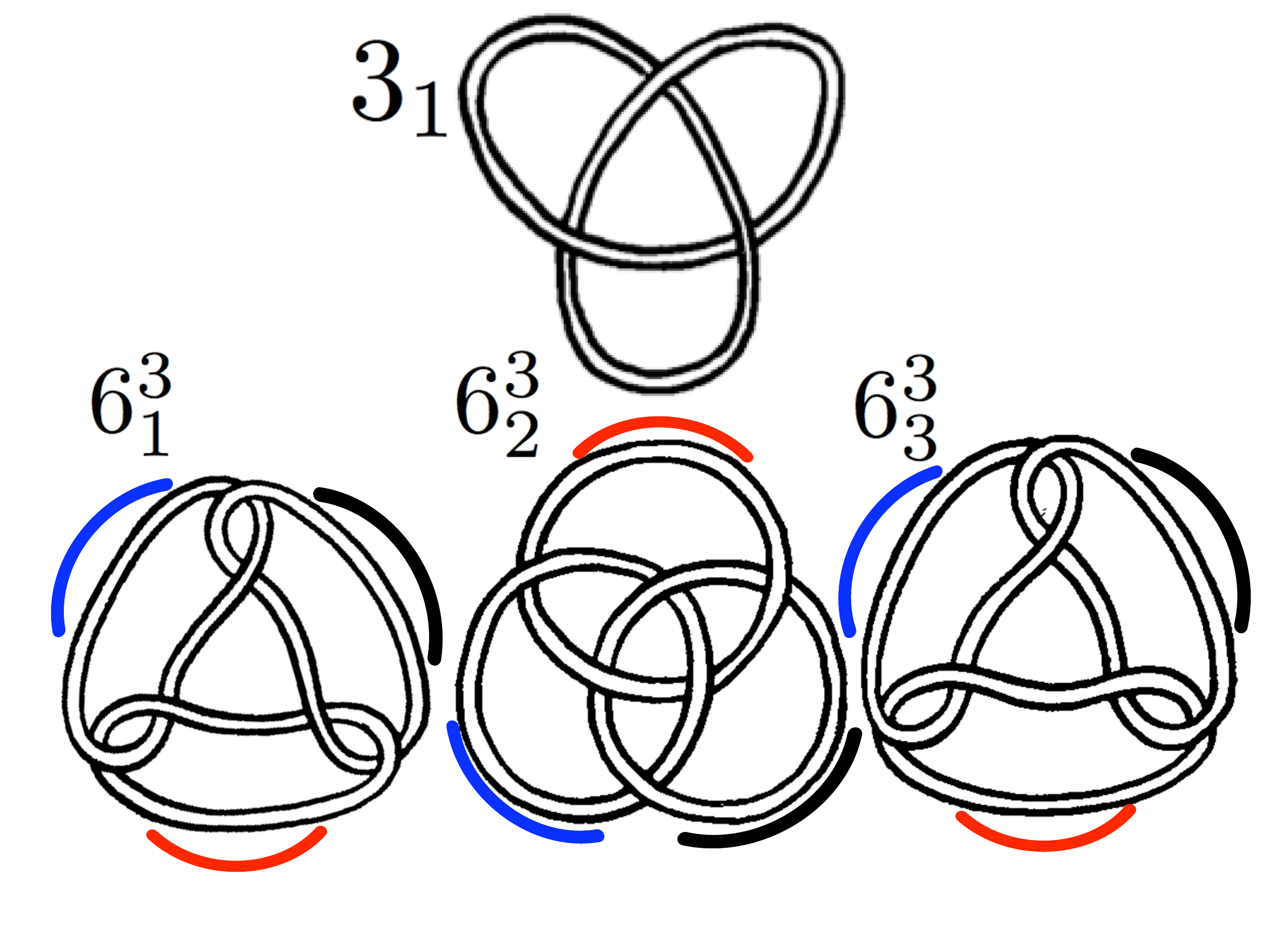}
\caption{
The trefoil knot is $3_1$. Some other simplest 3-string links (beyond Hopf links $2^2_1 \# 2^2_1$) are given:
$6^3_1$,
$6^3_2$ (Borromean rings),
$6^3_3$. From the {spin-statistics} relation of a closed string discussed in Sec.\ref{Sec:physicsST},
where the {topological spin} of certain knot/link configurations ($0_1$ for the monodromy flux $b=0$ and $2^2_1$ for $b \neq 0$) is equivalent to the {exchange statistics} of 
certain knot/link configurations ($0^2_1$ for $b=0$ and $2^2_1 \# 2^2_1$  for $b \neq 0$) under Eq.(\ref{Spin-Statistics3D}). 
Therefore, we may further conjecture that the 
{topological spin} of a trefoil knot $3_1$ may relate to the {braiding statistics} of 
$6^3_1$,
$6^3_2$,
$6^3_3$.
}
\label{fig:link_6} 
\end{figure}

The configuration we study so far with three strings is the Hopf link $2^2_1 \# 2^2_1$. 
We propose that using more general three strings pattern, such as the link
$$\cN^3_m$$  
or its connected sum
to study topological states. ($\cN^3_m$ is in Alexander-Briggs notation, here $3$ means that there are three closed loops, $\cN$ means the crossing number, 
and $m$ is the label for different kinds for $\cN^3$ linking.) For example, three-string braiding can include links of $6^3_1$, $6^3_2$, $6^3_3$ in Fig.\ref{fig:link_6}.
Configurations in Fig.\ref{fig:link_6} are potentially promising 
for studying the braiding statistics of strings to classify or characterize topological orders.


To examine whether the \emph{multi-string braiding} is topologically well-defined,  we propose a way to check that 
(such as the braiding processes in Fig.\ref{fig:3strings_2D_3D_Ab_to_nAb},\ref{fig:link_6}):\\

\frmN{
``\emph{The path that one (red) loop A winds around another (blue) loop B along the time evolution  
is nontrivial in the complement space of the B and the base (black) loop C. 
Namely, the path of A needs to be a nontrivial element of the fundamental group for the complement space of $B$ and $C$. 
Thus the path needs to be homotopically nontrivial}.''} 

Before concluding this subsection, another final remark is that
in Sec.\ref{spin-statistics}, we mention generalizing the framed worldline picture of particles in Fig.\ref{fig:spin_statistics} to the framed worldsheet picture of closed-strings.
(ps. The framed worldline is like a worldsheet, the framed worldsheet is like a worldvolume.)
Thus, it may be interesting to study how incorporating the framing of particles and strings (with worldline/worldsheet/worldvolume) 
can provide richer physics and textures into the knot-link pattern.

\subsection{Cyclic identity for Abelian and non-Abelian strings}  \label{summary:cyclic} 

In Sec.\ref{TypeII,III twist} and Sec.\ref{sec:TypeIV4cocycle}, we discuss \emph{cyclic identity} for Abelian and non-Abelian strings 
particularly for 3+1D twisted gauge theories. 
We find 
Eq.(\ref{eq:cyclic_nAb_S_eq}), \\ 
``{\emph{Cyclic identity of 3D's $\sfS^{xyz}$ matrix}} of Eq.(\ref{eq:Sxyzdab}) $\sfS^{xyz}_{(\alpha, a,b)(\beta, c,d)} \equiv\frac{1}{|G|}  \sfS^{\alpha,\beta}_{d,a,b} \delta_{b, c}$'': 
\bea \label{eq:cyclic_nAb_S_eq_Sum} 
&&\boxed{\sfS^{\alpha,\alpha}_{a,b,d} \cdot \sfS^{\alpha,\alpha}_{b,d,a} \cdot \sfS^{\alpha,\alpha}_{d,a,b} \cdot |{\dim(\alpha)}|^{-3} =1}.
\eea
For the Abelian case, the dimension of Rep is simply ${\dim(\alpha)}=1$, which reduces to Eq.(\ref{eq:cyclic_S_eq}).

On the other hand, we find that there is also another cyclic identity, based on 2D's $\mathsf{S}^{xy}_b =\sfS^{xy}_{(\alpha,a,b)({\beta,c,d})} \equiv \frac{1}{|G|} \sfS^{2\tD\;\alpha,\beta}_{a,c\;(b)} \delta_{b,d}$ matrix of Eq.(\ref{eq:Sxyb}),
written in terms of $\sfS^{2\tD\;\alpha,\beta}_{a,c\;(b)}$, at least for Abelian strings of Type II, III 4-cocycle twists, namely \\
``{\emph{Cyclic identity of 2D's $\sfS^{xy}$ matrix}}'':
\bea \label{eq:cyclic_S2D}
\boxed{\sfS^{2\tD\;0,0}_{a_i,c_k\;(b_j)}  \cdot \sfS^{2\tD\;0,0}_{c_k, b_j \;(a_i)} \cdot \sfS^{2\tD\;0,0}_{b_j, a_i \;(c_k)}=1}.
\eea
This Eq.(\ref{eq:cyclic_S2D}) cyclic identity has two additional criteria:
(1) Here $\alpha=\beta=0$ means that all strings must have zero charges. 
(2) In addition, the $\prod_i Z_{N_i}$ flux labels $a_i,b_j,c_k$ must satisfy 
that $a_i=|a| \hat{e}_i$, $b_j=|b| \hat{e}_j$, $c_k=|c| \hat{e}_k$, as a multiple of a single unit flux,
each only carries one of $\prod_i Z_{N_i}$ fluxes.
Note that $\hat{e}_j \equiv (0,\dots,0, 1, 0,\dots,0)$ is defined to be a unit vector with a nonzero component in the $j$-th component for the $Z_{N_j}$ flux.
Eq.(\ref{eq:cyclic_S2D}) is true even in the non-canonical basis, such as the case for the $b$-flux sector in Table \ref{tableSxyZN123}.
Thus, the fact whether in the canonical basis\cite{1303.0829} 
or not does not affect the identity Eq.(\ref{eq:cyclic_S2D}), at least for the example 
of Abelian Type II, III 4-cocycles. 

This 2D's $\sfS^{xy}_b$ cyclic identity in Eq.(\ref{eq:cyclic_S2D}) is indeed the cyclic relation of Ref.\onlinecite{WL1437}.
From the fact that we associate 2D's $\sfS^{xy}_b$ matrix to the dimensional reduction of string braiding in Fig.\ref{fig:3strings_2D_3D_xy}, 
it 
shows that the Abelian statistical angle $\theta_{a_i, c_k, (b_j)}$ can be defined, up to a basis,\cite{1303.0829} as 
\bea
\sfS^{2\tD\;0,0}_{a_i,c_k\;(b_j)}   =\exp(\ti \; \theta_{a_i, c_k, (b_j)} ).
\eea
Thus Eq.(\ref{eq:cyclic_S2D}) implies a cyclic relation for Abelian statistical angles:
\bea
\theta_{a_i, c_k, (b_j)} + \theta_{c_k, b_j, (a_i)} + \theta_{b_j, a_i, (c_k)} = 0\;\; (\text{mod} \;\; 2\pi).
\eea

In contrast, the 3D cyclic relation works for both Abelian and non-Abelian strings, and
it does not restrict on zero charge but only for equal charges $\alpha=\beta$.
More importantly, Eq.(\ref{eq:cyclic_nAb_S_eq_Sum}) allows any flux for each $a,b,c$,
instead of limiting to a single unit flux or a multiple of a single unit flux in Eq.(\ref{eq:cyclic_S2D}).


\subsection{Main results}

We have studied string and particle excitations in 3+1D twisted discrete gauge theories, which belong to a class of topological orders. 
These 3D theories are gapped topological systems with topology-dependent ground state degeneracy.
The twisted gauge theory contains its data of gauge group $G$ and 4-cocycle twist $\omega_4 \in \mathcal{H}^4(G,\R/\Z)$ of $G$'s fourth cohomology group.
Such a data provides many {\bf types} of theories, however several types of theories belong to the same {\bf class} of a topological order.
To better characterize and classify topological orders, we use the mapping class group on $\mathbb{T}^3$ torus, by $\text{MCG}(\mathbb{T}^d)= \text{SL}(d,\Z)$,
to extract the $\text{SL}(3,\Z)$ modular data $\sfS^{xyz}$ and $\sfT^{xy}$ in the ground state sectors,
which however reveal information of gapped excitations of particles and strings. 
We have posed five main questions {\bf Q1}-{\bf Q5} and other sub-questions throughout our work, and have addressed each of them in some depth.
We summarize our results and approaches below,
and make comparisons with some recent works:\\

\noindent
(1) {\bf Dimensional Reduction}: By inserting a gauge flux $b$ into a compactified circle $z$ of 3D topological order $\cC^{3\text{D}}$, we can realize 
$\mathcal{C}^{3\text{D}} = \oplus_b \cC_b^{2\text{D}}$, where $\cC^{3\text{D}}$ becomes a direct sum of degenerate states of 2D
topological orders $\cC_b^{2\text{D}}$ in different flux $b$ sectors. We should emphasize that this dimensional reduction is 
{\bf not} {\it real space decomposition along the z direction}, {\bf but} {\it the decomposition in the Hilbert space of ground states} (excitations basis such as the canonical basis of Eq.(\ref{eq:3Dbasis})).
We propose this decomposition in Eq.(\ref{eq:C3DtoC2D}) will work for a generic topological order without a gauge group description. 
In the most general case, $b$ becomes certain basis label of Hilbert space. The recent study of Ref.\onlinecite{MW14} implements 
the dimensional reduction idea on the \emph{normal gauge theories} described by 3D Kitaev $Z_N$ toric code and 3D quantum double models \emph{without cocycle twists}
using the \emph{spatial Hamiltonian} approach.
In our work, we consider more generic \emph{twisted gauge theories} with a lattice realization in
the twisted 3D quantum double models under the framework of Dijkgraaf-Witten theory.\cite{Dijkgraaf:1989pz}
We apply both the \emph{spatial Hamiltonian} approach and the \emph{spacetime path integral} approach. 
\\

\noindent
(2) {\bf Modular Data}: We find explicit formula representations of $\text{SL}(3,\Z)$ modular data 
$\sfS^{}$ and $\sfT^{}$ using (i) path integral and cocycle approach, and (ii) Representation(Rep) theory approach.
The Rep theory approach is convenient, and perhaps contains more general and simplified expressions.
While the recent work either focus on Abelian statistics\cite{{WL1437},{JMR1462}} or focus on normal gauge theories,\cite{{MW14}} our formula embodies
generic non-Abelian twisted gauge theories thus is most powerful.\\

\noindent
(3) {\bf Classification and Characterization}:  We use the modular data $\sfS^{}$ and $\sfT^{}$ to characterize the braiding statistics of some 2D and 3D topological orders.
We can further use the modular data $\sfS^{}$ and $\sfT^{}$ taking into account excitation-relabeling to classify (or partially classify)  topological orders.
Explicit 2D examples are $G=(Z_2)^3$ twisted gauge theories, and 3D examples are $G=(Z_2)^4$ twisted gauge theories. 
Some of our results are compared with the mathematics literature in the Supplemental Material (Appendix \ref{App:cocycles}). 
Some of 2D results are compared to twisted quantum double models $D^{\omega}(G)$.\\

Our result can also facilitate the study of symmetric protected topological states (SPTs) protected by a global symmetry $G_s$.\cite{Chen:2011pg} 
By gauging the $G_s$ symmetry of SPTs, one can use the induced dynamical gauged theory to study the braiding
of excitations and to characterize SPTs.\cite{{Levin:2012yb},Chenggu,{WL1437},Ye:2013upa}\\

\begin{table}[!t]
\begin{tabular}{|c||c| c|} 
 \hline
Braiding statistics - \; & \;  $b=0$ braiding  & \; $b\neq0$  braiding\\  
$(G,\omega_4)$ of $\cC^{3\text{D}}_{G,\omega_4}= \oplus_b \cC_b^{2\text{D}}$ \; & \;  2-strings $0^2_1$ & \; 3-strings $2^2_1 \# 2^2_1$\\ \hline
$(G_{Abel}, 1)$  \; & \; Abelian st & \; Abelian  st\\ 
 $(G_{Abel}, \text{without } \omega_{4,\tIV})$\; & \; Abelian  st& \;   Abelian  st\\  
 $(G_{Abel}, \text{with } \omega_{4,\tIV})$ \; & \; Abelian  st& \;  non-Abelian  st\\ \hline
 $(G_{nAbel}, 1)$ \; & \;  non-Abelian  st  & \; (non)Abelian  st  \\ 
$(G_{nAbel}, \omega)$\;  & \;  non-Abelian  st & \; non-Abelian  st\\ 
\hline
 \end{tabular}
\caption{
Braiding statistics being Abelian or non-Abelian in terms of $(G,\omega_4)$, a gauge group $G$ and a cocycle twist $\omega_4$ of 
a 3D topological order $\cC^{3\text{D}}_{G,\omega_4}$. Here $G_{Abel}$ means Abelian $G$, $G_{nAbel}$ means non-Abelian $G$,
and st means statistics. A normal gauge theory has $\omega_4=1$ with no cocycle twist.
The (non)Abelian st means it can be either non-Abelian or pure Abelian statistics 
(e.g. Any $b \neq 0$ sector of an untwisted $S_3$ gauge theory has pure Abelian statistics, due to $S^3$ centralizers of non-indentity elements are Abelian.
But some $b \neq 0$ sector of untwisted $D_4$ and $Q_8$ gauge theories have non-Abelian statistics.)
The $b=0$ 2-strings $0^2_1$ braiding is the process of Fig.\ref{fig:3strings_2D_3D_Ab_to_nAb} (a). 
The $b \neq 0$ 3-strings $2^2_1 \# 2^2_1$ braiding is the process of Fig.\ref{fig:3strings_2D_3D_Ab_to_nAb} (b)}.
\label{table:braid_summary}
\end{table}

\noindent 
(4) {\bf Physics of string and particle braiding}: We provide the physics meaning of the {\bf topological spin} and  {\bf spin-statistics} relation for a closed string.
We also interpret the 3-string braiding statics firstly studied in Ref.\onlinecite{WL1437} from a new perspective - 
a dimensional reduction with $b$ flux monodromy.
We find that {\bf with Type IV 4-cocycle twist for the twisted gauge theory}, by threading a third string through two-string unlink $0^2_1$ 
into three-string Hopf links $2^2_1 \# 2^2_1$ configuration, {\bf Abelian two-string statistics is promoted to non-Abelian three-string statistics}. 
In Ref.\onlinecite{MW14},  the sort-of opposite effect of ours is found:
where the normal (untwisted) non-Abelian 3D topological order has found with non-Abelian statistics in $b=0$ sector, but there may have Abelian statistics in $b \neq 0$ sector.
Incorporate this understanding, 
We have a more unified picture organized in Table \ref{table:braid_summary},
for the string-braiding statistics of twisted/untwisted Abelian/non-Abelian gauge theories as topological orders.
Since the string deformation on the lattice can blur the Abelian U(1) phase,
our non-Abelian string-braiding statistics provides a better alternative for a robust physical observable than Abelian string-braiding statistics\cite{{WL1437},{JMR1462}}
to be tested numerically or experimentally in the future.
Last but not least, we propose to {use more general patterns, such as $\cN^3_m$ (or $\cN^l_m \# \dots$) knots/links of Alexander-Briggs
to study the three-string (or multi-string) braiding statistics. }\\

 \section{Acknowledgements} 
JW would like to thank Tian Lan, as well as Ling-Yan Hung and Yidun Wan for helpful conversations and warm encouragements.
JW is grateful to Louis H.~Kauffman for a long blackboard discussion about the theory of knots and links at Perimeter Institute, along with Yidun Wan. 
JW also thanks M.~de Wild Propitius for email correspondence in the early 2013 
and expressing his interests in our work. 
JW acknowledges the use of computational resources Compute and Titan at Perimeter Institute.
We thank Zheng-Cheng Gu for collaboration on a related work Ref.\onlinecite{Wang:2014pma}.\\

During the {\it Symmetry in Topological Phases} workshop at Princeton University, we become aware that the authors of Ref.\onlinecite{WL1437}
are working on the braiding statistics of 3+1D gapped phases, which intersect some of our studies, but also further inspire our work.
In the long process of preparing this manuscript, we notice two recent works appear in Ref.\onlinecite{{WL1437},{JMR1462}}
dealing with the Abelian braiding statistics of twisted gauge theories. 
Also a recent preprint\cite{Thorngren:2014pza} considers the surface topological order of SPTs with loop braiding statistics.\\

This research is supported by NSF Grant No.
DMR-1005541, NSFC 11074140, 
NSFC 11274192, BMO Financial Group 
and the John Templeton Foundation. 
JW has been supported in part by the U.S. Department of Energy under cooperative research agreement Contract Number DE-FG02-05ER41360,
and by Isaac Newton Chair Fellowship.
Research at Perimeter Institute is supported by the Government of Canada through Industry Canada and by the Province of Ontario through the Ministry of Research, 
Economic Development \& Innovation.

\begin{widetext}
\end{widetext}

\onecolumngrid
\appendix

{\bf Supplemental Material}

\section{Group Cohomology and Cocycles} \label{App:cocycles} 

\subsection{Cohomology group}

Here we review the cohomology group $\mathcal{H}^{d+1}(G,\R/\Z)=\mathcal{H}^{d+1}(G,\tU(1))$ by $\R/\Z=\tU(1)$, 
as the ${(d+1)}$th-cohomology group of G over G module U(1).
Each class in $\mathcal{H}^{d+1}(G,\R/\Z)$ corresponds to a distinct $(d+1)$-cocycles.
The $n$-cocycle is a $n$-cochain 
additionally
satisfying the $n$-cocycle-conditions $\delta \omega=1$. 
The $n$-cochain is a mapping $\omega_{}^{}(A_1,A_2,\dots,A_n)$:  $G^n \to \tU(1)$ (which inputs
$A_i \in G$, $i=1,\dots, n$, and outputs a $\tU(1)$ phase).
The $n$-cochains satisfy the group multiplication rule:
\bea
(\omega_{1}\cdot\omega_{2})(A_1,\dots,A_n)= \omega_{1}^{}(A_1,\dots,A_n)\cdot \omega_{2}^{}(A_1,\dots,A_n), 
\eea
thus form a group.
The coboundary operator $\delta$ 
\bea
\delta \sfc(g_1, g_2, \dots,  g_{n+1}) \equiv  
\sfc(g_2, \dots,  g_{n+1})  \sfc(g_1, \dots,  g_{n})^{(-1)^{n+1}} 
\cdot \prod_{j=1}^{n} \sfc(g_1, \dots, \; g_j g_{j+1},\; \dots,  g_{n+1})^{(-1)^{j}}, 
\eea
defines the $n$-cocycle-condition $\delta \omega=1$ (a pentagon relation in 2D).
We check the distinct $n$-cocycles are not equivalent by $n$-coboundaries.  
The $n$-cochain forms a group $\text{C}^n$,
the $n$-cocycle forms a subgroup $\text{Z}^n$ of $\text{C}^n$,
and the $n$-coboundary further forms a subgroup $\text{B}^n$ of $\text{Z}^n$ (since $ \delta^2=1$). Overall, it shows
$
\text{B}^n \subset \text{Z}^n \subset \text{C}^n
$.
The $n$-cohomology group is exactly a relation of a kernel $\text{Z}^n$ (the group of $n$-cocycles) modding out an image $\text{B}^n$ (the group of $n$-coboundary):
\bea
\cH^n(G,\tU(1))= \text{Z}^n /\text{B}^n.
\eea

To derive the expression of $\cH^d(G,\tU(1))$ in terms of groups explicitly, we apply some key formulas:\\

\twocolumngrid
\noindent
{\bf{(1). K\"unneth formula}}:\\
We denote $R$ as a ring, $\M,\M'$ as the R-modules, $X,X'$ are some chain complex.
The K\"unneth formula shows the cohomology of a chain complex $X \times X'$ 
in terms of the cohomology of a chain complex $X$ and another chain complex $X'$.
For topological cohomology $H^d$, we have
\begin{align}
\label{kunn}
&\ \ \ \ H^d(X\times X',\M\otimes_R \M')
\nonumber\\
&\simeq \Big[\oplus_{k=0}^d H^k(X,\M)\otimes_R H^{d-k}(X',\M')\Big]\oplus
\nonumber\\
&\ \ \ \ \ \
\Big[\oplus_{k=0}^{d+1}
\text{Tor}_1^R(H^k(X,\M),H^{d-k+1}(X',\M'))\Big]  .
\end{align}
\begin{align}
&\ \ \ \ H^d(X\times X',\M)
\nonumber\\
&\simeq \Big[\oplus_{k=0}^d H^k(X,\M)\otimes_{\Z} H^{d-k}(X',\Z)\Big]\oplus
\nonumber\\
&\ \ \ \ \ \
\Big[\oplus_{k=0}^{d+1}
\text{Tor}_1^{\Z}(H^k(X,{\M}),H^{d-k+1}(X',\Z))\Big].\label{eq:kunnZ}
\end{align}
The above is valid for both topological cohomology $H^d$ and group cohomology $\cH^d$ (for $G'$ is a finite group):
\begin{align}
&\ \ \ \ \cH^d(G\times G',\M)
\nonumber\\
&\simeq \Big[\oplus_{k=0}^d \cH^k(G,\M)\otimes_{\Z} \cH^{d-k}(G',\Z)\Big]\oplus
\nonumber\\
&\ \ \ \ \ \
\Big[\oplus_{k=0}^{d+1}
\text{Tor}_1^{\Z}(\cH^k(G,\M),\cH^{d-k+1}(G',\Z))\Big].
\end{align}

\noindent
{\bf{(2). Universal coefficient theorem(UCT)}}
 can be derived from K\"unneth formula, Eq.(\ref{eq:kunnZ}), by taking $X=0$ or $Z_1$ or a point thus only $H^{0}(X',\M)=\M$ survives, 
\bea
\label{eq:ucf}
 H^d(X',\M)
&\simeq  \M \otimes_{\Z} H^d(X',\Z)
\oplus
\text{Tor}_1^{\Z}(\M,H^{d+1}(X',\Z)).\;\;\;\;\;\;\;\;
\eea
Using UCT, we can rewrite Eq.(\ref{eq:kunnZ}) as a decomposition below.\\

\noindent
{\bf{(3). Decomposition}}:
\begin{align}
\label{eq:kunnH}
H^d(X\times X',\M) \simeq \oplus_{k=0}^d H^k[X, H^{d-k}(X',\M)].
\end{align}
The above is valid for both topological cohomology and group cohomology:
\begin{align}
\label{eq:kunnG}
\cH^d(G\times G',\M) \simeq \oplus_{k=0}^d \cH^k[G, \cH^{d-k}(G',\M)],
\end{align}
provided that both $G$ and $G'$ are finite groups.

The expression of K\"unneth formula is in terms of
the tensor-product operation $\otimes_R$ and the torsion-product $\text{Tor}_1^R$ operation of a base ring $R$, which we write
$\boxtimes_R\equiv \text{Tor}_1^R$ as shorthand, their properties are:
\begin{align}
\label{tensorprd}
& \M \otimes_\Z \M' \simeq \M' \otimes_\Z \M ,
\nonumber\\
& \Z \otimes_\Z \M \simeq \M \otimes_\Z \Z =\M ,
\nonumber\\
& \Z_n \otimes_\Z \M \simeq \M \otimes_\Z \Z_n = \M/n\M ,
\nonumber\\
& \Z_n \otimes_\Z \tU(1) \simeq \tU(1) \otimes_\Z \Z_n = 0,
\nonumber\\
& \Z_m \otimes_\Z \Z_n  =\Z_{\gcd(m,n)} ,
\nonumber\\
&  (\M'\oplus \M'')\otimes_R \M = (\M' \otimes_R \M)\oplus (\M'' \otimes_R \M)   ,
\nonumber\\
& \M \otimes_R (\M'\oplus \M'') = (\M \otimes_R \M')\oplus (\M \otimes_R \M'')   ;
\end{align}
and
\begin{align}
\label{torsionprd}
& \text{Tor}_1^R(\M,\M') \equiv \M\boxtimes_R \M'  ,
\nonumber\\
& \M\boxtimes_R \M' \simeq \M'\boxtimes_R \M  ,
\nonumber\\
& \Z\boxtimes_\Z  \M = \M\boxtimes_\Z  \Z = 0,
\nonumber\\
& \Z_n\boxtimes_\Z \M = \{m\in \M| nm=0\},
\nonumber\\
& \Z_n\boxtimes_\Z \tU(1) = \Z_n,
\nonumber\\
& \Z_m\boxtimes_\Z \Z_n = \Z_{\<m,n\>} ,
\nonumber\\
& \text{Tor}_1^\Z(U(1), U(1)) = 0 ,
\nonumber\\
& \M'\oplus \M''\boxtimes_R\M = \M'\boxtimes_R \M\oplus\M''\boxtimes_R \M,
\nonumber\\
& \M\boxtimes_R\M'\oplus \M'' = \M\boxtimes_R\M'\oplus\M\boxtimes_RB.
\end{align}
For other details, we suggest to read Ref.\onlinecite{Chen:2011pg} and Reference therein.

\onecolumngrid

\begin{widetext}
\end{widetext}

\onecolumngrid

We summarize some useful facts in Table \ref{table:Hfact}, and some derived results in Table \ref{table:Hgroup}. 

\begin{center}
\begin{table}[!h]
\begin{tabular}{|c||c|c|c|}
\hline
0&$\cH^0(G,M)=M$& $\cH^0(G,Z)=\Z$& $\cH^0(G,\tU(1))=\tU(1)$ \\[0mm]  \hline
1&$\cH^1(G,M)$& $\cH^1(G,Z)=\Z_1$  &$\cH^1(G,\tU(1))=G$ \text{(1D Rep of group)} \\[0mm]  \hline
2 &$\cH^2(G,M)$ & $\cH^2(G,Z)=\cH^1(G,\tU(1))$ & $\cH^2(G,\tU(1))$ \text{(Projective Rep of group)}\\ \hline
3 &$\cH^3(G,M)$ & $\cH^3(G,Z)=\cH^2(G,\tU(1))$ &\\  \hline
$d\geq 2$ & $\cH^d(G,M)$ & $\cH^d(G,Z)=\cH^{d-1}(G,\tU(1))$ & \\
\hline
\end{tabular}
\caption{Some facts about the cohomology group. For a finite Abelian group $G$, we have $\cH^2(G,Z)=\cH^1(G,\tU(1))=G$.}
\label{table:Hfact}
\end{table}
\end{center}

\begin{table}[!h]
\begin{tabular}{|c||c|c|c|c|c|c|c|c|c|}
\hline
      &Type I& Type II & Type III &Type  IV &Type  V  &Type  VI & $\dots$ & $\dots$ & \\[0mm]  \hline
      &$\Z_{N_i}$& $\Z_{N_{ij}}$& $\Z_{N_{ijl}}$ &  $\Z_{N_{ijlm}}$ &  $\Z_{{\gcd} \otimes^5_{i}(N^{(i)})}$ 
      &  $\Z_{{\gcd} \otimes^6_{i}(N_i)}$ & $\Z_{{\gcd} \otimes^m_{i}(N_i)}$ & $\Z_{{\gcd} \otimes^{d-1}_{i}N_i}$ & $\Z_{{\gcd} \otimes^d_{i}N^{(i)}}$\\[0mm]  \hline
$\cH^1(G,\tU(1))$ &$1$& $$& $$ & $$ & $$ & $$ & $$ & $$ & $$ \\[0mm]  \hline
$\cH^2(G,\tU(1))$&$0$  &$1$ & $$ & $$ & $$ & $$ & $$ & $$ & $$\\[0mm]  \hline
$\cH^3(G,\tU(1))$ &$1$ & $1$ & $1$ & $$ & $$ & $$ & $$ & $$ & $$\\ \hline
$\cH^4(G,\tU(1))$ &$0$ & $2$ & $2$ & $1$ & $$ & $$ & $$ & $$ & $$\\ \hline
$\cH^5(G,\tU(1))$ & $1$ & $2$ & $4$ & $3$ & $1$ & $$ & $$ & $$ & $$\\ \hline
$\cH^6(G,\tU(1))$ & $0$ & $3$ & $6$ & $7$ & $4$ & $1$ & $$ & $$ & $$\\ \hline
$\cH^d(G,\tU(1))$ & $\frac{(1-(-1)^d)}{2}$ & $\frac{d}{2}-\frac{(1-(-1)^d)}{4}$ & $\dots$ & $\dots$ & $\dots$ & $\dots$ & $\dots$ & $d-2$ & $1$\\[1mm] \hline
\end{tabular}
\caption{ The table shows the exponent of the $\Z_{{\gcd} \otimes^m_{i}(N_i)}$ class in $\cH^d(G,\tU(1))$ for $G=\prod_{i=1}^n  \Z_{N_i}$. 
We define a shorthand of $\Z_{{\gcd} (N_i,N_j)}\equiv  \Z_{N_{ij}} \equiv  \Z_{{\gcd} \otimes^2_{i}(N_i)}$, etc, also for other higher gcd.
Our definition of the Type $m$  is from its number $m$ of cyclic gauge groups in the gcd class $\Z_{{\gcd} \otimes^m_{i}(N_i)}$. 
The number of exponents can be systematically obtained by adding all the numbers of 
the previous column from the top row 
to a row before the wish-to-determine number.
For example, our table shows that we derive that
$\cH^3(G,\R/\Z) = \underset{1 \leq i < j < l \leq n}{\prod}  \Z_{N_i} \times \Z_{N_{ij}} \times  \Z_{N_{ijl}}$ and
$\cH^4(G,\R/\Z) = \underset{1 \leq i < j < l < m \leq n}{\prod}   (\Z_{N_{ij}})^2 \times  (\Z_{N_{ijl}})^2 \times \Z_{N_{ijlm}}$, etc.
}.
\label{table:Hgroup}
\end{table}

\subsection{Derivation of cocycles}

To derive Table \ref{table:Hgroup}, we find that by doing the K\"unneth formula decomposition carefully for a generic finite Abelian group $G=\prod_i Z_{N_i}$, 
some corresponding structure becomes transparent. See Table \ref{table:cocyclefact}.
\begin{center}
\begin{table}[!h]
\makebox[\textwidth][c]{
\begin{tabular}{|c||c|c|c|}
\hline
(d+1)\text{dim} & $\cH^{d+1}(G,\tU(1))$ & K\"unneth formula in $\cH^{d+1}(G,\tU(1))$ & path integral forms in ``fields''  \\ 
 \hline\hline
0+1\tD& $\Z_{n_1}$ &$\cH^{1}(\Z_{n_1},  \tU(1)) $& $[\exp(\ti k_{..} \int A_1) ]$   \\[0mm]  \hline
1+1\tD & $\Z_{n_{12}}$ &$ \cH^{1}(\Z_{n_1}, \tU(1)) \boxtimes_\Z \cH^{1}(\Z_{n_2}, \tU(1)) $& $[\exp(\ti k_{..}  \int A_1 A_2) ]$\\[0mm]  \hline
2+1\tD & $\Z_{n_{1}}$  &$ \cH^{3}(\Z_{n_1}, \tU(1))$ & $[\exp(\ti k_{..}  \int A_1dA_1) ]$    \\ \hline
2+1\tD & $\Z_{n_{12}}$ &$\cH^{1}(\Z_{n_1}, \tU(1)) \otimes_\Z \cH^{1}(\Z_{n_2}, \tU(1))$ & $[\exp(\ti k_{..}  \int A_1dA_2) ]$  \\  \hline
2+1\tD & $\Z_{n_{123}}$ &$[\cH^1(\Z_{n_1},\tU(1))\boxtimes_\Z \cH^1(\Z_{n_2}, \tU(1))]\boxtimes_\Z \cH^1(\Z_{n_3}, \tU(1))$ & $[\exp(\ti k_{..}  \int A_1 A_2 A_3) ]$ \\  \hline
3+1\tD & $\Z_{n_{12}}$ &$ \cH^{1}(\Z_{n_1}, \tU(1)) \boxtimes_\Z \cH^{3}(\Z_{n_2}, \tU(1))  $ & $[\exp(\ti k_{..}  \int A_1 A_2 d A_2) ]$ \\  \hline
3+1\tD & $\Z_{n_{12}}$ &$ \cH^{1}(\Z_{n_2}, \tU(1)) \boxtimes_\Z \cH^{3}(\Z_{n_1}, \tU(1))  $ & $[\exp(\ti k_{..}  \int A_2 A_1 d A_1) ]$ \\  \hline
3+1\tD &  $\Z_{n_{123}}$ &$ [ \cH^{1}(\Z_{n_1}, \tU(1)) \otimes_\Z \cH^{1}(\Z_{n_2}  , \tU(1)) ] \boxtimes_\Z  \cH^1(\Z_{n_3}, \tU(1))  $ & $[\exp(\ti k_{..}  \int (A_1 dA_2) A_3) ]$ \\  \hline
3+1\tD & $\Z_{n_{123}}$ &$ { [ \cH^{1}(\Z_{n_1}, \tU(1)) \boxtimes_\Z \cH^{1}(\Z_{n_2}, \tU(1)) ]     \otimes_\Z  \cH^1(\Z_{n_3}, \tU(1)) }   $ & $[\exp(\ti k_{..}  \int (A_1 A_2) d A_3) ]$ \\  \hline
3+1\tD &$\Z_{n_{1234}}$ &$  \big[[\cH^1(\Z_{n_1}, \tU(1))  \boxtimes_\Z \cH^1(\Z_{n_2}, \tU(1))]  \boxtimes_\Z  \cH^1(\Z_{n_3}, \tU(1)) \big] \boxtimes_\Z  \cH^1(\Z_{n_4}, \tU(1))$ & $[\exp(\ti k_{..}  \int A_1 A_2 A_3 A_4) ]$ \\  \hline
\hline
\end{tabular} 
}
\caption{Some derived facts about the cohomology group and its cocycles.}
\label{table:cocyclefact}
\end{table}
\end{center}
From the known field theory fact, we know that 2+1D twisted gauge theories from 
$\cH^{3}(G,\tU(1))= \underset{1 \leq i < j < l \leq m}{\prod}  \Z_{N_i} \times \Z_{N_{ij}} \times  \Z_{N_{ijl}}$, 
their $\Z_{n_{i}}$ classes are captured by a path integral
$\simeq \exp(\ti k_{..} \int A_i dA_i)$ up to some normalization factor. 
(Here we omit the wedge product, denoting $A_i dA_i \equiv A_i \wedge dA_i$. We also schematically denote the quantization factor $k_{..}$, the details
of $k_{..}$ level quantizations can be found in Ref.\onlinecite{Wang:2014pma}.)
The $\Z_{n_{jl}}$ classes are captured by a path integral
$\simeq \exp(\ti k_{..} \int A_j dA_l)$, where $A$ is a 1-form gauge field. We deduce that the K\"unneth formula decomposition in $\cH^{d+1}(G,\tU(1))$ 
with 
the torsion product $\text{Tor}_1^R \equiv \boxtimes_R $ 
suggests a wedge product  $\wedge$ structure in the corresponding field theory, 
while  the tensor product $\otimes_\Z$ suggests appending 
an extra exterior derivative $\wedge d$ structure in the corresponding field theory. For example, 
$ \cH^{1}(\Z_{n_1}, U(1)) \boxtimes_\Z \cH^{1}(\Z_{n_2}, U(1))  \to [\exp(\ti \int A_1 \wedge A_2) ]$, and $\cH^{1}(\Z_{n_1}, U(1))  \to [\exp(\ti \int A_1) ]$,
then $\cH^{1}(\Z_{n_1}, U(1)) \otimes_\Z \cH^{1}(\Z_{n_2}, U(1))  \to [\exp(\ti \int A_1 \wedge d A_2) ]$. 
Such an organization also shows the corresponding form of cocycles for 3+1D in Table \ref{table1} and 2+1D 
in Table \ref{tableA1}. 
For example: The relation $A_1 \to a_1$, maps a 1-form field to a gauge flux $a_1$ (or a group element).
The relation $dA_2 \to (b_2+c_2-[b_2+c_2])$, maps an exterior derivative to the operation taking on different edges/vertices on the spacetime complex. 
We use this fact to see through whether two cocycles are the same forms or whether they are up to coboundaries.
We comment that such a path integral so far is only suggestive, but not yet being strongly evident enough to formulate a consistent field theoretic path integral. 
Thus we coin them with a speculative quotation mark in {path integral} forms in ``fields.''
The more systematic formulation in terms of field theoretic \emph{partition functions} will be reported elsewhere in the following work in Ref.\onlinecite{Wang:2014pma} from the perspective of 
symmetric protected topological states (SPTs).

\subsection{Dimensional reduction from a slant product} 

In general, for dimensional reduction of cochains, we can use the slant product mapping $n$-cochain $\sfc$ to $(n-1)$-cochain $i_g \sfc$:
\bea
i_g \sfc(g_1, g_2, \dots,  g_{n-1}) \equiv  \sfc(g, g_1, g_2,\dots,g_{n-1})^{(-1)^{n-1}} \cdot \prod_{j=1}^{n-1} \sfc(g_1,\dots,g_j, (g_1\dots g_j)^{-1} \cdot g \cdot  (g_1\dots g_j),\dots,g_{n-1})^{(-1)^{n-1+j}}. \;\;\;\;\;\;\;\;\;\;
\eea

Here we focus on the Abelian group  $G$.
For example in 2+1D, we have $3$-cocycle to $2$-cocycle: 
\bea \label{eq:CA(B,C)}
\sfC_A(B,C) \equiv i_A \omega(B,C) =\frac{\omega(A,B,C) \omega(B,C,A) }{ \omega(B,A,C) }
\eea
In 3+1D, we have $4$-cocycle to $3$-cocycle: 
\bea
\sfC_A(B,C,D) \equiv i_A \omega(B,C,D) =\frac{\omega(B,A,C,D) \omega(B,C,D,A) }{ \omega(A,B,C,D) \omega(B,C,A,D) }
\eea
In order to study the projective representation (the second cohomology group $\mathcal{H}^2$) from 4-cocycles, we do the slant product again: 
\bea
&&\sfC^{(2)}_{AB}(C,D) \equiv i_B \sfC_A(C,D) =\frac{ \sfC_A(B,C,D)  \sfC_A(C,D,B) }{ \sfC_A(C,B,D)}\\
&&=\frac{\omega(B,A,C,D) \omega(B,C,D,A) }{ \omega(A,B,C,D) \omega(B,C,A,D) } \cdot \frac{ \omega(A,C,B,D) \omega(C,B,A,D) }{\omega(C,A,B,D) \omega(C,B,D,A) }
\cdot \frac{\omega(C,A,D,B) \omega(C,D,B,A) }{ \omega(A,C,D,B) \omega(C,D,A,B) }
\eea

\subsection{2+1D topological orders of $\mathcal{H}^3(G,\R/\Z)$} \label{2D_TO}

\subsubsection{3-cocycles}

Here we organize the known fact about the third cohomology group $\mathcal{H}^3(G,\R/\Z)$ with $G=\prod^k_{i=1} Z_{N_i}$:
\bea \label{eq:3rd-group-cohomology}
\cH^3(G,\R/\Z) = \prod_{1 \leq i < j < l \leq m}  \Z_{N_i} \times \Z_{N_{ij}} \times  \Z_{N_{ijl}}. \nonumber
\eea
\begin{center}
\begin{table} [!h]
\begin{tabular}{|c||c|c|c|c|}
\hline
\;$\mathcal{H}^3(G,\R/\Z)$\; &  3-cocycle name &  3-cocycle form  & Induced  $\sfC_a(b,c)$\\[0mm]  \hline \hline 
$\Z_{N_{i}}$ & Type I  $ k_{{ \tI(i)}}^{} $  & 
$\omega_{3,\text{I}}^{(i)}(a,b,c)    =\exp \Big( \frac{2 \pi \ti k_{i}  }{N_{i}^{2}} \; a_{i}(b_{i} +c_{i} -[b_{i}+c_{i}]) \Big) $ & 
$\exp \Big( \frac{2 \pi \ti k_{i}  }{N_{i}^{2}} \;  a_{i}(b_{i} +c_{i} -[b_{i}+c_{i}]) \Big)$ \\[2mm] \hline
$\Z_{N_{ij}}$ & Type II $ k_{{ \tII(ij)}}^{} $ & 
$  \omega_{3,\text{II}}^{(ij)}(a,b,c) = \exp \Big( \frac{2 \pi \ti k_{ij} }{N_{i}N_{j}}  \;a_{i}(b_{j} +c_{j} - [b_{j}+c_{j}]) \Big)$ & 
$\exp \Big( \frac{2 \pi \ti k_{ij} }{N_{i}N_{j}}  \; a_{i}(b_{j} +c_{j} - [b_{j}+c_{j}]) \Big) $\\[2mm] \hline
$\Z_{N_{ijl}}$ & Type III $k_{{ \tIII(ijl)}}^{} $ &  $\omega_{3,\text{III}}^{(ijl)} (a,b,c) = \exp \Big( \frac{2 \pi \ti k_{ijl}  }{N_{ijl}} \;  a_{i}b_{j}c_{l} \Big)$ &
$\exp \Big( \frac{2 \pi \ti k_{ijl}  }{N_{ijl}} \; ( a_{i}b_{j}c_{l} - b_{i} a_{j}c_{l} +b_{i} c_{j} a_{l} ) \Big) $ \\[1.mm] \hline
\end{tabular}
\caption{ 
The cohomology group $\mathcal{H}^3(G,\R/\Z)$ and 3-cocycles $\omega_3$ for a generic finite Abelian group $G=\prod^n_{i=1} Z_{N_i}$.
The first column shows the classes in $\mathcal{H}^3(G,\R/\Z)$. The second column shows the topological term indices for 2+1D {\it twisted gauge theory}. 
(When all indices $k_{\dots}=0$, it becomes the {\it normal untwisted gauge theory}.)
The third column shows explicit 3-cocycle function $\omega_{3}^{}(a,b,c)$:  $(G)^3 \to \tU(1)$. Here $a=(a_1,a_2,\dots,a_k)$, with $a\in G$ and $a_i \in Z_{N_i}$. 
Same notations for $b,c,d$. The last column shows induced 2-cocycles from the slant product $\sfC_a(b,c)$ using Eq.(\ref{eq:CA(B,C)}).
}
\label{tableA1}
\end{table}
\end{center}

We will study the the 2D's $\text{MCG}(\mathbb{T}^2)= \text{SL}(2,\Z)$ modular data: $\sfS$, $\sfT$ using Rep theory approach.

\subsubsection{Projective Rep and $\sfS$, $\sfT$ for Abelian topological orders} \label{subsubsec:Z22DAbeClass}

This section will simply review some known facts for the later convenience of new results. Much of the discussions can be absorbed from 
Ref.\onlinecite{deWildPropitius:1995cf,{Hu:2012wx},{Hung:2012dx},{Coste:2000tq}}.
Firstly we study the Abelian topological orders from Type I, II 3-cocycles $\omega_3$ of Table \ref{tableA1} for 2+1D topological orders.
We can determine the 
$\sfC^{}_{a}$ projective representation (Rep) and $\tilde{\rho}_\alpha^{a} (b)$:
\bea
\label{eq:CaRep}
\widetilde{\rho}_{\alpha}^{a}(b)\widetilde{\rho}_{\alpha}^{a}(c)=\sfC^{}_{a}(b,c)\widetilde{\rho}_{\alpha}^{a}(bc). 
\eea
Given $Z_a$ is the centralizer of $a \in G$, $\sfC_a$ determines the projective Rep of $Z_a$. Each $\sfC_a$ classifies a class of projective Rep named 
$\sfC_a$-{representations} $\widetilde{\rho}:Z_a{\rightarrow}\text{GL}(Z_a)$.
In Type I, II $\omega_3$, the irreducible $\sfC_A$-representations $\widetilde{\rho}^g_{\alpha}$ of $Z_g$ are in the one-to-one correspondence to 
the irreducible linear representations. The linear Rep originates from the normal untwisted $\prod_i Z_{N_i}$ gauge theory/toric code is: $\exp(  {2\pi \ti}{} (  \sum_i \frac{1}{N_i} \alpha_i h_i  ))$. 
It has pure-charge ($\alpha_i$)-pure-flux ($h_i$) coupling formulated by a BF theory in any dimension (a mutual Chern-Simons theory in 2+1D). 
The full $\sfC_a$-{representations} is:
\bea
\label{}
\widetilde{\rho}^g_{\alpha}(h)=
\exp\big(  {2\pi \ti}{} (  \sum_i \frac{1}{N_i} \alpha_i h_i  )\big) \exp\big(  {2\pi \ti}{} \sum_i \frac{1}{N_i^2} p_i g_i h_i  ) \big) 
\exp\big(  {2\pi \ti}{} \sum_{i,j} \frac{1}{N_i N_j} p_i g_i h_j  ) \big).
\eea
We will interpret $(\alpha_1, g_1,\alpha_2,g_2,\alpha_3,g_3)$ and $(\beta_1, h_1,\beta_2,h_2,\beta_3,h_3)$ as 
the charges $\alpha, \beta$ and fluxes $a,b$ of particles in a doubled basis $| \alpha, g \rangle $, $| \beta, h \rangle $.
The generic $\sfT$ matrix formula of modular $\text{SL}(2,\Z)$ data is\cite{deWildPropitius:1995cf,{Hu:2012wx}} 
\be
\sfT_{({\alpha, A})(\beta, B)}=\sfT_{({\alpha, A})} \delta_{\alpha,\beta} \delta_{A,B}=\frac{\text{Tr} \widetilde{\rho}^{g^A}_{\alpha}(g^A)}{\text{dim}({\alpha})}.
\ee
We obtain: 
\be
 \sfT_{({\alpha, A})}=
\exp\big(  {2\pi i}{} (  [\sum_i \frac{1}{N_i} \alpha_i a_i ] 
+  \sum_{j=1,2,3} \frac{1}{N_j^2} p_j \;(a_j^2) + \sum_{ij=12,23,13} \frac{1}{N_i N_j} p_{ij}\;(a_i a_j))\big), 
\ee
which $ \sfT_{({\alpha, A})}=e^{\ti \Theta_\alpha^A }$ describe the exchange statistics of two identical particles or the topological spin of the same particle.
On the other hand, the generic $\sfS$ matrix formula in 2+1D reads from\cite{deWildPropitius:1995cf,{Hu:2012wx}} 
\be
\sfS_{(\alpha,a)(\beta,b)}=\frac{1}{|G|}\sum_{\substack{{g\in C^a,h\in C^b}\\{gh=hg}}}
  \text{Tr}\widetilde{\rho}^{g}_{\alpha}(h)^*
  \text{Tr}\widetilde{\rho}^{h}_{\beta}(g)^*
  \ee
  yields 
\bea
\label{modSAb}
&& {{  \sfS_{(\alpha, a)(\beta, b)}(p_j,p_{ij})  
= \frac{1}{|G|} \exp\big(-2\pi \ti (\frac{1}{N_i} [\sum_i^2 \alpha_i b_i + \beta_i a_i] 
+ 2 \sum_{j=1,2,3} \frac{1}{N_i^2}  p_j \;(a_j b_j) + \sum_{ij=12,23,13} \frac{1}{N_i N_j}  p_{ij}\;(a_i b_j+ b_i a_j))\big) }}. \;\;\;\;\;\;\;\;\;\;
\eea
One can use a $K$-matrix Chern-Simons theory of an action $\mathbf{S}=\frac{1}{4\pi}\int K_{IJ} a_I \wedge da_J$
to encode the information of $| \alpha, g \rangle $, $| \beta, h \rangle $ into quasiparticles vectors $l,l'$ respectively, and formulate a $K$ with
$
\sfS_{l,l'}(p_j,p_{ij})  = \frac{1}{|G|} \exp(-2\pi \ti l^T K^{-1} l').
$
We can use $\sfS,\sfT$ to study {\bf the classifications of classes of topological orders}.
For example, for $G=(Z_2)^2$ twisted theories, simply using $\sfT$ under basis(particles)-relabeling, 
we find the diagonal eigenvalues of $\sfT$ can be labeled by 
$(\tN_1, \tN_{-1}, \tN_{\ti}, \tN_{-\ti})$,
as numbers of eigenvalues for $\sfT=1,-1,\ti,-\ti$.  
We show that using the data show in Table \ref{table:Z22Abclass} 
is enough to match the classes found in Ref.\onlinecite{Chenggu}.
We 
denote $(\tn_{\pm \ti}, \tn_{\pm 1}, \tn_{1})$ 
as the numbers for (the pair of ${\pm \ti}$, the pair of ${\pm 1}$, individual 1).
Note that $\tN_1+ \tN_{-1}+ \tN_{\ti}+ \tN_{-\ti}=$$2 \tn_{\pm \ti}+ 2 \tn_{\pm 1} + \tn_{1}=$ GSD$_{\mathbb{T}^2}=|G|^2$.
There are 8 {\bf  types} of 3-cocycles but there are only 4 {\bf classes} in Table \ref{table:Z22Abclass}. The number in the bracket $[.]$ of $\omega_3[.]$ indicates
the number of $+\ti$ (or equivalently the number of a pair of $\pm \ti$, paired due to the {\it twisted quantum doubled model} nature).
\begin{table}[!h]
\begin{tabular}{|c||c| c| c|} 
\hline
Class \; & \; $(\tN_1, \tN_{-1}, \tN_{\ti}, \tN_{-\ti})$  & \; $(\tn_{\pm \ti}, \tn_{\pm 1}, \tn_{1})$  & \; Number of Types\\ \hline
$\omega_3[0]$ \; & \; $(10,6,0,0)$ & \; $(0,6,4)$ & \;  1\\ \hline
$\omega_3[{2}]$ \; & \; $(8,4,2,2)$ & \; $(2,4,4)$  & \;  3\\ \hline
$\omega_3[{4}]$ \; & \; $(6,2,4,4)$ & \; $(4,2,4)$ & \;  3\\ \hline
$\omega_3[{6}]$ \; & \; $(4,0,6,6)$& \;  $(6,0,4)$  & \; 1 \\ \hline 
\end{tabular}
\caption{Phases of $\mathcal{H}^3((Z_2)^2,\R/\Z)=(\Z_2)^3$. 8 {\bf  types} of 3-cocycles but there are only 4 {\bf classes}.   }
\label{table:Z22Abclass}
\end{table}

For another example, $G=(Z_2)^3$ twisted theories, we find that, in Table \ref{table:Z23Abclass}, by classifying 
and identifying the modular $\sfS,\sfT$ data,
the {\bf 64 Abelian types} 3-cocycles (all with Abelian statistics) in $\mathcal{H}^3(G,\R/\Z)$ are truncated to only {\bf 4 classes}.

\begin{table}[!h]
\begin{tabular}{|c||c| c| c|} 
\hline
Class \; & \; $(\tN_1, \tN_{-1}, \tN_{\ti}, \tN_{-\ti})$  & \; $(\tn_{\pm \ti}, \tn_{\pm 1}, \tn_{1})$  & \; Number of Types\\ \hline
$\omega_3[0]$ \; & \; $(36,28,0,0)$ & \; $(0,28,8)$ & \;  1\\ \hline
$\omega_3[{8}]$ \; & \; $(28,20,8,8)$ & \; $(8,20,8)$  & \;  21\\ \hline
$\omega_3[{16}]$ \; & \; $(20,12,16,16)$ & \; $(16,12,8)$ & \;  35\\ \hline
$\omega_3[{24}]$ \; & \; $(12,4,24,24)$& \;  $(24,4,8)$  & \; 7 \\ \hline 
\end{tabular}
\caption{Phases of $\mathcal{H}^3((Z_2)^3,\R/\Z)=(\Z_2)^7$. Among 128 types of 3-cocycles, 
64 {\bf  types} of 3-cocycles with Abelian statistics but there are only 4 {\bf classes}.  }
\label{table:Z23Abclass}
\end{table}

\subsubsection{Projective Rep and $\sfS$, $\sfT$ for non-Abelian topological orders} \label{subsubsec:ProjRepZ2cubenAb}

For 2+1D $G=(Z_2)^3$ twisted gauge theories of $\mathcal{H}^3((Z_2)^3,\R/\Z)=(Z_2)^7$, with 128 types of theories, we have shown 
that the 64 types of theories with 
Abelian statistics (from 
64 types of 3-cocycles without Type III twist) are truncated to 4 classes in Table \ref{table:Z23Abclass}.
Here we will consider the remaining 64 types 3-cocycles with Type III twist in $\mathcal{H}^3((Z_2)^3,\R/\Z)$. 
Although the gauge group $G$ is Abelian, the Type III cocycle twist promotes the theory to have non-Abelian statistics. 
Our basic knowledge and formalism are rooted in Ref.\onlinecite{deWildPropitius:1995cf}, where the dual $D_4$ and $Q_8$ gauge theories are found for certain
Type III twist.
Here we 
generalize Ref.\onlinecite{deWildPropitius:1995cf}'s result to all kinds of 3-cocycles twists.

Our expression is the generalized case where 3-cocycles are based on Type III's but can include (or not include) Type I, II 3-cocycles. 
There are 8 Abelian charged particles with zero flux, and 
14 non-Abelian charged particles (which projective Rep $\widetilde{\rho}_{\alpha}^{a}(b)$ is 2 dimensional, described by a rank-2 matrix) with nonzero fluxes as dyons.
For $a,b,c \in G=(Z_2)^3$, 
we will label $8$ elements in $G=(Z_2)^3$ by $(0,0,0)$, $(1,0,0)$, $(0,1,0)$, $(0,0,1)$, $(1,1,0)$, $(1,0,1)$, $(0,1,1)$, $(1,1,1)$.
We denote the above $8$ elements as the abbreviation: $F(0),F(1),F(2),F(3),F(4),F(5),F(6),F(7)$ accordingly.
Let us recall:
$\widetilde{\rho}^{g_a}_\alpha(g_b)$ 
contains $\alpha$ meaning the representation as charges,
also $g_b$ meaning the flux,
and $g_a$ indicating in general the {conjugacy class (i.e. flux) as basis.
In short, our notation leads to $\widetilde{\rho}^{g_a}_\alpha(g_b)=\widetilde{\rho}^{\text{conjugacy class(flux) as basis}}_{\text{representation(charge)}} \text{(flux)}$.\\

\noindent
{$\bullet$\;{ { \bf{$1 \cdot 8=8$ particles: $F(0),(\alpha_1,\alpha_2,\alpha_3)$}  } }  }\\

When the flux is zero flux, $a=F(0)$ is the conjugacy class $C^{F(0)}$. There are 8 linear irreducible representations as charges. These charges can be labeled by $(\alpha_1,\alpha_2,\alpha_3)$ with $(\alpha_1,\alpha_2,\alpha_3) \in (Z_2)^3$, $\alpha_1,\alpha_2,\alpha_3 \in \{0,1\}$.  
So we have
\bea
\widetilde{\rho}^{F(0)}_{F(0),(\alpha_1,\alpha_2,\alpha_3)}(b)=\widetilde{\rho}^{F(0)}_{F(0),(\alpha_1,\alpha_2,\alpha_3)}(b_1,b_2,b_3)=\exp(\frac{2\pi i}{m^2} m \big(\sum_{j=1,2,3} \alpha_j b_j \big)).
\eea

\noindent
{$\bullet$\; \bf{$7 \cdot 2=14$ particles: $F(j),\pm$}}

The other remained 7 kinds of fluxes are $a=F(j)$ for $j=1,\dots, 7$. There are two kinds of representations for each. We can denote these two representations as $+$ or $-$. So these together give $14$ more type of particles. Totally there are $1 \cdot 8+  7 \cdot 2 =22$ quasi-particle excitations as the $\GSD$ on $\mathbb{T}^2$ torus.
Generally, the representation is 
$
\widetilde{\rho}^{F(j)}_{F(j),\pm}(F(l))
$
for some inserting flux $F(l)$. This is a 2-dimensional representation. The identity always assigns to $F(0)$, namely
$
\widetilde{\rho}^{F(j)}_{F(j),\pm}(F(0))= {\begin{pmatrix} 
1 & 0 \\
0 & 1 
\end{pmatrix}}
$. 
We will list down three more elements $\widetilde{\rho}^{F(j)}_{F(j),\pm}(F(1))$, $\widetilde{\rho}^{F(j)}_{F(j),\pm}(F(2))$, $\widetilde{\rho}^{F(j)}_{F(j),\pm}(F(3))$. The other 
remaining $\widetilde{\rho}^{F(j)}_{F(j),\pm}(F(l))$ for $l=4,\dots, 7$ can be determined by
Eq.(\ref{eq:CaRep}). The representations are adjusted by a 1-dimensional projective Rep by Type I $\omega_I$, Type II $\omega_{II}$ 3-cocycles: 
with topological level quantized coefficients as $p_1,p_2,p_3$  of Type I and 
$p_{12},p_{13},p_{23}$ of Type II. Under the Type I, Type II twists, the Type III Rep adjusts to:
\bea
\widetilde{\rho}^{F(j)=a}_{F(j)=a,\pm}(b) \to \widetilde{\rho}^{F(j)}_{F(j),\pm}(b) e^{i \frac{\pi}{2} ( \underset{j<l}{\underset{j,l\in\{1,2,3\}}{\sum}} p_l a_l b_l+p_{ln} a_l b_n )}.
\eea

\noindent
{$\bullet$\; \bf{$2$ particles: $F(1),\pm$}}

$j=1$, here $(a_1,a_2,a_3)=F(1)=(1,0,0)$,
\bea
\widetilde{\rho}^{F(j)}_{F(j),\pm}(F(1))= \pm {\begin{pmatrix} 
1 & 0 \\
0 & 1 
\end{pmatrix}} e^{i \frac{\pi}{2} (p_1 a_1)},\;\;\; 
\widetilde{\rho}^{F(j)}_{F(j),\pm}(F(2))= {\begin{pmatrix} 
0 & 1 \\
1 & 0 
\end{pmatrix}} e^{i \frac{\pi}{2} (p_2 a_2 + p_{12} a_1)},\;\;\;
\widetilde{\rho}^{F(j)}_{F(j),\pm}(F(3))= {\begin{pmatrix} 
1 & 0 \\
0 & -1 
\end{pmatrix}} e^{i \frac{\pi}{2} (p_3 a_3 + p_{13} a_1+p_{23} a_2)} \nonumber
\eea

\noindent
{$\bullet$\; \bf{$2$ particles: $F(2),\pm$}}

$j=2$, here $(a_1,a_2,a_3)=F(2)=(0,1,0)$,
\bea
\widetilde{\rho}^{F(j)}_{F(j),\pm}(F(1))= 
{\begin{pmatrix} 
1 & 0 \\
0 & -1 
\end{pmatrix}}
e^{i \frac{\pi}{2} (p_1 a_1)},\;\;\; 
\widetilde{\rho}^{F(j)}_{F(j),\pm}(F(2))=
\pm {\begin{pmatrix} 
1 & 0 \\
0 & 1 
\end{pmatrix}} 
 e^{i \frac{\pi}{2} (p_2 a_2 + p_{12} a_1)},\;\;\;
\widetilde{\rho}^{F(j)}_{F(j),\pm}(F(3))= {\begin{pmatrix} 
0 & 1 \\
1 & 0 
\end{pmatrix}}
 e^{i \frac{\pi}{2} (p_3 a_3 + p_{13} a_1+p_{23} a_2)} \nonumber
\eea

\noindent
{$\bullet$\; \bf{$2$ particles: $F(3),\pm$}}

$j=3$, here $(a_1,a_2,a_3)=F(3)=(0,0,1)$,
\bea
\widetilde{\rho}^{F(j)}_{F(j),\pm}(F(1))= 
 {\begin{pmatrix} 
0 & 1 \\
1 & 0 
\end{pmatrix}}
e^{i \frac{\pi}{2} (p_1 a_1)},\;\;\; 
\widetilde{\rho}^{F(j)}_{F(j),\pm}(F(2))=
{\begin{pmatrix} 
1 & 0 \\
0 & -1 
\end{pmatrix}}
 e^{i \frac{\pi}{2} (p_2 a_2 + p_{12} a_1)},\;\;\;
\widetilde{\rho}^{F(j)}_{F(j),\pm}(F(3))=
\pm {\begin{pmatrix} 
1 & 0 \\
0 & 1 
\end{pmatrix}} 
 e^{i \frac{\pi}{2} (p_3 a_3 + p_{13} a_1+p_{23} a_2)} \nonumber
\eea

\noindent
{$\bullet$\; \bf{$2$ particles: $F(4),\pm$}}

$j=4$, here $(a_1,a_2,a_3)=F(4)=(1,1,0)$,
\bea
\widetilde{\rho}^{F(j)}_{F(j),\pm}(F(1))= 
 {\begin{pmatrix} 
0 & 1 \\
1 & 0 
\end{pmatrix}}
e^{i \frac{\pi}{2} (p_1 a_1)},\;\;\; 
\widetilde{\rho}^{F(j)}_{F(j),\pm}(F(2))=
\pm {\begin{pmatrix} 
0 & 1 \\
1 & 0 
\end{pmatrix}} 
 e^{i \frac{\pi}{2} (p_2 a_2 + p_{12} a_1)},\;\;\;
\widetilde{\rho}^{F(j)}_{F(j),\pm}(F(3))=
{\begin{pmatrix} 
1 & 0 \\
0 & -1 
\end{pmatrix}}
 e^{i \frac{\pi}{2} (p_3 a_3 + p_{13} a_1+p_{23} a_2)} \nonumber
\eea

\noindent
{$\bullet$\; \bf{$2$ particles: $F(5),\pm$}}

$j=5$, here $(a_1,a_2,a_3)=F(5)=(1,0,1)$,
\bea
\widetilde{\rho}^{F(j)}_{F(j),\pm}(F(1))= 
\pm {\begin{pmatrix} 
0 & 1 \\
1 & 0 
\end{pmatrix}}
e^{i \frac{\pi}{2} (p_1 a_1)},\;\;\; 
\widetilde{\rho}^{F(j)}_{F(j),\pm}(F(2))=
 {\begin{pmatrix} 
1 & 0 \\
0 & -1 
\end{pmatrix}} 
 e^{i \frac{\pi}{2} (p_2 a_2 + p_{12} a_1)},\;\;\;
\widetilde{\rho}^{F(j)}_{F(j),\pm}(F(3))=
{\begin{pmatrix} 
0 & 1 \\
1 & 0 
\end{pmatrix}}
 e^{i \frac{\pi}{2} (p_3 a_3 + p_{13} a_1+p_{23} a_2)} \nonumber
\eea

\noindent
{$\bullet$\; \bf{$2$ particles: $F(6),\pm$}}

$j=6$, here $(a_1,a_2,a_3)=F(6)=(0,1,1)$,
\bea
\widetilde{\rho}^{F(j)}_{F(j),\pm}(F(1))= 
 {\begin{pmatrix} 
1 & 0 \\
0 & -1 
\end{pmatrix}} 
e^{i \frac{\pi}{2} (p_1 a_1)},\;\;\; 
\widetilde{\rho}^{F(j)}_{F(j),\pm}(F(2))=
 {\begin{pmatrix} 
0 & 1 \\
1 & 0 
\end{pmatrix}}
 e^{i \frac{\pi}{2} (p_2 a_2 + p_{12} a_1)},\;\;\;
\widetilde{\rho}^{F(j)}_{F(j),\pm}(F(3))=
\pm{\begin{pmatrix} 
0 & 1 \\
1 & 0 
\end{pmatrix}}
 e^{i \frac{\pi}{2} (p_3 a_3 + p_{13} a_1+p_{23} a_2)} \nonumber
\eea

\noindent
{$\bullet$\; \bf{$2$ particles: $F(7),\pm$}}

$j=7$, here $(a_1,a_2,a_3)=F(7)=(1,1,1)$, (note in particular this Rep, our choice $\mp$ differs from Ref.\onlinecite{deWildPropitius:1995cf}.)
\bea
\widetilde{\rho}^{F(j)}_{F(j),\pm}(F(1))= 
\mp {\begin{pmatrix} 
0 & 1 \\
1 & 0 
\end{pmatrix}}
e^{i \frac{\pi}{2} (p_1 a_1)},
\widetilde{\rho}^{F(j)}_{F(j),\pm}(F(2))=
\mp {\begin{pmatrix} 
0 & -i \\
i & 0 
\end{pmatrix}} 
 e^{i \frac{\pi}{2} (p_2 a_2 + p_{12} a_1)},
\widetilde{\rho}^{F(j)}_{F(j),\pm}(F(3))=
\mp {\begin{pmatrix} 
1 & 0 \\
0 & -1 
\end{pmatrix}}
 e^{i \frac{\pi}{2} (p_3 a_3 + p_{13} a_1+p_{23} a_2)} \nonumber
\eea

With the above projective Rep $\widetilde{\rho}_{\alpha}^{a}(b)$, we can derive the analytic form of modular data $\sfS,\sfT$ in 2D. Here for $G=(Z_2)^3$,
\bea
\mathsf{T}^A_{\alpha}=  e^{\ti \frac{\pi}{2} ( \underset{l<m}{\underset{l,m\in\{1,2,3,4\}}{\sum}} p_{l} a_l{}^2  +p_{lm} a_l a_l )} (\pm)_a  (\ti)^{\eta_{a,a}}  \;  
\to \;\;\; \mathsf{T}^A_{\alpha}=\pm 1 \text{ or }  \pm \ti
\eea
\bea \label{eq:eta2D}
{\eta_{g_1,g_2}} \equiv\left\{
    \begin{array}{lr}
      0,& \text{ if } \mathsf{C}^{}_{g_1}(g_2,g_2)=+1.\\
      1,& \text{ if } \mathsf{C}^{}_{g_1}(g_2,g_2)=-1.
    \end{array}
    \right.
 \eea
More explicitly, we compute $\mathsf{T}^A_{\alpha}$ in Table \ref{table:TmatrixZ2}:
\begin{table}[h]
\begin{tabular}{|c||c|}\hline
particle \; & \;  $\mathsf{T}^a_{\alpha}$ \\ \hline
$((\alpha_1,\alpha_2,\alpha_3),F(0))$ \; & \;  1\\ \hline
$(\pm, F(1))$, $(\pm, F(2))$, $(\pm, F(3))$ \; & \;  $\pm \ti^{p_1}$, $\pm \ti^{p_2}$, $\pm \ti^{p_3}$ \\ \hline
$(\pm, F(4))$, $(\pm, F(5))$, $(\pm, F(6))$ \; & \;  $\pm \ti^{p_1+p_2+p_{12}}$, $\pm \ti^{p_1+p_3+p_{13}}$, $\pm \ti^{p_2+p_3+p_{23}}$ \\ \hline
$(\pm, F(7))$\; & \;  $\pm \ti  \cdot \ti^{p_1+p_2+p_3+p_{12}+p_{13}+p_{23}}$ \\ 
\hline  
 \end{tabular} \label{table:TmatrixZ2}
\caption{ The modular $\mathsf{T}^a_{\alpha}$ matrix for 2D twisted $(Z_2)^3$ theories with non-Abelian statistics. The table contains all $64$ non-Abelian theories in $\cH^3((Z_2)^3, \R/\Z)$.}
\end{table}

With the modular $\mathsf{S}^{xy}=\mathsf{S}^{xy}_{(\alpha,a)(\beta,b)}$ matrix (of 64 types of 2D twisted $(Z_2)^3$ theories with non-Abelian statistics):
\begin{equation} \label{eq:SxyZ2cube}
\mathsf{S}=\frac{1}{|G|}\bBigg@{6.5}( \mkern-10mu
\begin{tikzpicture}[baseline=-.65ex]
\matrix[
  matrix of math nodes,
  column sep=1ex,
] (m)
{
1&2(-1)^{ b_1 \alpha_1+b_2 \alpha_2+b_3 \alpha_3 }& 2(-1)^{ b_1 \alpha_1+b_2 \alpha_2+b_3 \alpha_3 }\\
2(-1)^{ a_1 \beta_1+a_2 \beta_2+a_3 \beta_3 } & 
\delta_{a,b} 4 \cdot (-1)^{\eta_{a,a}}   \cdot (-1)^{\underset{j<l}{\underset{j,l=1,2,3}{\sum}}p_j a_j + p_{jl} a_j a_l  }  &-\delta_{a,b} 4 (-1)^{\eta_{a,a}}  \cdot (-1)^{\underset{j<l}{\underset{j,l=1,2,3}{\sum}}p_j a_j + p_{jl} a_j a_l  } \\
2(-1)^{ a_1 \beta_1+a_2 \beta_2+a_3 \beta_3 } & -\delta_{a,b} 4 (-1)^{\eta_{a,a}} \cdot (-1)^{\underset{j<l}{\underset{j,l=1,2,3}{\sum}}p_j a_j + p_{jl} a_j a_l  }  & \delta_{a,b} 4 (-1)^{\eta_{a,a}}  \cdot (-1)^{\underset{j<l}{\underset{j,l=1,2,3}{\sum}}p_j a_j + p_{jl} a_j a_l  } \\
};
\draw[dashed]
  ([xshift=10.15ex]m-1-1.north east) -- ([xshift=0.5ex]m-3-1.south east);
  \draw[dashed]
  ([xshift=10.ex]m-1-2.north east) -- ([xshift=0.8 ex]m-3-2.south east);
\draw[dashed]
  ([xshift=-9.ex] m-1-1.south west) -- ([xshift=4.ex]m-1-3.south east);
  \draw[dashed]
  ([xshift=-.5ex] m-2-1.south west) -- ([xshift=-1.ex]m-2-3.south east);
\node[above] at ([yshift=.2ex]m-1-1.north) {$\scriptstyle (\beta_j,\;0)$};  
\node[above] at ([yshift=.2ex]m-1-2.north) {$\scriptstyle (+,\;b_j)$};
\node[above] at ([yshift=.2ex]m-1-3.north) {$\scriptstyle (-,\;b_j)$};
\node[right,overlay] at ([xshift=8.8ex]m-1-3.east) {$\scriptstyle (\alpha_j,\;0)$};
\node[right,overlay] at ([xshift=-.5ex]m-2-3.east) {$\scriptstyle (+,\;a_j)$};
\node[right,overlay] at ([xshift=.5ex]m-3-3.east) {$\scriptstyle (-,\;a_j)$};
\end{tikzpicture}\mkern-16mu
\bBigg@{6.5})
\end{equation}

In Eq.(\ref{eq:SxyZ2cube}), the factor $(-1)^{\eta_{a,a}}$ is derived from a computation of $(\ti)^{\eta_{a,b}} \cdot (\ti)^{\eta_{b,a}} \delta_{a,b} = (-1)^{\eta_{a,a}} \delta_{a,b}$. 
From Eq.(\ref{eq:eta2D}), we notice that ${\eta_{a,a}}=1$ is nonzero only when $a=(1,1,1)=F(7)$ for the $(Z_2)^3$ flux.  

\subsection{Classification of 2+1D twisted $(Z_2)^3$ gauge theories, $D^\omega((Z_2)^3)$ and $\mathcal{H}^3((Z_2)^3,\R/\Z)$.}

Those twisted $(Z_2)^3$ gauge theories dual to $D_4$, $Q_8$ non-Abelian gauge theories are firstly discovered by Ref.\onlinecite{deWildPropitius:1995cf}.
Here we will present {the other three classes which cannot be dual to any non-Abelian gauge theory, but only to be a  twisted (Abelian or non-Abelian) gauge theory itself.}
We again label the diagonal eigenvalues of $\sfT$ by 
$(\tN_1, \tN_{-1}, \tN_{\ti}, \tN_{-\ti})$,
their number of eigenvalues for $\sfT=1,-1,\ti,-\ti$. 
We also use shorthand $(\tn_{\pm \ti}, \tn_{\pm 1}, \tn_{1})$ instead, 
which stands for the numbers for (the pair of ${\pm \ti}$, the pair of ${\pm 1}$, individual 1) in the diagonal of $\sfT$.
Note that $\tN_1+ \tN_{-1}+ \tN_{\ti}+ \tN_{-\ti}=$$2 \tn_{\pm \ti}+ 2 \tn_{\pm 1} + \tn_{1}=$ GSD$_{\mathbb{T}^2}=22$.
There are 64 {\bf  types} of 3-cocycles 
corresponding to theories with non-Abelian statistics but there are only 5 inequivalent {\bf classes} in Table \ref{table:Z22Abclass}. The number in the bracket $[.]$ of $\omega_3[.]$ indicating
the number of $+\ti$ (or equivalently the number of a pair of $\pm \ti$, paired due to the {\it quantum doubled model} nature.).

\begin{table}[!h]
\begin{tabular}{|c||c|c|c|c|} 
\hline
Class \; & \;  $(\tn_{\pm \ti}, \tn_{\pm 1}, \tn_{1})$ & \; $(\tN_1, \tN_{-1}, \tN_{\ti}, \tN_{-\ti})$  & \;  Twisted quantum double $D^\omega(G)$ & \; Number of Types\\ \hline
$\omega_3[1]$ \; & \; (1,6,8)  & \;  (14,6,1,1) & \; $D^{\omega_3{[1]}}(Z_2{}^3)$, $D(D_4)$ & \;  7\\ \hline
$\omega_3[{3d}]$ \; & \; (3,4,8) & \; (12,4,3,3)  & \; $D^{\omega_3{[3d]}}(Z_2{}^3)$, $D^{\gamma^4}(Q_8)$ & \;  7\\ \hline
$\omega_3[{3i}]$ \; & \; (3,4,8)  & \; (12,4,3,3) & \; $D^{\omega_3[{3i}]}(Z_2{}^3)$, $D^{}(Q_8)$, $D^{\alpha_1}(D_4)$, $D^{\alpha_2}(D_4)$ & \;  28\\ \hline
$\omega_3[{5}]$ \; & \; (5,2,8)  & \; (10,2,5,5)& \; $D^{\omega_3{[5]}}(Z_2{}^3)$, $D^{\alpha_1\alpha_2}(D_4)$ & \; 21 \\ \hline
$\omega_3[{7}]$ \; & \;  (7,0,8) & \; (8,0,7,7) & \; $D^{\omega_3{[7]}}(Z_2{}^3)$ & \; 1 \\ \hline
 \end{tabular}
\caption{ $D^\omega(G)$ is the twisted quantum double of $G$ with a cocycle twist $\omega$ of $G$'s cohomology group. 
Here we consider a 3-cocycle twist $\omega_3$ in $\mathcal{H}^3((Z_2)^3,\R/\Z)=(\Z_2)^7$, which $\omega_3$ contains a factor of Type III 3-cocycle.
We compute the second and the fourth columns, and then compare them with the mathematics literature Ref.\onlinecite{Goff2006} to match for the third column.
We find that the 64 types of non-Abelien theories are truncated to 5 classes. 
}
\label{table:Z2cubeclassApp}
\end{table}

Although $\omega_3[{3d}]$ and $\omega_3[{3i}]$ share the same $\sfT$ matrix data, but they can still be distinguished by the linear dependency of the
fluxes which carry three pairs of eigenvalues $\ti$. (And, of course, they can be distinguished by the more-involved $\sfS$ matrix.) 
There are 7 types in the $\omega_3[{3d}]$ class, whose $\pm \ti$ are generated by linear-dependent fluxes. 
Another 28 types in the $\omega_3[{3i}]$ class, whose $\pm \ti$ are generated by linear-independent fluxes. 
In this notation of linear (in)dependency, we have $\omega_3[{1}]=\omega_3[{1i}]$, $\omega_3[{5}]=\omega_3[{5d}]$, $\omega_3[{7}]=\omega_3[{7d}]$.
Such a concept is also used in the mathematic literature in Ref.\onlinecite{Goff2006},
where they study the Frobenius-Schur indicators, Frobenius-Schur exponents and the support of cocycle twist, supp $\omega$; and use these data
to classify twisted quantum double model $D^\omega(G)$. 
Remarkably, we find that using our data is enough to match the classes found in the math literature\cite{Goff2006} in the quantum double and module category framework.

These altogether with Sec.\ref{subsubsec:Z22DAbeClass} 
form a complete data set of $\mathcal{H}^3((Z_2)^3,\R/\Z)=(\Z_2)^7$,
where {\bf 128 types of 3-cocycles fall into 4 distinct classes of Abelian topological orders in Table \ref{table:Z22Abclass} and 5 distinct classes of
non-Abelian topological orders in Table \ref{table:Z2cubeclassApp} }.
Totally there are 9 distinct classes of topological orders within twisted $(Z_2)^3$ gauge theories.
We note that $\omega_3[{3i}]$, $\omega_3[{5}]$, $\omega_3[{7}]$ can only be twisted gauge theories, not dual to any untwisted non-Abelian gauge theory.

\subsection{3+1D topological orders of $\mathcal{H}^4(G,\R/\Z)$}

This section continues the discussion and notations from $\mathcal{H}^3(G,\R/\Z)$ of 2+1D to $\mathcal{H}^4(G,\R/\Z)$ of 3+1D topological orders. 
Now we fill in some more information about the data of the projective Rep.

\subsubsection{Projective Rep and $\sfS$, $\sfT$ for Abelian topological orders} \label{subsubsec:Z23DAbeClass}
The data of $\widetilde{\rho}^{ab}_{\text{}\alpha}(c)$ is organized below for $G=Z_{N_1} \times Z_{N_2} \times Z_{N_3}$ of the cohomology group $\mathcal{H}^4(G,\R/\Z)$.
Its modular $\sfS$, $\sfT$ matrices for this Rep have been presented in Table \ref{tableSxyzZN123}, \ref{tableTxyZN123}, \ref{tableSxyZN123}.
In the main text, we provide an example of classifying 3D topological orders from 3+1D $(Z_2)^2$ twisted gauge theories of 4 types (from $\mathcal{H}^4((Z_2)^2,\R/\Z)=(Z_2)^2$), and find out that 4 types are truncated to only 2 distinct classes of topological orders.

\begin{center}
\begin{table} [!h]
\begin{tabular}{|c||c|c| }
\hline
\;$\mathcal{H}^4(G,\R/\Z)$\; &  4-cocycle &  $\widetilde{\rho}^{a, b}_{\text{}\alpha}(c)$  \\[0mm]  \hline \hline 
$\Z_{N_{12}}$ & Type II 1st   & 
$\widetilde{\rho}^{(1st)a, b}_{\text{II},\alpha}(c) =\exp \big( \sum_k \frac{2 \pi \ti  }{ N_k   }  \; \alpha_k c_k  \big) \cdot \exp \big( \frac{2 \pi \ti p_{{ \text{II}(12)}}^{(1st)} }{ (N_{12} \cdot N_2  )   }  \;  (a_2 b_1-a_1 b_2 ) c_2  \big)$  \\[2mm] \hline
$\Z_{N_{12}}$ & Type II 2nd & 
$\widetilde{\rho}^{(2nd)a, b}_{\text{II},\alpha}(c) =\exp \big( \sum_k \frac{2 \pi \ti  }{ N_k   }  \; \alpha_k c_k  \big) \cdot 
\exp \big( \frac{2 \pi \ti p_{{ \text{II}(12)}}^{(2nd)} }{ (N_{12} \cdot N_1  )   }  \;  (a_1 b_2-a_2 b_1 ) c_1  \big)$  \\[2mm] \hline
$\Z_{N_{123}}$ & Type III 1st & 
${\widetilde{\rho}^{(1st)a, b}_{\text{III},\alpha}(c) =\exp \big( \sum_k \frac{2 \pi \ti  }{ N_k   }  \; \alpha_k c_k  \big) \cdot  \exp \big(  \frac{2 \pi \ti p_{{ \text{III}(123)}}^{(1st)} }{ (N_{12} \cdot N_3 )  }  \;  (a_2 b_1-a_1 b_2 )  c_3 \big)}$  \\[2mm] \hline
$\Z_{N_{123}}$ & Type III 2nd  & 
$\widetilde{\rho}^{(2nd)a, b}_{\text{III},\alpha}(c) =\exp \big( \sum_k \frac{2 \pi \ti  }{ N_k   }  \; \alpha_k c_k  \big) \cdot   \exp \big(  \frac{2 \pi \ti p_{{ \text{III}(123)}}^{(2nd)} }{ (N_{31} \cdot N_2 )  }  \;  (a_1 b_3-a_3 b_1 )  c_2 \big)$   \\[2mm] \hline
\end{tabular}
\caption{ $\widetilde{\rho}^{a, b}_{\text{}\alpha}(c)$ for a 3+1D twisted gauge theory with
$G=Z_{N_1} \times Z_{N_2} \times Z_{N_3}$ of $\mathcal{H}^4(G,\R/\Z)$.
We derive $\widetilde{\rho}_{\alpha}^{a,b}(c)$ from the equation introduced in the main text,
$\widetilde{\rho}_{\alpha}^{a,b}(c)\widetilde{\rho}_{\alpha}^{a,b}(d)=\sfC^{(2)}_{a,b}(c,d)\widetilde{\rho}_{\alpha}^{a,b}(cd)$,
presenting the {projective representation}, 
because the induced 2-cocycle belongs to the second cohomology group $\cH^2(G,\R/\Z)$.
The $\widetilde{\rho}_{\alpha}^{a,b}(c)$: $(Z_a, Z_b)$ ${\rightarrow}$ $\text{GL}\left(Z_a,Z_b\right)$ can be written
as a general linear matrix. 
}
\label{tableProjRep3DN123}
\end{table}
\end{center}

\subsubsection{Projective Rep and $\sfS$, $\sfT$ for non-Abelian topological orders} \label{subsubsec:Z23DnAbeClass}

Below we will present the data of twisted gauge theories for those with non-Abelian statistics in 
$\mathcal{H}^4(G=(Z_2)^4,\R/\Z)$ labeled by 4-cocycles $\omega_4$. 
Among $\mathcal{H}^4((Z_2)^4,\R/\Z)=(Z_2)^{21}$ types of theories, there are $2^{20}$ types of them endorsed with non-Abelian statistics.
In some case, we will write the formula in terms of a slightly generic $G=(Z_n)^4$, for a prime $n$.

Analogous to Sec.\ref{subsubsec:ProjRepZ2cubenAb}, we recall  that the 3D triple basis renders: 
$
\widetilde{\rho}^{g^a,g^b}_\alpha(g^c)=\widetilde{\rho}^{\text{conjugacy class(flux,flux) as basis}}_{\text{representation(charge)}} \text{(flux)}.
$
So we understand that the representation $\widetilde{\rho}(c)$ is constrained by the flux $a,b$.
We will consider Type IV $\omega_{4,\tIV}$ twisted theories, but
we include $\omega_{4,\tIV}$ further multiplied by Type II $\omega_{4,\tII}$, Type III $\omega_{4,III}$ 4-cocycles. 
Thus, the representation also relates to 
their topological terms
$p_{lm}$ of Type II $\omega_{4,\tII}$ labeling $(Z_2)^{2{4 \choose 2}}=(Z_2)^{12}$ types of theories, 
$p_{lmn}$ of Type III $\omega_{4,\tIII}$ labeling $(Z_2)^{2{4 \choose 3}}=(Z_2)^8$  types of theories. 
Totally all these 4-cocycles multiplied by $\omega_{4,\tIV}$ yields the $2^{20}$ types of theories endorsed with non-Abelian statistics.
Under the Type II, Type III twists, the Type IV Rep is adjusted to:
\bea \label{eq:Red3DnAb}
\widetilde{\rho}^{a,b}_{a,b,(\pm,\pm)}(c)= \widetilde{\rho}^{F(j_1)=a,F(j_2)=b}_{F(j_1)=a,F(j_2)=b,(\pm,\pm)}(c) 
\cdot e^{i \frac{\pi}{2} ( \underset{l<m<n}{\underset{l,m,n\in\{1,2,3,4\}}{\sum}} p_{lm} f_{{}_{lm}}(a,b,c) +p_{lmn} f_{{}_{lmn}}(a,b,c) )}.
\eea
Note that the trace $\Tr[\widetilde{\rho}^{a,b}_{a,b,(\pm,\pm)}(c)]$ is nonzero only when (1) $c=a$, $c=b$ or $c=ab$ with
 $\Tr[\widetilde{\rho}^{a}_{a,b,(\pm,\pm)}(c)] \neq 0$,
or (2) $c=F(0)$ zero flux, i.e. $\Tr[\widetilde{\rho}^{a,b}_{a,b,(\pm,\pm)}(F(0))] \neq 0$. Other cases have zero traces.
Among the degeneracy sectors on the ${\mathbb{T}^3}$ torus, we have
$\text{GSD}_{\mathbb{T}^3}= \big(n^8+n^9-n^5\big)+\big(n^{10} -n^{7} -n^{6} +n^{3}\big)$ (ground state bases in terms of particles and string quasi-excitations), which is $1576$ for $n=2$. 
We can use $|G|^2=(n^4)^2=256$ (doubled) fluxes to do the first labeling. Note the fluxes form a doubled basis $(a,b)$ in $| \alpha, a, b\rangle$. 
Among 256 fluxes, there are $n^4+n^5-n=46$ fluxes carrying Abelian excitations, while the remained $(n^8-(n^4+n^5-n))=210$ are non-Abelian excitations.
(Beware: the bases carry two fluxes and one charge, these bases should {\bf not} be confused with string and particle types.)
We may organize the ground state bases in terms of two kinds, which correspond to Abelian and non-Abelian excitations:

\noindent
\frm{ {$\bullet$\; {$(n^4+n^5-n) \cdot n^4=46 \times 16 =736$ Abelian excitations: $F(j_{ab}),(\alpha_1,\alpha_2,\alpha_3,\alpha_4)$}
} }

Here $a=F(j_{ab})$ can be zero fluxes, or nonzero fluxes by satisfying the following conditions:
\bea
 a_1 b_2 = a_2 b_1, 
a_1 b_3 = a_3 b_1, 
a_1 b_4 = a_4 b_1, 
a_2 b_3 = a_3 b_2, 
 a_2 b_4 = a_4 b_2, 
a_3 b_4 = a_4 b_3 \pmod{N} 
\eea
There are $(n^4+n^5-n)$ independent solutions for these sets of $a$, $b$.
The conjugacy class $C^{F(j_{ab})}$ stands for fluxes. 
There are $n^4$ representation as charges; 
these can be labeled by $(\alpha_1,\alpha_2,\alpha_3,\alpha_4)$ with $(\alpha_1,\alpha_2,\alpha_3,\alpha_4) \in (Z_2)^4$, and $Z_2= \{0,1\}$. We will write
$(\alpha_1,\alpha_2,\alpha_3,\alpha_4)=\alpha$.
Eq.(\ref{eq:Red3DnAb}) becomes 
\bea
\widetilde{\rho}^{F(j_{ab})}_{F(j_{ab}),(\alpha_1,\alpha_2,\alpha_3,\alpha_4)}(c)=\widetilde{\rho}^{F(0)}_{F(0),(\alpha_1,\alpha_2,\alpha_3,\alpha_4)}(c_1,c_2,c_3,c_4)=\exp \big(\sum_{k=1}^4 \frac{2\pi i}{N_k}\alpha_k c_k \big).
\eea

For $n=2$, there are $(2^4+2^5-2)=46$ (doubled) fluxes contributing Abelian excitations.

\noindent
\frm{$\bullet$\; {$(n^8-(n^4+n^5-n)) \cdot n^2=210 \times 4 =840$ non-Abelian excitations: $F(j_{non.ab}),(\pm,\pm)$ }
}

For $n=2$, there are $(n^8-(n^4+n^5-n))=210$ (doubled) fluxes contributing non-Abelian excitations.
Each of them carries 2-dimensional Rep with two pairs of $(\pm,\pm)$ charge Rep. Thus the number of doubled fluxes multiplied by 4
yields $840$ excitations. 
It is 
equivalent to count the $\sfC^{(2)}_{a,b}(c,d)$ class that they belong to. There are six $c_ld_m$ terms in the Type IV 4-cocycles:
\bea
\sfC^{(2)}_{a,b}(c,d)&=&\exp \big( \frac{2 \pi i
p_{{ IV}{(1234)}}^{}}{ N_{ijlm} } \; 
(a_4 b_3 -a_3 b_4 ) c_1 d_2+(a_2 b_4 -a_4 b_2 ) c_1 d_3+(a_4 b_1 -a_1 b_4 ) c_2 d_3 \nonumber\\
&+&(a_3 b_2 -a_2 b_3 ) c_1 d_4+ (a_1 b_3 -a_3 b_1 ) c_2 d_4+(a_2 b_1 -a_1 b_2 ) c_3 d_4 \big). 
\eea
Below each solution will be multiplied by $6$, due to ${3 \choose 2} \times 2$, that $3$ terms $a,b,ab$ can choose $2$ as the generator basis for $a$, $b$. 
Those terms have $\Tr[\widetilde{\rho}^{a}_{a,b,(\pm,\pm)}(c)] \neq 0$ for $c=0,a,b,ab$. And the permutation of $a,b$ results in an extra multiple of $2$. 
We organize the solutions to the following six {\bf styles}.
Each style may contain dimensionally reduced 3-cocycles, as ``{Type III 3-cocycle like}'' or ``mixed-Type III 3-cocycles.''
Here ``\emph{Type III 3-cocycle like}'' means that the dimensional reduced 2D theory has an induced 3-cocycle which is a Type III 3-cocycle \emph{within} a subgroup $(Z_2)^3$.
``\emph{Mixed-Type III 3-cocycle}'' means that the dimensional reduced 2D theory has an induced 3-cocycle which contains several Type III 3-cocycles spanning the full group $(Z_2)^4$.
The {\bf six styles} of solutions are:

\noindent
{$\bullet$\; {$\sfC^{(2)}_{a,b}(c,d)$ contains 1 $cd$ term: ${6 \choose 1}\times 6=36$ non-Abelian fluxes  - {\bf Style 1} (Type III 3-cocycle like)}
}:\\

\noindent
{$\bullet$\; {$\sfC^{(2)}_{a,b}(c,d)$ contains 2 $cd$ term: $ ({6 \choose 2}-3) \times 6= 72$ non-Abelian fluxes - {\bf Style 2} (Type III 3-cocycle like)}
}:\\
We have ${6 \choose 2}$ subtract 3, due to it is impossible to have nonzero coefficients $cd$ terms of $\sfC^{(2)}_{a,b}(c,d)$ for both of the following terms together:\\

(1) $c_3 d_4$ and $c_1 d_2$ terms, 
(2) $c_2 d_4$ and $c_1 d_3$ terms, 
(3) $c_2 d_3$ and $c_1 d_4$ terms.\\

\noindent
{$\bullet$\; {$\sfC^{(2)}_{a,b}(c,d)$ contains 3 $cd$ term: $ {4 \choose 3} \times 6+ {4 \choose 3} \times 6= 48$ non-Abelian fluxes
- Style 3 (Type III 3-cocycle like), Style 4 (mixed-Type III 3-cocycles)}
}:\\

{\bf Style 3} (Type III 3-cocycle like) ${4 \choose 3} \times 6$:\\
${4 \choose 3}$ out of 6 have nonzero coefficients for: 
(1) $c_2 d_3$, $c_2 d_4$, $c_3 d_4$. 
(2) $c_1 d_3$, $c_1 d_4$, $c_3 d_4$. 
(3) $c_1 d_2$, $c_1 d_4$, $c_2 d_4$. 
(3) $c_1 d_2$, $c_1 d_3$, $c_2 d_3$. 
Each type has 6 possible choices for $a$,$b$.\\

{\bf Style 4} (mixed-Type III 3-cocycles) ${4 \choose 3} \times 6$:\\
${4 \choose 3}$ out of 6 have nonzero coefficients for:
(1) $c_1 d_2$, $c_1 d_3$,  $c_1 d_4$. 
(2) $c_1 d_2$, $c_2 d_3$,  $c_2 d_4$. 
(3) $c_1 d_3$, $c_2 d_3$,  $c_3 d_4$. 
(4) $c_1 d_4$, $c_2 d_4$,  $c_4 d_4$. 
Each type has 6 possible choices for $a$,$b$.\\

\noindent
{$\bullet$\; {$\sfC^{(2)}_{a,b}(c,d)$ contains 4 $cd$ term: $ ({6 \choose 4}-{4 \choose 3} \cdot 3)\times 6 =3 \times 6= 18$ non-Abelian fluxes
- {\bf Style 5} (mixed-Type III 3-cocycles)}
}

Among 15 terms (with 4 $cd$) in ${6 \choose 4}=15$, there are only 3 terms allowed.\\
\noindent
(1) $c_1 d_2$, $c_2 d_3$, $c_1 d_4$, $c_3 d_4$, 
(2) $c_1 d_3$, $c_2 d_3$, $c_1 d_4$, $c_2 d_4$,  
(3) $c_1 d_2$, $c_1 d_3$, $c_2 d_4$, $c_3 d_4$.\\ 
There are terms from ${4 \choose 3 } \cdot 3 =12$ is not allowed, like
$c_1 d_2$, $c_1 d_3$, $c_2 d_3$, $c_1 d_4$. (i.e. choose 3 elements as ${4 \choose 3 }$ and choose one of the three, thus times $3$, to pair with the remained unchosen.) \\

\noindent
{$\bullet$\; {$\sfC^{(2)}_{a,b}(c,d)$ contains 5 $cd$ term: $ {6 \choose 5} \times 6= 36$ non-Abelian fluxes
- {\bf Style 6} (mixed-Type III 3-cocycles)}
}

\noindent
(1) $c_1 d_2$, $c_1 d_3$, $c_1 d_4$, $c_2 d_3$, $c_2 d_4$,\;
(2) $c_1 d_2$, $c_1 d_3$, $c_1 d_4$, $c_2 d_3$, $c_3 d_4$,\;
(3) $c_1 d_2$, $c_1 d_3$, $c_1 d_4$, $c_2 d_4$, $c_3 d_4$.\;\\
(4) $c_1 d_2$, $c_1 d_3$, $c_2 d_3$, $c_2 d_4$, $c_3 d_4$,\;
(5) $c_1 d_2$, $c_1 d_4$, $c_2 d_3$, $c_2 d_4$, $c_3 d_4$,\;
(6) $c_1 d_3$, $c_1 d_4$, $c_2 d_3$, $c_2 d_4$, $c_3 d_4$

Those Style 1, 2, 3 are pure Type III 3-cocycle $\omega_3$ like, which $\widetilde{\rho}^{a,b}_{a,b,(\pm,\pm)}(c)$ can be deduced from Sec.\ref{subsubsec:ProjRepZ2cubenAb}'s $G=(Z_2)^3$ result.
Style 4, 5, 6 are mixed Type III 3-cocycle in the full $G=(Z_2)^4$ group, so one needs to assign the Rep $\widetilde{\rho}^{a,b}_{a,b,(\pm,\pm)}(c)$  in slightly different manners. But it turns out that rank-2 matrices are always sufficient to
encode the irreducible projective representation of $\mathsf{C}^{(2)}_{ab}(c,d)$. After finding the $\widetilde{\rho}^{a,b}_{a,b,(\pm,\pm)}(c)$, we 
analytically derive their non-Abelian $\sfS^{xyz}$, $\sfT^{xy}$ of 3D presented in the main text, in Table \ref{TxyZ2f}, Eq.(\ref{eq:SnAb3DZ2f}), Eq.(\ref{eq:SnAb3DZ2fAr}).

\begin{widetext}
\end{widetext}

\twocolumngrid

\section{
$\sfS^{xyz}$ and $\sfT^{xy}$ calculation in terms of the gauge group $G$ and 4-cocycle $\om_4$} \label{AppendixST} 

\subsection{Unimodular Group and SL$(N,\Z)$} \label{subsec:SLunit}

In the case of the unimodular group, 
there are the unimodular matrices of rank $N$ forms GL$(N,\Z)$.
 $\sfS_\sfU$ and $\sfT_\sfU$ have determinants $\det(\sfS_\sfU)=-1$ and $\det(\sfT_\sfU)=1$ for
any general $N$:
\bea
&&\sfS_\sfU=\begin{pmatrix}
0 & 0 & 0 & \dots & (-1)^{N}\\
1 & 0 & 0 & \dots & 0 \\
0 & 1 & 0 & \dots & 0\\
\vdots & \vdots & \ddots & \dots & \vdots\\
0 & 0 & 0 & \dots & 0
\end{pmatrix},\;\; \\
&&\sfT_\sfU=\begin{pmatrix}
1 & 1 & 0 & \dots & 0\\
0 & 1 & 0 & \dots & 0\\
0 & 0 & 1 & \dots & 0\\
\vdots & \vdots & \vdots  & \ddots & \vdots\\
0 & 0 & 0  & \dots & 1
\end{pmatrix}.
\eea
Note that $\det(\sfS_\sfU)=-1$ in order to generate both determinant 1 and $-1$ matrices.

For the SL$(N,\Z)$ modular transformation, we denote their generators as $\sfS$ and $\sfT$ for a general $N$
with $\det(\sfS)=\det(\sfT)=1$:
\bea
&&\sfS=\begin{pmatrix}
0 & 0 & 0 & \dots & (-1)^{N-1}\\
1 & 0 & 0 & \dots & 0 \\
0 & 1 & 0 & \dots & 0\\
\vdots & \vdots & \ddots & \dots & \vdots\\
0 & 0 & 0 & \dots & 0
\end{pmatrix},\;\; \\
&&\sfT=\sfT_\sfU.
\eea

Here for simplicity, let us denote $\sfS^{xyz}$ as $\sfS_{3\tD}$, $\sfS^{xy}$ as $\sfS_{2\tD}$, $\sfT^{xy} = \sfT_{3\tD}= \sfT_{2\tD}$. Recall
SL$({3},\mathbb{Z})$ is fully generated by generators $\sfS_{3\tD}$ and $\sfT_{3\tD}$. 
\bea
\sfS_{3\tD}=\begin{pmatrix}
0 & 0 & 1\\
1 & 0 & 0 \\
0 & 1 & 0
\end{pmatrix},\;\; 
\sfT_{3\tD}=\begin{pmatrix}
1 & 1 & 0\\
0 & 1 & 0\\
0 & 0 & 1
\end{pmatrix}, \;\; 
\sfS_{2\tD}=\begin{pmatrix}
0 & -1 & 0\\
1 & 0 & 0\\
0 & 0 & 1
\end{pmatrix}.  \nonumber 
\eea

\bea
\sfS_{2\tD}=(\sfT_{3\tD}^{-1} \sfS_{3\tD})^3 (\sfS_{3\tD}\sfT_{3\tD})^2  \sfS_{3\tD} \sfT_{3\tD}^{-1}.
\eea

By dimensional reduction (note $\sfT_{2\tD}=\sfT_{3\tD}$), we expect that,
\bea
&&\sfS_{2\tD}^4=(\sfS_{2\tD}\sfT_{3\tD})^6=1,\\ 
&&(\sfS_{2\tD}\sfT_{3\tD})^3=e^{\frac{2\pi \ti}{8} c_-}\sfS_{2\tD}^2=e^{\frac{2\pi \ti}{8} c_-} C.
\eea
$c_-$ carries the information of central charges. 
We can express
\bea
\sfR\equiv \begin{pmatrix} 0& 1& 0\\ -1& 1& 0\\ 0& 0& 1 \end{pmatrix}=(\sfT_{3\tD} \sfS_{3\tD})^2  \sfT_{3\tD}^{-1}  \sfS_{3\tD}^2 \sfT_{3\tD}^{-1} \sfS_{3\tD} \sfT_{3\tD} \sfS_{3\tD}.
\;\;\;\;\;\;\;\;
\eea
One can check that 
\bea
&& \sfS_{3\tD} \sfS_{3\tD}^\dagger=\sfS_{3\tD}^3=\sfR^6=(\sfS_{3\tD} \sfR)^4=(\sfR \sfS_{3\tD})^4=1,\\
&& (\sfS_{3\tD}\sfR^2 )^4=(\sfR^2  \sfS_{3\tD})^4=(\sfS_{3\tD} \sfR^3)^3=(\sfR^3 \sfS_{3\tD})^3=1,\;\;\;\;\;\;\;\;\;\; \\ 
&& (\sfS_{3\tD}\sfR^2 \sfS_{3\tD})^2 \sfR^2 = \sfR^2(\sfS_{3\tD}\sfR^2 \sfS_{3\tD})^2 \text{ (mod 3)}.
\eea
Such expressions are known in the mathematic literature, part of them are listed in Ref.\onlinecite{Coxeter}.

\subsection{Rules for the path integral for the spacetime complex of cocycles} 

For the branching of a spacetime-complex or a simplex,
we define any arrow goes from a small number to a large number, the number ordering is $1<2<3<4<\cdots<0'<1'<2'<{2^*}'<3'<4'<5'<6'<{6^*}'<\cdots$. 
The time evolves along the fourth direction from the left to the right, 
or from a smaller number to a larger number. 
Also we may denote two ways: $[01]. [12] =[02]$ or equivalently $g_{01}. g_{12}=g_{02}$, 
If $[01]=g$ and $[12]=h$, then $[02]=g h$.

\subsection{Explicit expression of $\sfS^{xyz}$ in terms of $(G,\om_4)$}

\begin{figure}[tb] 
\begin{center} 
\includegraphics[scale=0.5]{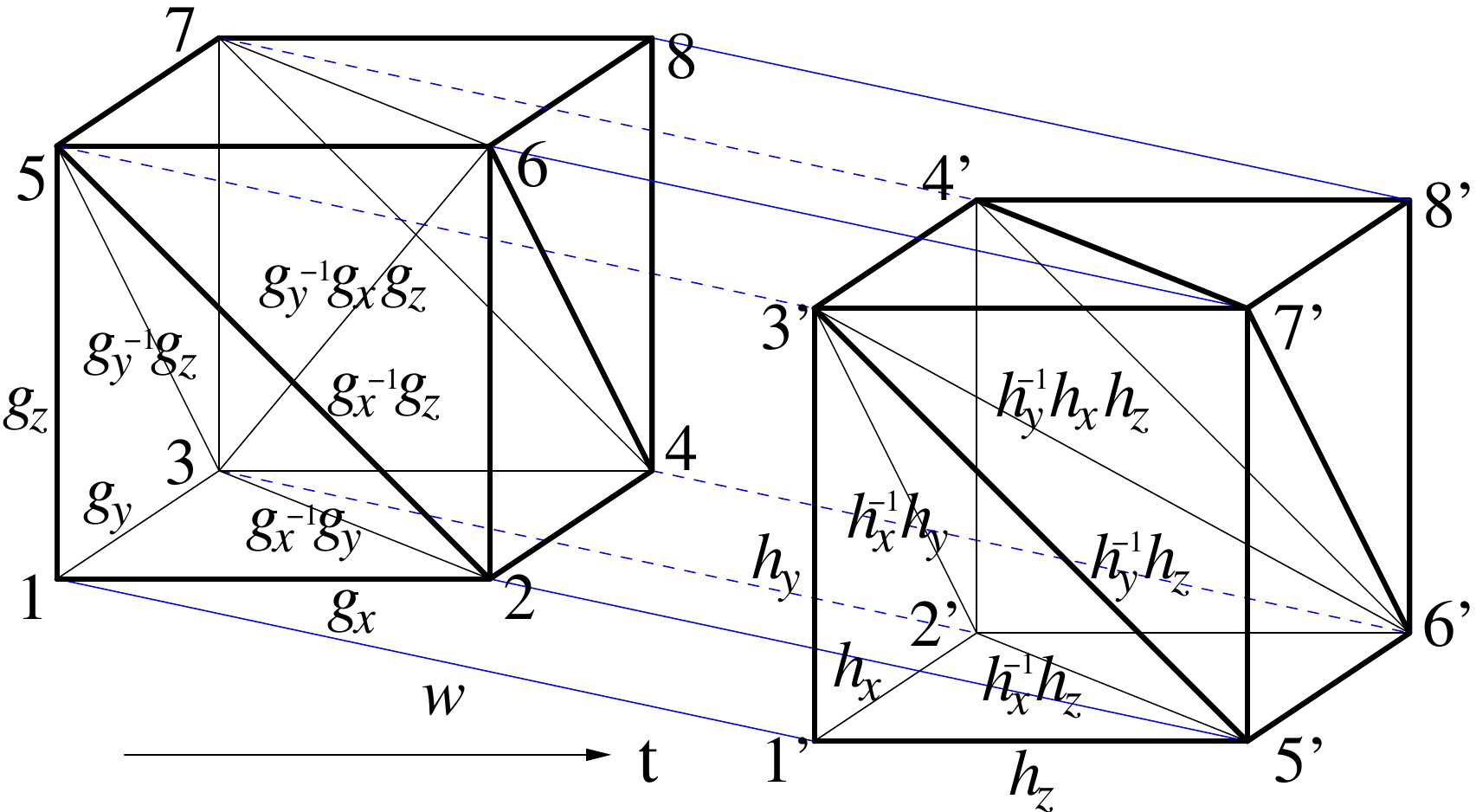} \end{center}
\caption{
The space-time complex $\mathbb{T}^3\times I$, where $I=[0,1]$ is the time direction.
The figure shows $\mathbb{T}^3\times \{0\}$ and $\mathbb{T}^3\times \{1\}$.  The blue lines
illustrate how the two $\mathbb{T}^3$ are connected for $t\in (0,1)$.  Note that the two
$\mathbb{T}^3$'s differ by a rotation $\sfS^{xyz}$. In other words, when time forms a loop, the
two $\mathbb{T}^3$ are glued together by $1\to 1'$, $2\to 2'$, $3\to 3'$, $4\to 4'$,
$5\to 5'$, $6\to 6'$, $7\to 7'$, and $8\to 8'$.
} 
\label{strn0} 
\end{figure}

The $\sfS^{xyz}$-matrix can be computed from the amplitude
$A^{xyz}(g_x,g_x,g_z,h_x,h_y,h_z;\tw)$ of the path integral on spacetime complex
$\mathbb{T}^3\times I$ (see Fig.\ref{strn0}).  Each $\mathbb{T}^3$ is divided into six
tetrahegrons.  The  amplitude $A^{xyz}(g_x,g_x,g_z,h_x,h_y,h_z;\tw)$ is the
product of the four amplitutes $A_i$ for the four shapes $M_i$, $i=1,\cdots,
4$, which are given in Fig.\ref{strn1}--\ref{strn4}).

\begin{figure}[tb] 
\begin{center} 
\includegraphics[scale=0.4]{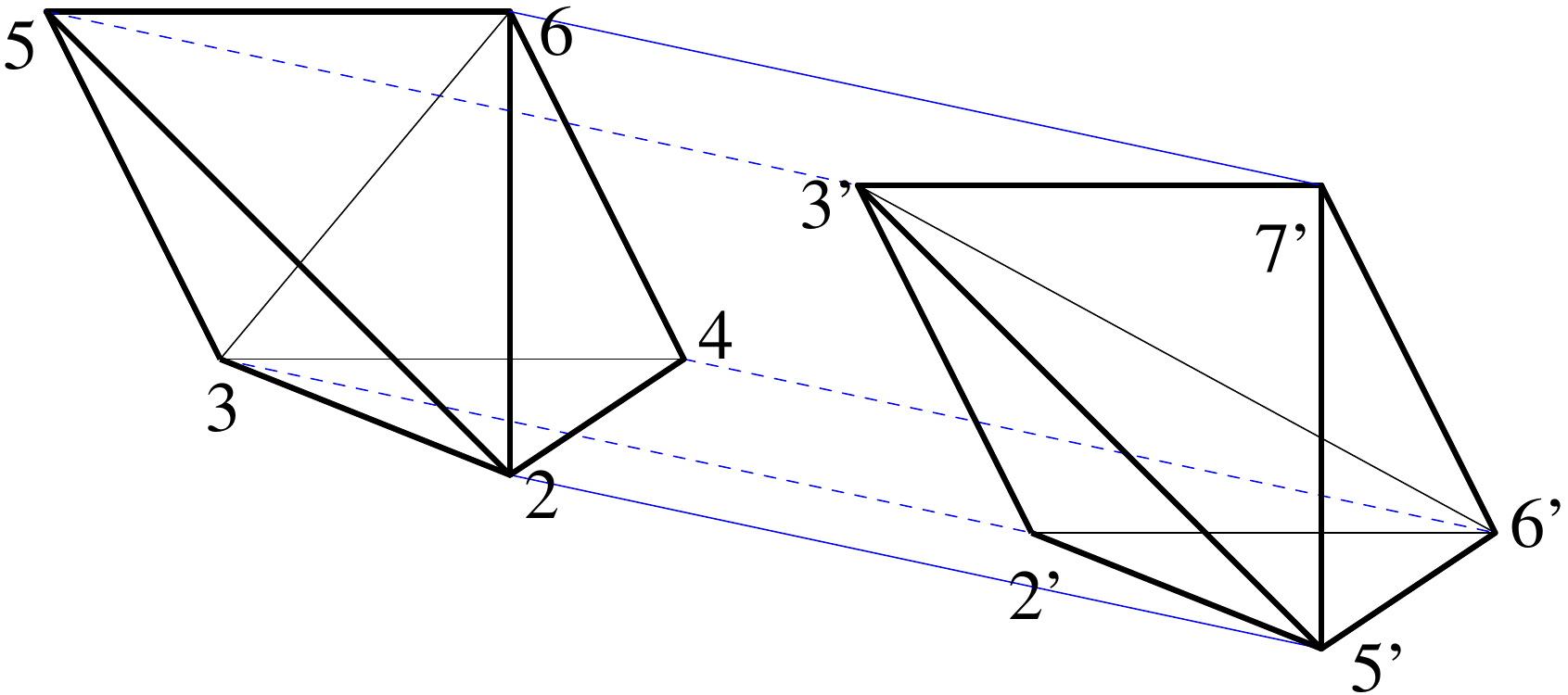} \end{center}
\caption{
The complex $M_1$.
} 
\label{strn1} 
\end{figure}

\begin{figure}[tb] 
\begin{center} 
\includegraphics[scale=0.4]{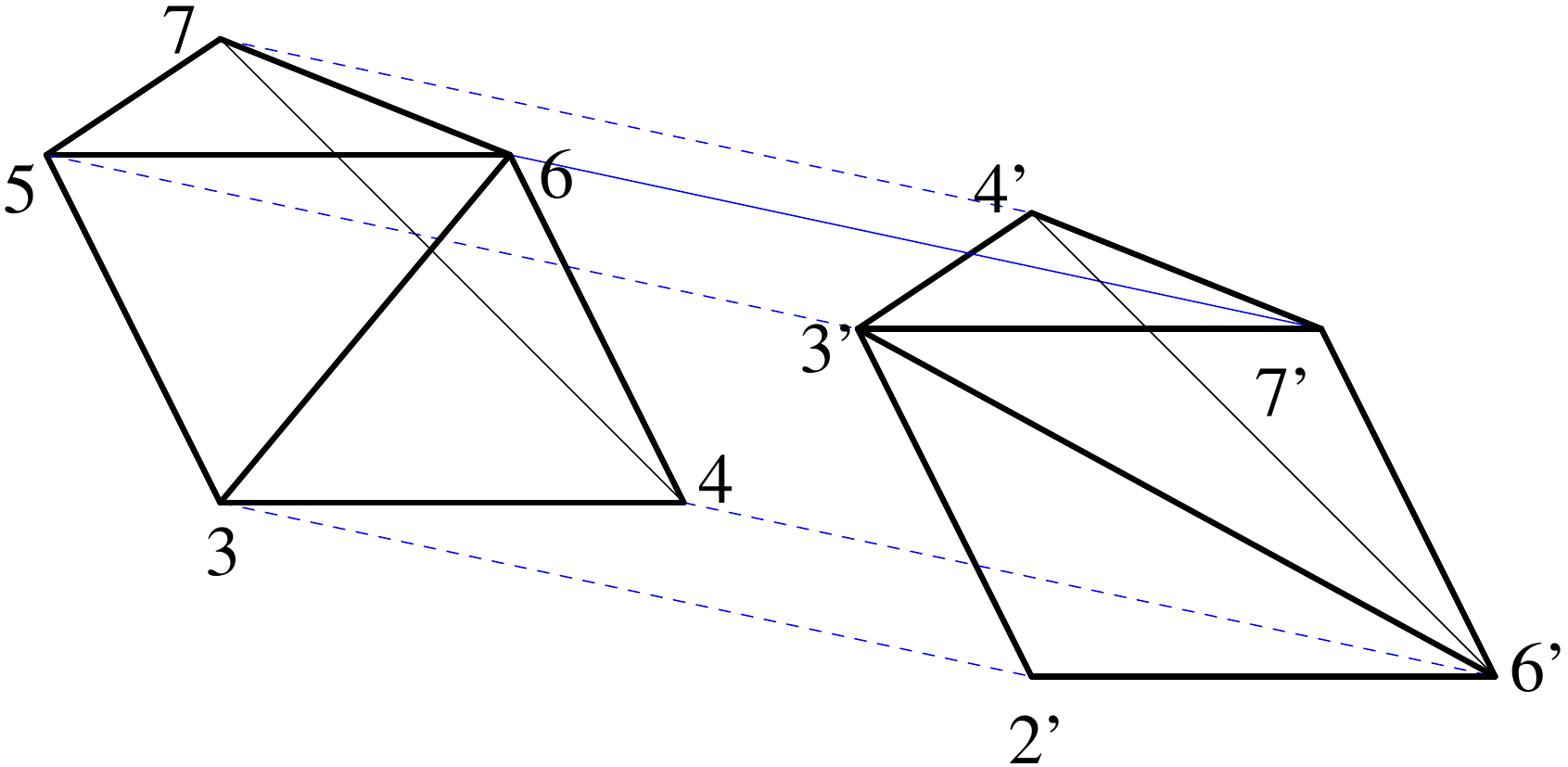} \end{center}
\caption{
The complex $M_2$.
} 
\label{strn2} 
\end{figure}

\begin{figure}[tb] 
\begin{center} 
\includegraphics[scale=0.4]{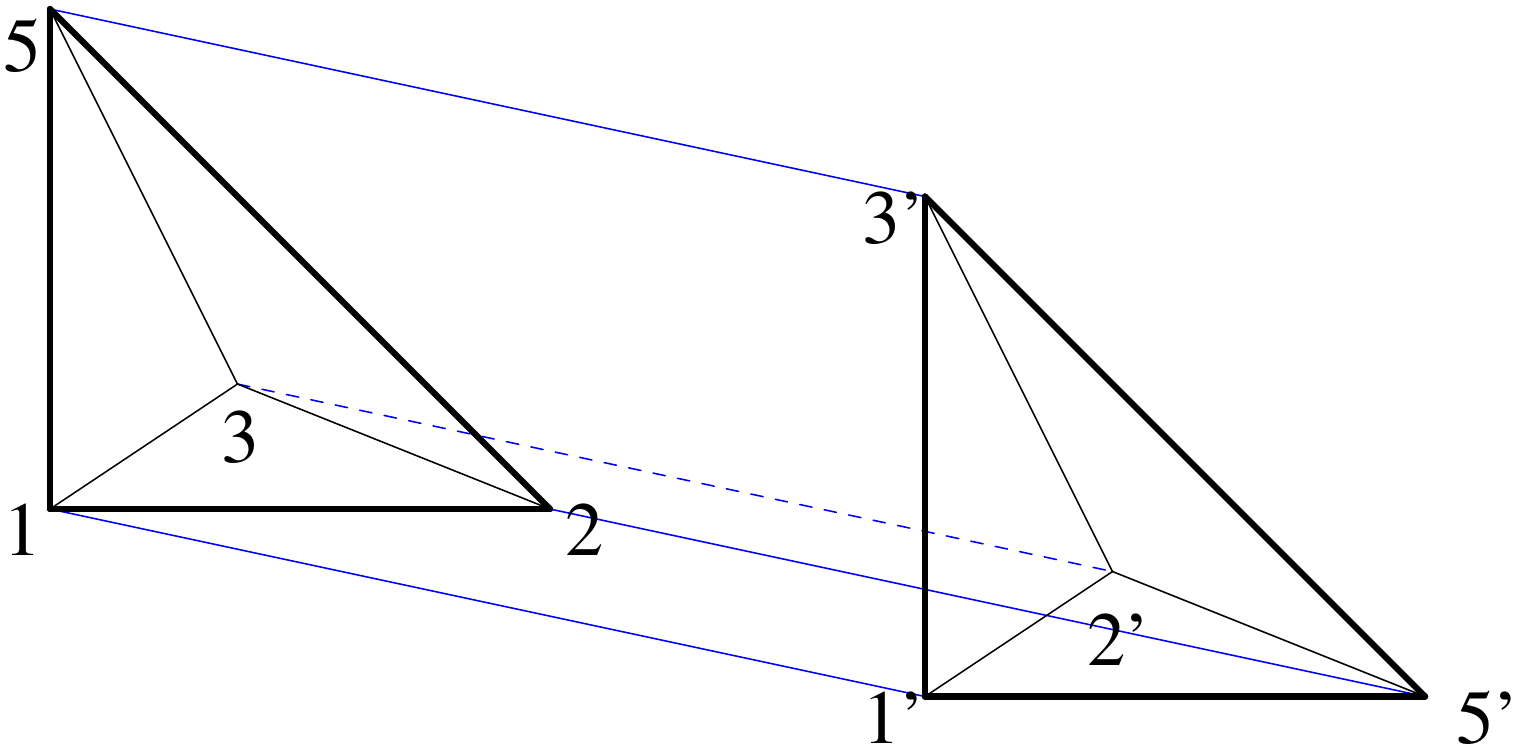} \end{center}
\caption{
The complex $M_3$.
} 
\label{strn3} 
\end{figure}

\begin{figure}[tb] 
\begin{center} 
\includegraphics[scale=0.4]{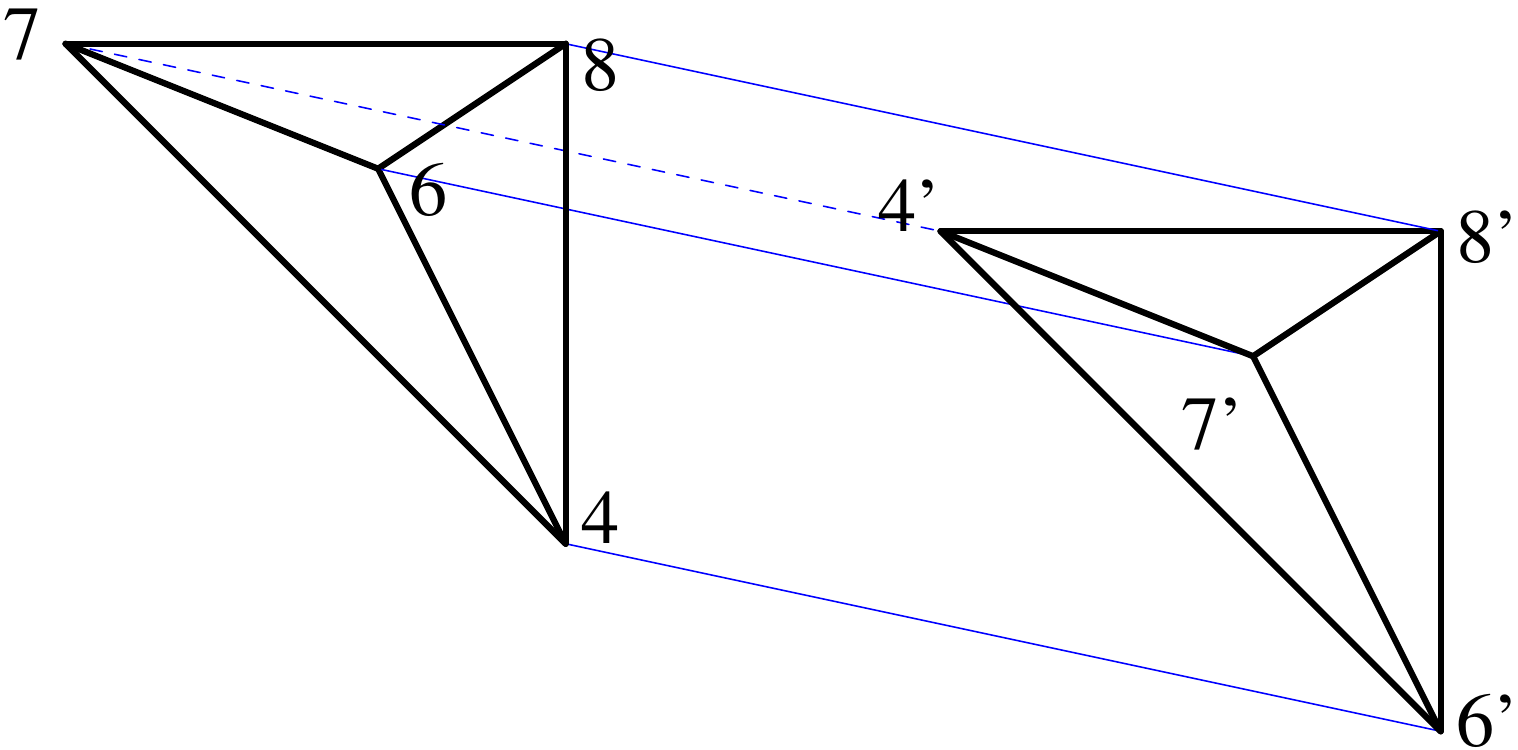} \end{center}
\caption{
The complex $M_4$.
} 
\label{strn4} 
\end{figure}

Each shape $M_i$ can be divided into several 4-simplices.
So the amplitude $A_i$ for each $M_i$ is the product of several cocycles
on the simplices.
We find that, 
for $M_3$:
\begin{align}
 A_3=\frac{
\omega_4(g_{12},g_{23},g_{35},g_{51'}) 
\omega_4^{-1}(g_{35},g_{51'},g_{1'2'},g_{2'5'}) 
}{
\omega_4(g_{23},g_{35},g_{51'},g_{1'5'}) 
\omega_4(g_{51'},g_{1'2'},g_{2'3'},g_{3'5'}) 
}
\end{align}
for $M_4$:
\begin{align}
 A_4=\frac{
\omega_4(g_{67},g_{78},g_{86'},g_{6'7'}) 
\omega_4(g_{84'},g_{4'6'},g_{6'7'},g_{7'8'}) 
}{
\omega_4(g_{46},g_{67},g_{78},g_{86'}) 
\omega_4(g_{78},g_{84'},g_{4'6'},g_{6'7'}) 
}.
\end{align}

\begin{figure}[tb] 
\begin{center} 
\includegraphics[scale=0.4]{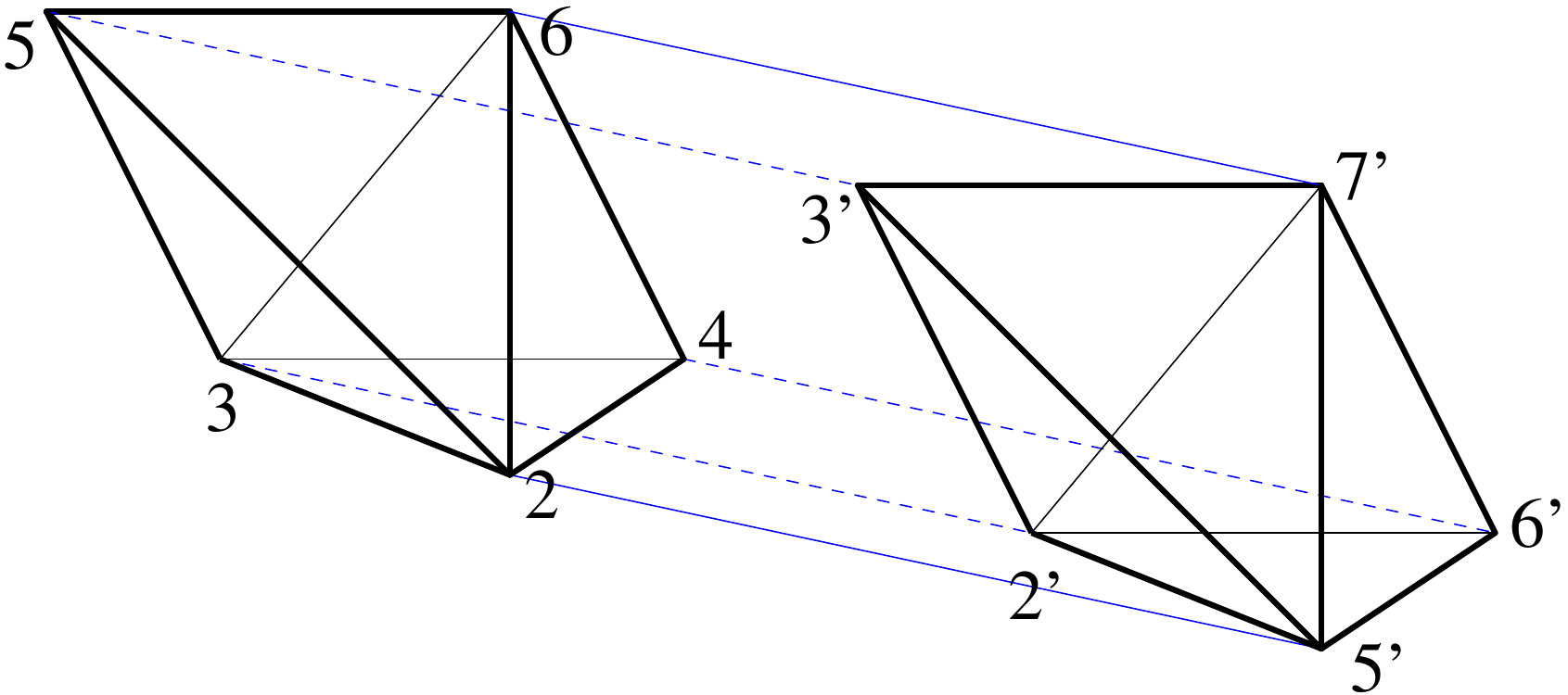} \end{center}
\caption{
The complex $M_1'$.
} 
\label{strn1a} 
\end{figure}

\begin{figure}[tb] 
\begin{center} 
\includegraphics[scale=0.4]{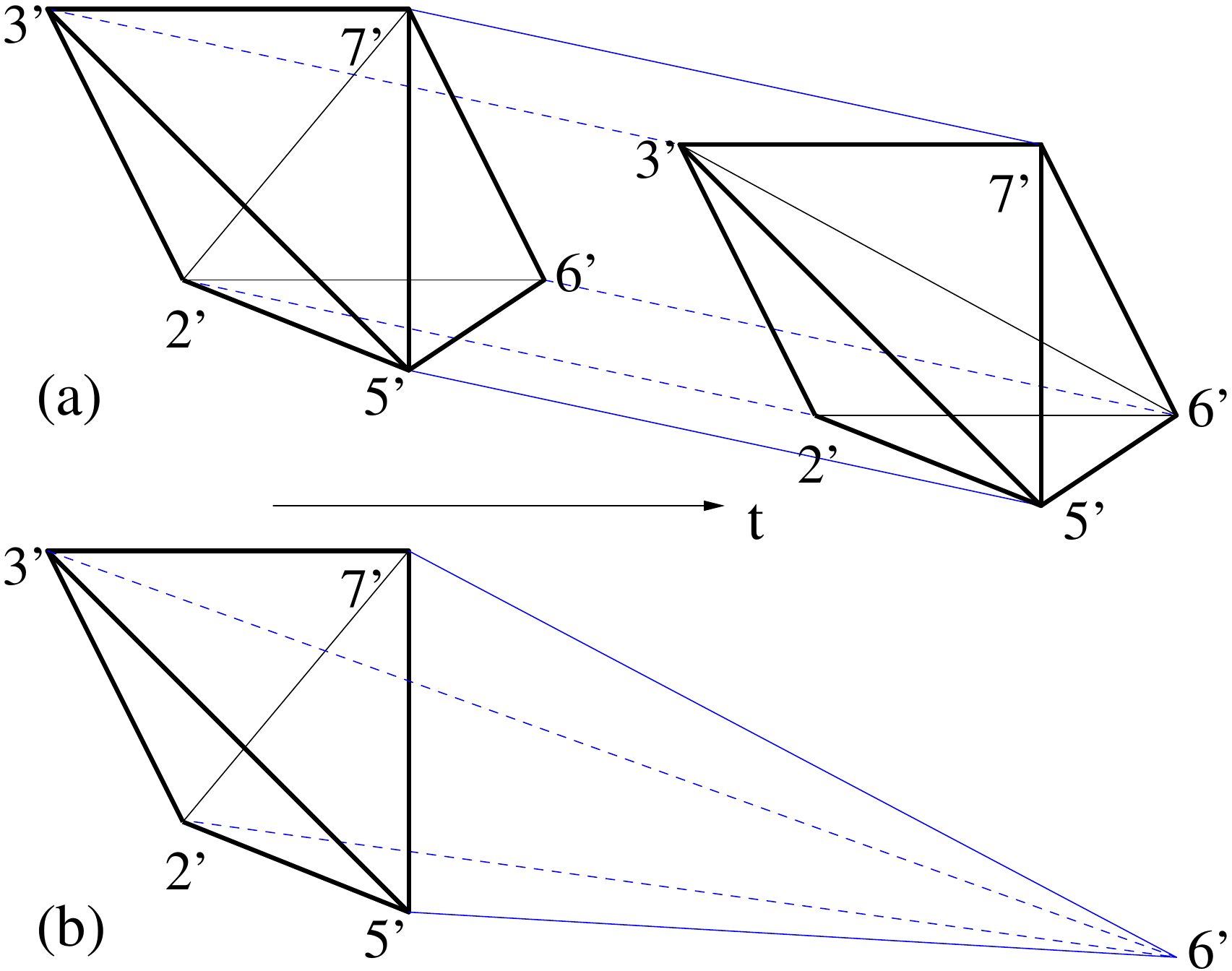} \end{center}
\caption{
The complex $M_1''$, which is formed by one 4-simplex.  Note
that all the vertices in (a) are on the same time slice but the (curved) edge
$(2'7')$ is on an earlier time slice and the (curved) edge $(3'6')$ is on
a later time slice.  To realize this using straight edges, we put the vertex
$6'$ on a later time slice, and this gives us a 4-simplex in (b).
} 
\label{strn1b} 
\end{figure}

\begin{figure}[tb] 
\begin{center} 
\includegraphics[scale=0.18]{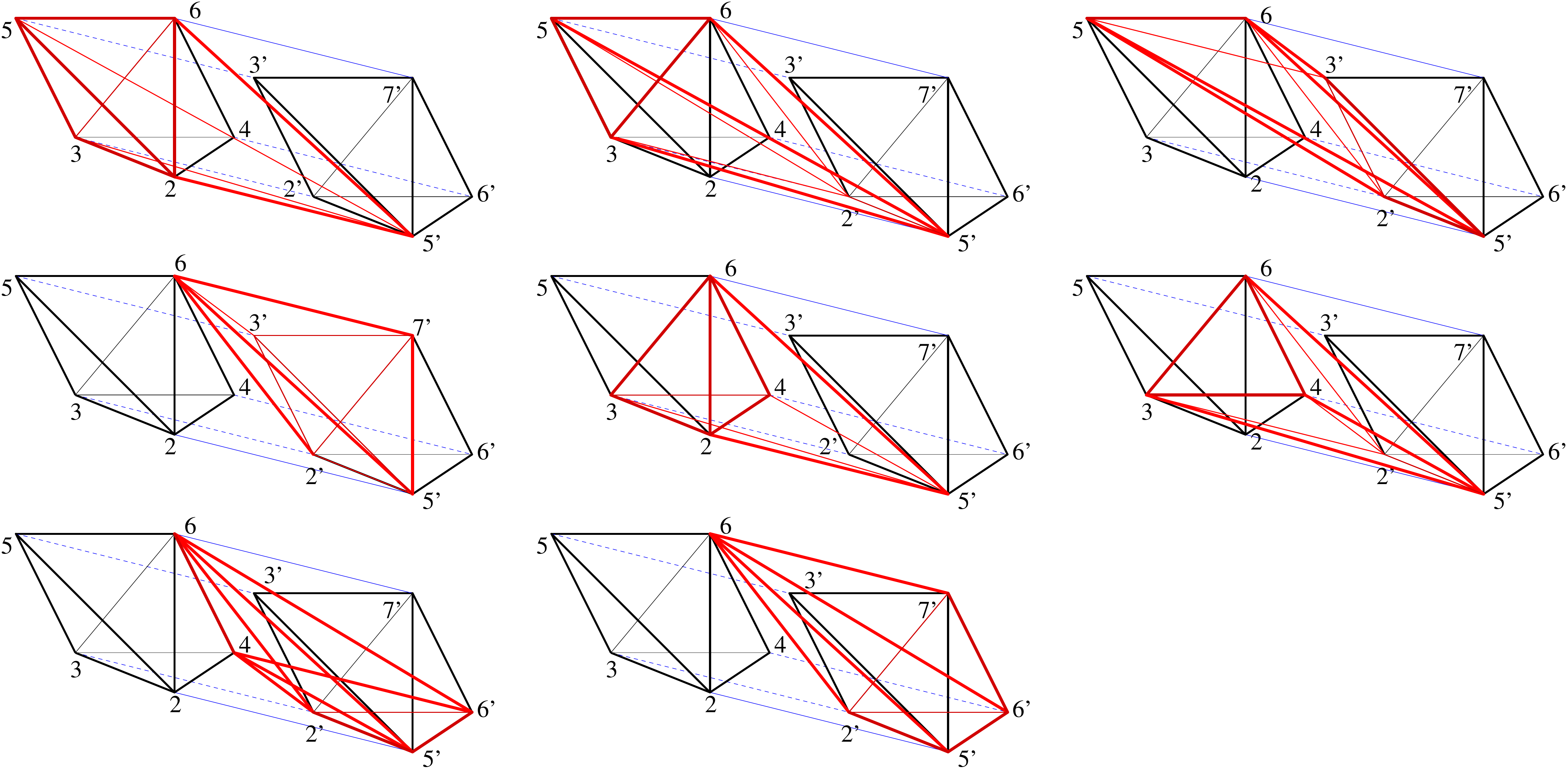} \end{center}
\caption{
The complex $M_1'$ is formed by eight 4-simplices.
} 
\label{strn1_8} 
\end{figure}

\begin{figure}[tb] 
\begin{center} 
\includegraphics[scale=0.4]{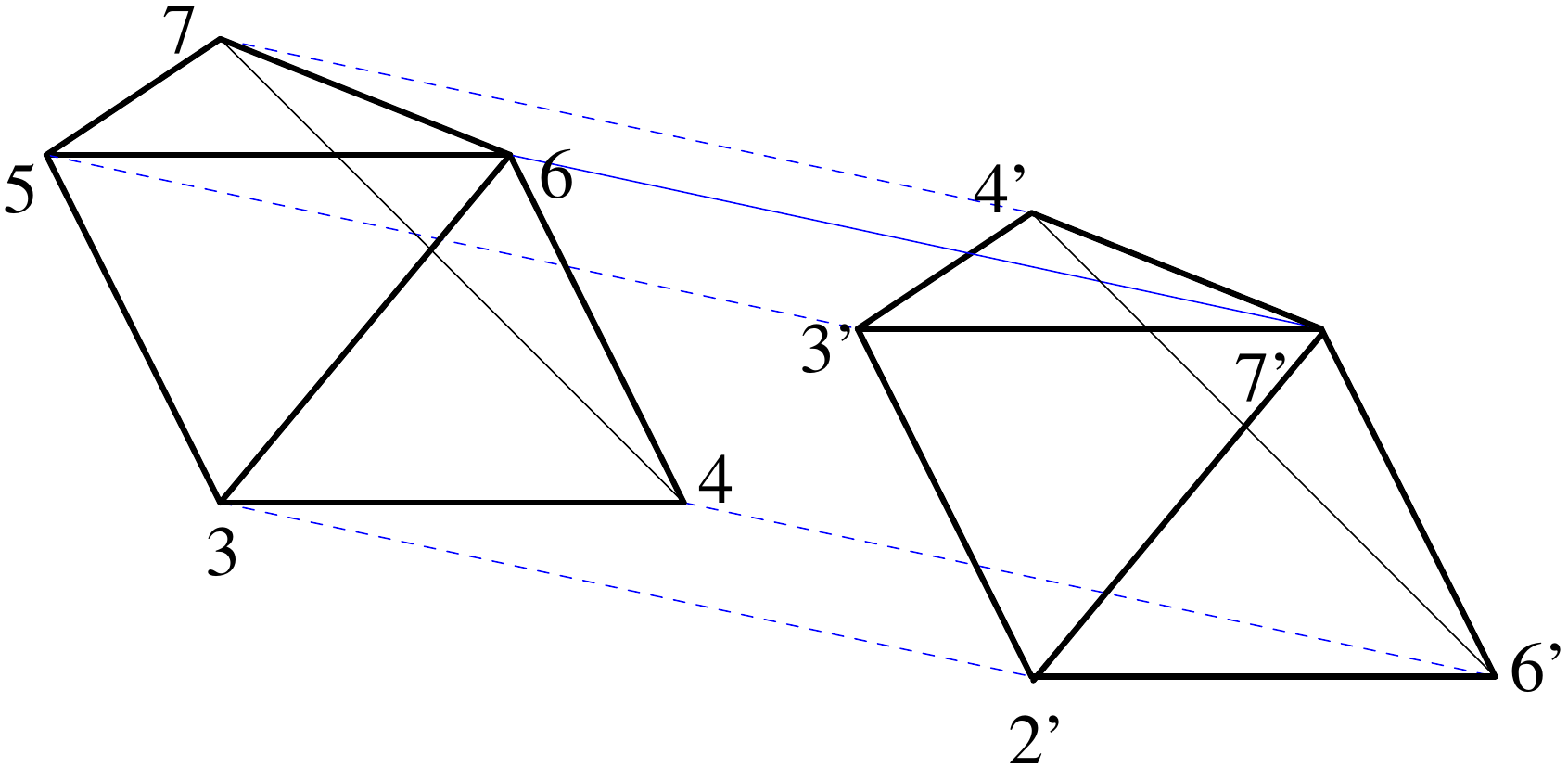} \end{center}
\caption{
The complex $M_2'$, which is formed by eight 4-simplices.
} 
\label{strn2a} 
\end{figure}

\begin{figure}[!h] 
\begin{center} 
\includegraphics[scale=0.4]{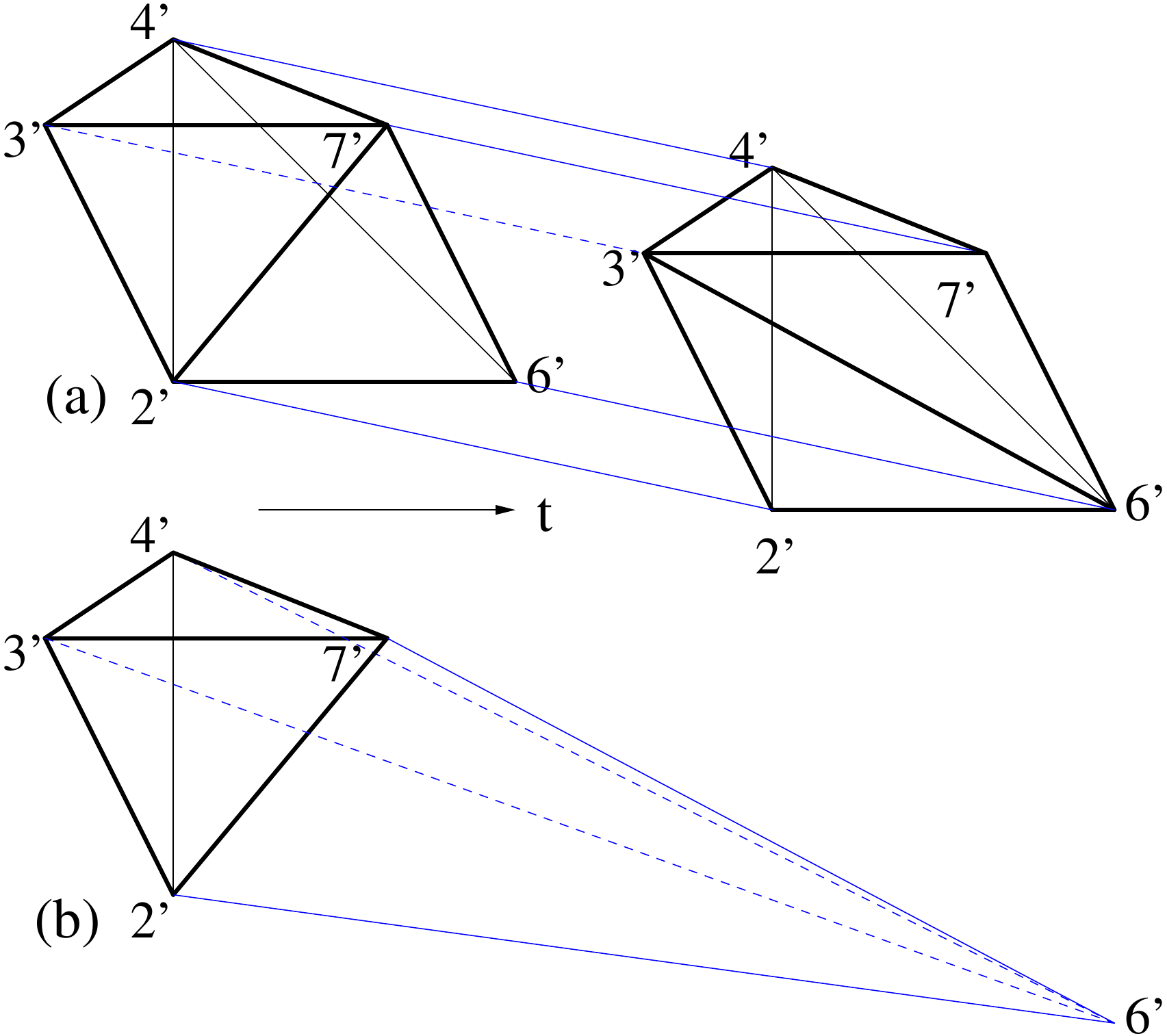} \end{center}
\caption{
The complex $M_2''$, which is formed by one 4-simplex.
Note
that all the vertices in (a) are on the same time slice but the (curved) edge
$(2'7')$ is on an earlier time slice and the (curved) edge $(3'6')$ is on
a later time slice.  To realize this using straight edges, we put the vertex
$6'$ on a later time slice, and this gives us a 4-simplex in (b).
} 
\label{strn2b} 
\end{figure}

To compute the amplitude for $M_1$, we may view $M_1$ and a composition of
$M_1'$ and $M_1''$ (see Fig. \ref{strn1a} and \ref{strn1b}).  The amplitude for
$M_1'$ is
\begin{align}
 A_1'=\frac{
\omega_4(g_{23},g_{35},g_{56},g_{65'}) 
\omega_4(g_{56},g_{62'},g_{2'3'},g_{3'5'}) 
}{
\omega_4^{-1}(g_{35},g_{56},g_{62'},g_{2'5'}) 
\omega_4(g_{62'},g_{2'3'},g_{3'5'},g_{5'7'}) 
} \times
\nonumber\\
\frac{
\omega_4^{-1}(g_{34},g_{46},g_{62'},g_{2'5'}) 
\omega_4^{-1}(g_{62'},g_{2'5'},g_{5'6'},g_{6'7'}) 
}{
\omega_4(g_{23},g_{34},g_{46},g_{65'}) 
\omega_4^{-1}(g_{46},g_{62'},g_{2'5'},g_{5'6'}) 
}.
\end{align}
The above eight cocycles come from eight 4-simplices as illustrated in Fig.
\ref{strn1_8}.  The amplitude for $M_1''$ is
\begin{align}
 A_1''=\omega_4^{-1}(g_{2'3'},g_{3'5'},g_{5'6'},g_{6'7'}).
\end{align}
and the total amplitude for $M_1$ is
\begin{align}
 A_1=A_1'A_1''.
\end{align}

Similarly, for $M_2$, we find that
\begin{align}
 A_2=A_2'A_2'',
\end{align}
where $A_2'$ is the amplitued for $M_2'$ (see Fig. \ref{strn2a})
\begin{align}
 A_2'=\frac{
\omega_4(g_{35},g_{56},g_{67},g_{72'}) 
\omega_4(g_{67},g_{72'},g_{2'3'},g_{3'7'}) 
}{
\omega_4(g_{56},g_{67},g_{72'},g_{2'3'}) 
\omega_4^{-1}(g_{72'},g_{2'3'},g_{3'4'},g_{4'7'}) 
}
\nonumber\\
\frac{
\omega_4(g_{46},g_{67},g_{72'},g_{2'6'}) 
\omega_4(g_{72'},g_{2'4'},g_{4'6'},g_{6'7'}) 
}{
\omega_4(g_{34},g_{46},g_{67},g_{72'}) 
\omega_4(g_{67},g_{72'},g_{2'6'},g_{6'7'}) 
}
\end{align}
and $A_2'$ is the amplitued for $M_2''$ (see Fig. \ref{strn2b})
\begin{align}
 A_2''=\omega_4(g_{2'3'},g_{3'4'},g_{4'6'},g_{6'7'}).
\end{align}

Here $g_{ij}$ is the group element on the edge $(ij)$.
We have
\begin{align}
g_{12} &=g_{34}=g_{56}=g_{78}=g_x,
\nonumber\\
g_{13} &=g_{24}=g_{57}=g_{68}=g_y,
\nonumber\\
g_{15} &=g_{26}=g_{37}=g_{48}=g_z,
\nonumber\\
g_{23}&=g_{67}=g_x^{-1}g_y,\ \ \ 
g_{35}=g_{46}=g_y^{-1}g_z,
\nonumber\\
g_{25}&=g_{47}=g_x^{-1}g_z,\ \ \ 
g_{36}= g_y^{-1}g_xg_z,
\end{align}
\begin{align}
h_{12} &=h_{34}=h_{56}=h_{78}=h_x,
\nonumber\\
h_{13} &=h_{24}=h_{57}=h_{68}=h_y,
\nonumber\\
h_{15} &=h_{26}=h_{37}=h_{48}=h_z,
\nonumber\\
h_{23}&=h_{67}=h_x^{-1} h_y,\ \ \ 
h_{35}=h_{46}=h_y^{-1} h_z,
\nonumber\\
h_{25}&=h_{47}=h_x^{-1} h_z,\ \ \ 
h_{36}= h_y^{-1}h_x h_z.
\end{align}
\begin{align}
 g_{51'}&=g_z^{-1}\tw, \ \ \ \
g_{62'}=g_z^{-1}g_x^{-1}g_y \tw, \ \ \ \
g_{84'}=\tw h_z^{-1},
\nonumber\\
g_{65'}&= g_{72'}= g_{86'}=\tw h_y^{-1}.
\end{align}
Also if the following conditions are not statisfied, the amplitude 
$A^{xyz}(g_x,g_x,g_z,h_x,h_y,h_z;\tw)$ will be zero:
\begin{align}
 g_x \tw&=\tw h_z,
& g_y\tw&=\tw h_x,
& g_z\tw&=\tw h_y,
\nonumber\\
g_x g_y&=g_y g_x,
&g_y g_z&=g_z g_y,
&g_z g_x&=g_x g_z,
\nonumber\\
h_x h_y&=h_y h_x,
&h_y h_z&=h_z h_y,
&h_z h_x&=h_x h_z,
\end{align}
Note the above has $g_x,g_y,g_z$ commute due to the identification on a $\mathbb{T}^3$ torus.

\subsection{Explicit expression of $\sfT^{xy}$ in terms of $(G,\om_4)$}

Similar to $\sfS^{xyz}$, we can triangulate $\sfT^{xy}$ on $\mathbb{T}^3 \times I$.
It is easier to start with a $\sfT^{xy}$ on $\mathbb{T}^2 \times I$ for 2D, which we denote $\sfT_{2\tD}(\tw)$
and triangulate in the following $3!+1=7$ tetrahedra (3-simplex).
Here we have the vertex ordering for the arrows: $1<2<3<4<5<6<7<8<1'<2'<{2^*}'<3'<5'<6'<{6^*}'<7'$.

\begin{widetext}
\bea
&&\sfT_{2\tD}(\tw)=\Ttorus =\TtorusONE \cdot \TtorusTWO  \cdot \TtorusThree \cdot \TtorusFour \cdot \TtorusFive \cdot \TtorusSix \cdot \TtorusSeven
\eea
The last extra piece is required to change the branching structure of the 3-simplex due to $\sfT^{xy}$ transformation.

For $\sfT_{3\tD}(\tw)$, we simply have 7 pieces of slant products. Each slant product contains four 4-simplices. So totally there are 28 pieces of 4-cocycles in $\sfT_{3\tD}(\tw)$.
\bea
&&\sfT_{3\tD}(\tw)=\TIIIDtorusA
=\TIIIDtorus 
= (\sfT_1)(\sfT_2)(\sfT_3)(\sfT_4)(\sfT_5)(\sfT_6)(\sfT_7). \;\;\;\;\;\;\;\;\;\;\;
\eea
\end{widetext}

The constraints given by $\sfT(\tw)$ are
\bea
\tw^{-1} g_x \tw &=& h_x,\\
\tw^{-1} g_x g_y \tw &=& h_y,\\
\tw^{-1} g_z \tw &=& h_z.
\eea
Below we explicitly write down seven $\sfT_i$, where we omit a $\tw$ arrow without drawing it,
which shall connect from the left 3-simplex to the right 3-simplex.
\bea
&&(\sfT_1)=\TtorusONE  \TtorusONETime   \\ 
&& = \omega_4([12],[23],[35],[51'])  \cdot \omega_4([23],[35],[56],[61']) \nonumber \\
&& \cdot \omega_4([35],[56],[67],[71']) \cdot \omega_4^{-1}([56],[67],[71'],[1'5']). \nonumber
\eea

\bea
&&(\sfT_2)=\TtorusTWO  \TtorusTWOTime \\ 
&&=\omega_4^{-1}([23],[36],[61'],[1'2']) 
\cdot \omega_4([36],[67],[71'],[1'2'])  \nonumber  
\\
&& \cdot \omega_4^{-1}([67],[71'],[1'2'],[2'5']) 
\cdot \omega_4([67],[72'],[2'5'],[5'6']). \nonumber 
\eea

\bea
&&(\sfT_3)=\TtorusThree  \TtorusThreeTime  \\ 
&&=\omega_4([37],[71'],[1'2'],[2'{2^*}']) \cdot \omega_4^{-1}([71'],[1'2'],[2'{2^*}'],[{2^*}'5'])  \nonumber\\ 
&&\cdot \omega_4^{-1}([72'],[2'{2^*}'],[{2^*}'5'],[5'6']) \nonumber\\
&& \cdot \omega_4^{-1}([7{2^*}'],[{2^*}'5'],[5'6'],[6'{6^*}']).  \nonumber 
\eea

\bea
&&(\sfT_4)=\TtorusFour  \TtorusFourTime \;\;\\ 
&&=\omega_4^{-1}([23],[34],[46],[62']) \cdot \omega_4^{-1}([34],[46],[67],[72'])\nonumber \\ 
&& \cdot \omega_4^{-1}([46],[67],[78],[82'])    \cdot \omega_4^{}([67],[78],[82'],[2'6']). \nonumber 
\eea

\bea
&&(\sfT_5)=\TtorusFive  \TtorusFiveTime \\ 
&&=\omega_4([34],[47],[72'],[2'{2^*}']) \cdot  \omega_4^{-1}([47],[78],[82'],[2'{2^*}'])\nonumber \\ 
&&\cdot\omega_4([78],[82'],[2'{2^*}'],[{2^*}'6']) \cdot\omega_4^{-1}([78],[8{2^*}'],[{2^*}'6'],[6'{6^*}']). \nonumber 
\eea

\bea
&&(\sfT_6)=\TtorusSix  \TtorusSixTime\\ 
&&=\omega_4^{-1}([48],[82'],[2'{2^*}'],[{2^*}'3'])\cdot \omega_4^{}([82'],[2'{2^*}'],[{2^*}'3'],[3'6'])\nonumber \\ 
&& \cdot \omega_4^{}([8{2^*}'],[{2^*}'3'],[3'6'],[6'{6^*}'])  \cdot \omega_4^{}([83'],[3'6'],[6'{6^*}'],[{6^*}'7']).\;\; \nonumber 
\eea

For the tricky $\sfT_7$, we shift $1'$ to a new later time slice $1''$, and shift $5'$ to a new later time slice $5''$:

\bea
&&(\sfT_7)=\TtorusSeven  \TtorusSevenTime\\ 
&& =\omega_4^{-1}([1'2'],[2'{2^*}'] ,[{2^*}'3'],[3'5'])  \nonumber \\
&& \cdot \omega_4^{}([2'{2^*}'] ,[{2^*}'3'],[3'5'],[5'6']) \nonumber \\
&& \cdot \omega_4^{-1}([{2^*}'3'],[3'5'],[5'6'],[6'{6^*}']) \nonumber\\
&& \cdot \omega_4^{}([3'5'],[5'6'],[6'{6^*}'],[{6^*}'7']). \nonumber 
\eea

One can also define the projection operator on $\mathbb{T}^3$ as
\bea
\sfP_{3\tD}(\tw)= (\sfT_1)(\sfT_2)(\sfT_3)(\sfT_4)(\sfT_5)(\sfT_6). 
\eea

Once obtaining the path integral of 4-cocycles, we can change the flux basis to the canonical basis, and follow the procedure outlined in the Appendix of Ref.\onlinecite{Hu:2012wx}
to derive the Rep theory formula given in our main text Sec.\ref{Rep}. 
An additional remark - an easier way to check the consistency for formulas of $\sfS$, $\sfT$ is to use the rules in Appendix\ref{subsec:SLunit} and to apply 
the discrete Fourier transformation of a finite group such as:
\bea
&&\frac{1}{|G|} \sum_{b,d,\beta}\text{tr}\widetilde{\rho}^{b,d}_{\beta}(a)^{}   \;  \text{tr}\widetilde{\rho}^{b,d}_{\beta}(e)^{*} =\delta_{a,e}, \\
&&  \frac{1}{|G|}  \sum_{a,b,d}
  \text{tr}\widetilde{\rho}^{a,b}_{\alpha}(d)^{*}  \; 
  \text{tr}\widetilde{\rho}^{a,b}_{\gamma}(d)^{} =\delta_{\alpha,\gamma}.
\eea
Use the properties of $\sfC^{(2)}_{a,b}(c,d)$ and the canonical basis $|\alpha, a, b\rangle$,
we can justify that our formulas satisfy the rules (up to some projective representation's complex phases).
See also Ref.\onlinecite{Wan:2014woa} for the derivation.

\bibliography{../../bib/wencross,../../bib/all,../../bib/publst,./local}

\end{document}